\documentclass{amsart}
\usepackage{amssymb}
\usepackage{amsfonts}

\newtheorem{theorem}{Theorem}
\theoremstyle{plain}

\newtheorem{corollary}{Corollary}

\newtheorem{definition}{Definition}

\newtheorem{remark}{Remark}

\numberwithin{equation}{section}
\input{tcilatex}

\begin{document}
\def\UrlFont{\rmfamily}

\title[Eventum Mechanics and Quantum Predictions]{Eventum Mechanics of
Quantum Trajectories:\\
Continual Measurements, Quantum Predictions and Feedback Control}
\author{Viacheslav P Belavkin}
\address{School of Mathematical Sciences\\
University of Nottingham, UK}
\email[A.~One]{vpb@maths.nottingham.ac.uk}
\urladdr{http://www.maths.nott.ac.uk/personal/vpb/}
\thanks{The support from the EC under the programme ATESIT (contract no
IST-2000-29681) and QBIC programme of Tokyo Science University where it was
completed is acknowledged.}
\date{February 7, 2007}
\dedicatory{}
\thanks{This paper is in final form and no version of it will be submitted
for publication elsewhere.}

\begin{abstract}
Quantum mechanical systems exhibit an inherently probabilistic nature upon
measurement which excludes in principle the singular direct observability
continual\ case. Quantum theory of time continuous measurements and quantum
prediction theory, developed by the author on the basis of an
independent-increment model for quantum noise and nondemolition causality
principle in the 80's, solves this problem allowing continual quantum
predictions and reducing many quantum information problems like problems of
quantum feedback control to the classical stochastic ones. Using explicit
indirect observation models for diffusive and counting measurements we
derive quantum filtering (prediction) equations to describe the stochastic
evolution of the open quantum system under the continuous partial
observation. Working in parallel with classical indeterministic control
theory, we show the Markov Bellman equations for optimal feedback control of
the \textit{a posteriori} stochastic quantum states conditioned upon these
two kinds of measurements. The resulting filtering and Bellman equation for
the diffusive observation is then applied to the explicitly solvable quantum
linear-quadratic-Gaussian (LQG) problem which emphasizes many similarities
and differences with the corresponding classical nonlinear filtering and
control problems and demonstrates microduality between quantum filtering and
classical control.
\end{abstract}

\maketitle
\tableofcontents

\section{Introduction}

With technological advances now allowing the possibility of continuous
monitoring and rapid manipulations of systems at the quantum level \cite%
{AASDM02,GSDM03}, there is an increasing awareness of the importance of
quantum feedback control in applications of quantum information such as the
dynamical problems of quantum error corrections and quantum computations.
The theory of quantum feedback control based upon a classical stochastic
process formally developed by the author in the 80's \cite{Bel83,Be85,Bel88}
has been recently applied in many contexts including state preparation \cite%
{Bel99,DJJ01,BEB05}, purification \cite{Jac03,Jac04}, risk-sensitive control 
\cite{Jam04,Jam05} and quantum error correction \cite{ADL02,GrW04}. It has
also been studied from the practical point of view of stability theory \cite%
{HSM05} which contains a useful introduction to quantum probability and
along with \cite{DHJMT00} gives a comprehensive discussion on the comparison
of classical and quantum control techniques.

The main ingredients of quantum control are essentially the same as in the
classical case. One controls the system by coupling to an external control
field which modifies the system in a desirable manner. The desired
objectives of the control can be encoded into a \emph{cost function} along
with any other stipulations or restrictions on the controls such that the
minimization of this cost indicates optimality of the control process. We
restrict ourselves to the more interesting case of quantum \emph{closed-loop}
or \emph{feedback} control based on the indirect continuous in time
observations, the quantum stochastic theory of which was initiated by
Belavkin in a series of increasing generality papers \cite%
{Be78,Bel79,Be80,Be87a,Bel88} as a quantum analogy to the classical
stochastic prediction theory which is based upon the nonlinear
(Stratonovich) filtering equations. This work was developed then in the
beginning of 90's \cite{Be90a,Be90b,Be91d,Bel92b,BeS92,Be93d,Bel94}, and
this development laid down the foundations for Eventum Mechanics, a new
quantum stochastic mechanics of continual observations giving a microscopic
theory of continuous reductions \cite{Bel92a,Be95b} and spontaneous
localizations \cite{Be95,Be_Mel96} for quantum diffusions \cite{Bel89,Be90},
quantum jumps \cite{Be89a,Be_Sta90}, and other mixed stochastic quantum
trajectories \cite{Be90e,BaB91,Be94e}.

In order to demonstrate the power of this new event enhanced quantum
mechanics, it was also applied right from the beginning \cite%
{Bel83,Be87,Bel88} to solve the typical problems of quantum feedback control
in parallel to the work on classical stochastic control with partial
observations first introduced by Stratonovich \cite{Str62,Str68} and
Mortensen \cite{Mor66}. Thus, the problem of optimal quantum feedback
control was separated into quantum filtering which provides optimal
estimates of the quantum state variables (density operators) and then a
classical optimal control problem based on the output of the quantum filter.
The classical noise which is filtered out by passing from the prior to the
posterior quantum states comes from the irreducible disturbance to the
quantum system during the observation (due to the interaction with
measurement apparatus). Unlike in classical case, this is an unavoidable
feature of quantum measurement since the state of an individual quantum
system is not directly observable. However, the lack of urgency for such a
theory and the novelty of the mathematical language at the time left this
work relatively undiscovered only to be rediscovered recently in the physics
and engineering community.

The purpose of this paper is to build on the original work of the author and
present an accessible account of the theory of quantum continual
measurements, quantum causality and predictions and optimal quantum feedback
control. Firstly we introduce the necessary concepts and mathematical tools
from modern quantum theory including quantum probability, continuous causal
(non-demolition) measurements, quantum stochastic calculus and quantum
filtering. Next the quantum Bellman equations for optimal feedback control
with diffusive and counting measurement schemes are informally derived. The
latter results were first stated in \cite{Bel88} without derivation and a
consideration for the diffusive case was recently given in \cite{GBS05}. We
conclude with an application of these results to the multi-dimensional
quantum Linear-Quadratic-Gaussian (LQG) problem and a discussion on the
comparison with the corresponding classical results. However we first start
from a model example of LQG quantum filtering and feedback control problem
which is important since it is one of the few exactly solvable control
problems which emphasizes the similarities between the corresponding
classical and quantum filtering and control theories. It allows us to set up
notations and clearly demonstrates not only the similarity but also the
difference of classical and quantum feedback control theories which can be
observed in \emph{microduality principle}, a more elaborated duality between
quantum linear Gaussian filtering and classical linear optimal control.

\subsection{Model example: quantum free particle}

The quantum linear filtering and optimal quadratic control problem with
quantum Gaussian noise was first studied and resolved by the author in a
series of quantum measurement and filtering papers \cite{Be78},\cite{Bel80},%
\cite{Bel92a} and based on this quantum feedback control papers \cite{Bel79},%
\cite{Be87a},\cite{Bel88}. The simplest example of a single quantum Gaussian
oscillator matched with a transmission line \cite{Bel80} as complex
one-dimensional channel was taken as a quantum feedback model in the
starting preprint \cite{Bel79}, eventually published in \cite{Bel99}.
However a more similar to the classical case quantum linear models require
at least two real dimensions instead of single complex, and we may now use
the multidimensional quantum LQG control solutions derived in the last
Section of this paper for application on higher dimensional systems which do
not have such complex representation. The optimal control of a continuously
observed quantum free particle with quadratic cost is the simplest such
example.

Let $\check{x}_{\bullet }=\left( \check{x}_{1},\check{x}_{2}\right) $ be a
pair of phase space operators $\check{x}_{1}=\check{q}$, $\check{x}_{2}=%
\check{p}$ for a quantum particle in one dimension, given by selfadjoint
operators of position $\check{q}$ and momentum $\check{p}$ satisfying the
canonical commutation relation (CCR)%
\begin{equation}
\left[ \check{q},\check{p}\right] :=\check{q}\check{p}-\check{p}\check{q}=%
\mathrm{i}\hbar \check{1}.  \label{eq QPccr}
\end{equation}%
Here $\check{1}$ is the identity operator in a Hilbert space $\mathfrak{h}$
of the CCR representation (\ref{eq QPccr}) and $\hbar $ is called Planck
constant, which could be for our purpose any positive constant $\hbar >0$.
Let us denote the row of initial expectations $\left\langle \check{x}%
_{j}\right\rangle $ of $\check{x}_{j}$ in a quantum Gaussian state by $%
x_{\bullet }=\left( q,p\right) $, and also denote the initial dispersions of 
$\check{q}$ and $\check{p}$ by $\sigma _{q}$ and $\sigma _{p}$ respectively
and the initial symmetric covariance $\func{Re}\left\langle \check{q}\check{p%
}\right\rangle -qp$ by $\sigma _{qp}=\sigma _{pq}$. The Hamiltonian $\check{p%
}^{2}/2\mu $ of free particle is perturbed by a controlling force using the
linear potential $\phi (t,\check{q})=\beta u\left( t\right) \check{q}$ with $%
u\left( t\right) \in \mathbb{R}$ as $H(u)=\check{p}^{2}/2\mu +\beta u\check{q%
}$ where $\mu >0$ is the mass of the particle. The particle is assumed to be
coupled not only to control which can be realized by a quantum coherent
(forward) channel, but also to a coherent observation (estimation) quantum
channel such that its open Heisenberg dynamics is described by quantum
Langevin equations as a case of (\ref{eq Langevin}): 
\begin{eqnarray}
\mathrm{d}Q\left( t\right) +\lambda Q\left( t\right) \mathrm{d}t &=&\frac{1}{%
\mu }P\left( t\right) \mathrm{d}t+\mathrm{d}W_{q}^{t},\ Q\left( 0\right) =%
\check{q}  \label{eq FPlangQ} \\
\mathrm{d}P\left( t\right) +\lambda P\left( t\right) \mathrm{d}t &=&\mathrm{d%
}V_{p}^{t}-\beta u\left( t\right) \mathrm{d}t,\;\;P\left( 0\right) =\check{p}
\label{eq FPlangP}
\end{eqnarray}%
Here $\lambda =\frac{1}{2}\left( \alpha \varepsilon +\beta \gamma \right) $
and $V_{q}^{t}=\alpha V_{e}^{t}+\beta V_{f}^{t}$, $W_{q}^{t}=-\varepsilon
W_{e}^{t}-\gamma W_{f}^{t}$ are given by two independent pairs $\left(
V_{\circ },W_{\circ }\right) $ (where $\circ =e,f$ stands for error and
force) of Wiener noises $V_{\circ }=\hbar \Im \left( A_{\circ }^{+}\right) $%
, $W_{\circ }=2\Re \left( A_{\circ }^{+}\right) $ due to the interaction
with the coupled estimation and feedback channels. Note that these noises do
not commute,%
\begin{equation}
W_{p}^{s}V_{q}^{r}-V_{q}^{r}W_{p}^{s}=\left( r\wedge s\right) \mathrm{i}%
\hbar \lambda I,
\end{equation}%
if $\lambda \neq 0$, which is necessary and sufficient condition for
preservation of the CCR (\ref{eq QPccr}) by the system (\ref{eq FPlangQ}), (%
\ref{eq FPlangP}). It can be easily found by substituting the solution%
\begin{equation*}
P\left( t\right) =\mathrm{e}^{-\lambda t}\check{p}+\int_{0}^{t}\mathrm{e}%
^{\left( s-t\right) \lambda }\left( \mathrm{d}V_{p}^{s}-\beta u\left(
s\right) \mathrm{d}s\right) 
\end{equation*}%
of the second equation (\ref{eq FPlangP}) into the first (\ref{eq FPlangQ})
that%
\begin{equation*}
\left[ Q\left( r\right) ,Q\left( s\right) \right] =\frac{\mathrm{i}\hbar }{%
\mu }\left\vert r-s\right\vert \mathrm{e}^{-\lambda \left\vert
r-s\right\vert }\neq 0\text{.}
\end{equation*}%
Therefore the family $\left\{ Q\left( t\right) \right\} $ is incompatible,
cannot be represented as a classical stochastic process and directly
observed. However it can be indirectly observed by continuous measuring of
the coupling operator $\alpha \check{q}$ with an error white noise in the
estimation channel as it was suggested in \cite{Bel88},\cite{BeS92}. To this
end we measure $W_{e}^{t}=2\Re \left( A_{e}^{+}\right) $ as an input process
evolved after an interaction with the particle into an output classical
process given by a commutative family $\left[ Y_{e}^{t}:t>0\right] $ in the
linear estimation channel 
\begin{equation}
\mathrm{d}Y_{e}^{t}=\alpha Q\left( t\right) \mathrm{d}t+\mathrm{d}W_{e}^{t},
\label{eq YforQ}
\end{equation}%
in which the input process appears as measurement error noise with
commutative independent increments $\mathrm{d}W_{e}^{t}$ representing the
standard Wiener process such that $\left( \mathrm{d}W_{e}\right) ^{2}=%
\mathrm{d}t$, but noncommuting with the perturbative force $V_{p}^{t}$
since, as it is explained in the last Section,%
\begin{equation}
\mathrm{d}W_{e}^{t}\mathrm{d}V_{p}^{t}=\frac{\alpha \hbar }{2\mathrm{i}}%
\mathrm{d}t,\;\mathrm{d}W_{e}^{t}\mathrm{d}W_{q}^{t}=-\varepsilon \mathrm{d}%
t.
\end{equation}%
Thus the measurement error noise $W_{e}^{t}$ satisfies the
error-perturbation CCR%
\begin{equation}
\left[ V_{p}^{r},W_{e}^{s}\right] =\left( r\wedge s\right) \mathrm{i}\hbar
\alpha I,
\end{equation}%
which is necessary and sufficient condition of quantum causality (or quantum
nondemolition condition) in the form%
\begin{equation}
\left[ Y_{e}^{r},Q\left( s\right) \right] =0=\left[ Y_{e}^{r},P\left(
s\right) \right] \;\forall r\leq s
\end{equation}%
requiring the statistical predictability of \emph{quantum hidden in the
future trajectories} $\left\{ X_{\bullet }\left( s\right) :s\geq t\right\} $
with respect to the \emph{classical observed in the past trajectories} $%
\left\{ Y_{e}^{r}:r\leq t\right\} $ for each $t$. From this we derive the
Heisenberg \emph{error-perturbation uncertainty principle} in the precise 
\emph{Belavkin inequality} form \cite{Bel80},\cite{Bel92a}%
\begin{equation}
\left( \mathrm{d}V_{p}^{t}\right) ^{2}\geq \left( \frac{\alpha \hbar }{2}%
\right) ^{2}\mathrm{d}t,\;\;\left( \mathrm{d}W_{e}^{t}\right) ^{2}=\mathrm{d}%
t
\end{equation}%
in terms of the perturbation $V_{p}^{t}$ in (\ref{eq FPlangP}) and standard
error $W_{e}^{t}$ in (\ref{eq YforQ}). Thus we have the case 
\begin{equation}
\mathbf{J}=\left( 
\begin{array}{cc}
0 & 1 \\ 
-1 & 0%
\end{array}%
\right) ,\;\mathbf{\Lambda }=\left( 
\begin{array}{cc}
\frac{\alpha }{2} & \frac{\mathrm{i}\varepsilon }{\hbar } \\ 
\frac{\beta }{2} & \frac{\mathrm{i}\gamma }{\hbar }%
\end{array}%
\right) 
\end{equation}%
of the general quantum linear open system considered in the last Section,
where $\mathbf{\Lambda }$ is the direct sum $\mathbf{\Lambda }_{e}\oplus 
\mathbf{\Lambda }_{f}$ of two rows $\mathbf{\Lambda }_{e},\mathbf{\Lambda }%
_{f}$ corresponding to $\mathbf{B}_{e}=\left( \alpha ,0\right) $, $\mathbf{E}%
=\left( 0,\varepsilon \right) $, $\mathbf{B}=\left( \beta ,0\right) $, $%
\mathbf{E}_{f}=\left( 0,\gamma \right) $. From this we compute the matrices (%
\ref{eq G}) and (\ref{eq H}) satisfying the microduality principle, which
are turned to be diagonal, 
\begin{equation*}
\mathbf{G=}\left( 
\begin{array}{cc}
\zeta _{q} & 0 \\ 
0 & \zeta _{p}%
\end{array}%
\right) ,\;\;\mathbf{H}=\left( 
\begin{array}{cc}
\eta _{q} & 0 \\ 
0 & \eta _{p}%
\end{array}%
\right) ,
\end{equation*}%
with eigenvalues $\zeta _{q}=\gamma ^{2}$, $\zeta _{p}=\left( \hbar
/2\right) ^{2}\left( \alpha ^{2}+\beta ^{2}\right) =\eta _{q}$, $\eta
_{p}=\varepsilon ^{2}$.

\subsection{Quantum feedback control example}

We can now apply the results obtained in the last Section to demonstrate
optimal quantum filtering and optimal feedback control and their
microduality on this model example. The optimal estimates of the position
and momentum based on a nondemolition observation of free quantum particle
via the continuous measurement of $Y_{t}$, originally derived in \cite{Bel88}%
,\cite{BeS92} in absence of control channel, are then given by the Belavkin
Kalman filter (\ref{eq qKalman}) in the form of linear stochastic equation 
\begin{eqnarray}
\mathrm{d}\hat{q}_{0}^{t}+\hat{q}_{0}^{t}\lambda \mathrm{d}t &=&\frac{1}{\mu 
}\hat{p}_{0}^{t}\mathrm{d}t+\left( \alpha \sigma _{q}\left( t\right)
-\varepsilon \right) \mathrm{d}\hat{W}_{e}^{t} \\
\mathrm{d}\hat{p}_{0}^{t}+\hat{p}_{0}^{t}\lambda \mathrm{d}t &=&\beta
u\left( t\right) \mathrm{d}t+\alpha \sigma _{qp}\left( t\right) \mathrm{d}%
\hat{W}_{e}^{t},
\end{eqnarray}%
where the estimation innovation process $\hat{W}_{e}^{t}$ describes the gain
of information due to measurement of $Y_{e}^{t}$ given by 
\begin{equation}
\mathrm{d}\hat{W}_{e}^{t}=\mathrm{d}Y_{e}^{t}-\alpha \hat{q}_{0}^{t}\mathrm{d%
}t,
\end{equation}%
and the error covariances satisfy the Riccati equations 
\begin{equation}
\begin{array}{lll}
\frac{\mathrm{d}}{\mathrm{d}t}\sigma _{q} & = & \zeta _{q}+2\left( \frac{1}{%
\mu }\sigma _{qp}+\sigma _{q}\delta \right) -(\alpha \sigma _{q})^{2} \\ 
\frac{\mathrm{d}}{\mathrm{d}t}\sigma _{qp} & = & \frac{1}{\mu }\sigma
_{p}-\left( \lambda -\delta \right) \sigma _{qp}-\alpha ^{2}\sigma
_{q}\sigma _{qp} \\ 
\frac{\mathrm{d}}{\mathrm{d}t}\sigma _{p} & = & \zeta _{p}-2\lambda \sigma
_{p}-(\alpha \sigma _{p})^{2},%
\end{array}
\label{eq FPFiltRicatti}
\end{equation}%
where we denote $\delta =\frac{1}{2}\left( \alpha \varepsilon -\gamma \beta
\right) $, with initial conditions 
\begin{equation*}
\sigma _{q}\left( 0\right) =\sigma _{q},\quad \sigma _{qp}\left( 0\right)
=\sigma _{qp},\quad \sigma _{p}\left( 0\right) =\sigma _{p}.
\end{equation*}

The Riccati equations for the error covariance in the filtered free particle
dynamics have an exact solution \cite{BeS92} with profound implications for
the ultimate quantum limit satisfying the Heisenberg uncertainty relations
for the accuracy of optimal quantum state estimation via the continuous
indirect quantum particle coordinate measurement.

The dual optimal control problem can be found by identifying the
corresponding dual matrices from the table (\ref{tb Duality}) which give the
quadratic control parameters 
\begin{eqnarray}
\check{c}(u) &=&\left( u-\check{z}\right) ^{2}+\eta _{q}\check{q}^{2}+\eta
_{p}\check{p}^{2}, \\
\check{s} &=&\omega _{q}\check{q}^{2}+\omega _{qp}(\check{p}\check{q}+\check{%
q}\check{p})+\omega _{p}\check{p}^{2}  \notag
\end{eqnarray}%
corresponding to the dual output process given by $\check{z}=\gamma \check{p}
$. For the linear Gaussian system (\ref{eq OptL}) gives the optimal control
strategy 
\begin{equation}
u\left( t\right) =\beta (\omega _{pq}\left( t\right) \hat{p}_{0}^{t}+\omega
_{p}\left( t\right) \hat{q}_{0}^{t})
\end{equation}%
where the coefficients are the solutions to the Riccati equations 
\begin{equation}
\begin{array}{lll}
-\frac{\mathrm{d}}{\mathrm{d}t}\omega _{q}\left( t\right) & = & \eta
_{q}-2\lambda \sigma _{q}-(\beta \sigma _{q})^{2} \\ 
-\frac{\mathrm{d}}{\mathrm{d}t}\omega _{qp}\left( t\right) & = & \frac{1}{%
\mu }\omega _{q}-\left( \lambda +\delta \right) \omega _{qp}-\beta
^{2}\omega _{p}\omega _{qp} \\ 
-\frac{\mathrm{d}}{\mathrm{d}t}\omega _{p}\left( t\right) & = & \eta
_{p}+2\left( \frac{1}{\mu }\omega _{qp}-\omega _{p}\delta \right) -(\beta
\omega _{p})^{2}%
\end{array}
\label{eq FPContRicatti}
\end{equation}%
with terminal conditions 
\begin{equation*}
\omega _{p}\left( T\right) =\omega _{p},\quad \omega _{qp}\left( T\right)
=\omega _{qp},\quad \omega _{q}\left( T\right) =\omega _{q}.
\end{equation*}%
Note that in this example, as well as identifying the dual matrices by
transposition and time reversal according to the duality table (\ref{tb
Duality}), one must also symplecticly interchange the phase coordinates $%
\left( \check{q},\check{p}\right) \leftrightarrow \left( \check{p},\check{q}%
\right) $. This is because the matrix of coefficients $\mathbf{A}$ is
non-symmetric and nilpotent, so it is dual to its transpose only when we
interchange the coordinates in the dual picture. Thus the optimal
coefficients $\{\omega _{p},\omega _{qp},\omega _{q}\}\left( t\right) $ in
the quadratic cost-to-go correspond to the minimal error covariances $%
\{\sigma _{q},\sigma _{qp},\sigma _{p}\}\left( T-t\right) $ in the dual
picture.

The minimal total cost for the experiment can be obtained from (\ref{eq
LQGmincost}) by substitution of these solutions 
\begin{equation}
\begin{array}{l}
\mathsf{S}=\omega _{q}(q^{2}+\sigma _{q})+2\omega _{qp}(qp+\sigma _{qp}) \\ 
\quad +\omega _{p}\left( 0\right) (p^{2}+\sigma _{p})+\int_{0}^{T}(\hbar
^{2}\omega _{p}\left( t\right) +\omega _{pq}^{2}\left( t\right) \sigma
_{q}\left( t\right) )\mathrm{d}t \\ 
\quad +\int_{0}^{T}(\omega _{p}^{2}\left( t\right) \sigma _{p}\left(
t\right) +2\omega _{qp}\left( t\right) \omega _{p}\left( t\right) \sigma
_{pq}\left( t\right) )\mathrm{d}t%
\end{array}
\label{eq FPtotalcost}
\end{equation}%
This demonstrates the linear microduality principle in the following
specified form of the table (\ref{tb Duality})%
\begin{equation*}
\begin{array}{c|c|c|c|c|c|c}
\text{Filtering }\check{q} & \lambda -\mu ^{-1} & \alpha & \varepsilon & 
\mathbf{J}^{\intercal }\mathbf{K} & \mathbf{GJ} & \mathbf{\Sigma J} \\ \hline
\text{Control\ }\check{p} & \lambda -\mu ^{-1} & \beta & \gamma & \mathbf{L}%
^{\intercal } & \mathbf{JH} & \mathbf{J\Omega }%
\end{array}%
\end{equation*}%
showing the complete symmetry under the time reversal and exchange of $%
\left( q,p\right) $, in which the coordinate observation is seen as
completely dual to the feedback of momentum.

\section{Quantum Dynamics with Trajectories}

This section highlights the differences between quantum and classical
systems and introduces the problem of quantum observation and its solution
in the framework of open dynamics. In orthodox quantum mechanics which
treats only closed quantum dynamics without observations, there is no such
problem. However, it is meaningless to consider quantum feedback control
without solution of this problem. After the appropriate setting of quantum
mechanics with observation is given, the measurement problem is then
restated as a statistical problem of quantum causality which can be resolved
by optimal dynamical estimation on the output of an open quantum system
called quantum filtering.

Quantum physics which deals with the unavoidable random nature of the
microworld requires a new, more general, noncommutative theory of stochastic
processes than the classical one based on Kolmogorov's axioms. The
appropriate quantum probability theory was developed through the 70s and 80s
by Accardi, Belavkin, Gardiner, Holevo, Hudson and Parthasarthy \cite%
{Acc_Fri_Lew82,Be78,Be80,Be85,GaZ00,Hol82,HuP84} amongst others.

The essential difference between classical and quantum systems is that
classical states, including the mixed states, are defined by probability
measures not on \emph{properties} but\emph{\ events}. This is because the
properties of classical systems are described by measurable subsets $\Delta
\subseteq \Omega $ forming a Boolean $\sigma $-algebra $\mathfrak{A}$ on the
space of classical pure states, the points $\omega \in \Omega $ of a phase
space. In principle they all can be tested simultaneously and identified
with the events represented by the indicator functions $1_{\Delta }(\omega )$
of $\Delta \in \mathfrak{A}$ on the \emph{universal observation} space $%
\Omega $. They are building blocks for classical random variables described
by essentially measurable functions with respect to a probability measure $%
\mathbb{P}$ on $\mathfrak{A}$. The algebra of all such complex functions $%
a:\Omega \rightarrow \mathbb{C}$ with pointwise operations is denoted by $%
\mathrm{A}$, while $\mathrm{L}^{p}\left( \Omega ,\mathbb{P}\right) $ with $%
p=1,2,\infty $ stands for the subspaces of absolutely integrable,
square-integrable and essentially bounded functions $f,g,b\in \mathrm{A}$
respectively. Note that the Banach space $\mathrm{M}=\mathrm{L}^{\infty
}\left( \Omega ,\mathbb{P}\right) $ is a commutative C*-algebra (see
Appendix 1.1) of the algebra $\mathrm{A}$ with involution $\ast :\mathrm{A}%
\ni a\mapsto a^{\ast }\in \mathrm{A}$ defined by the complex conjugation $%
a^{\ast }=\overline{a}$. Moreover, it is W*-algebra since $\mathrm{M}$ has
the preadjoint space $\mathrm{M}_{\star }=\mathrm{L}^{1}\left( \Omega ,%
\mathbb{P}\right) $ such that $\mathrm{M}_{\star }^{\star }=\mathrm{M}$ with
respect to the standard pairing%
\begin{equation}
\left\langle a|f\right\rangle :=\int_{\Omega }\overline{a\left( \omega
\right) }f\left( \omega \right) \mathbb{P}\left( \mathrm{d}\omega \right)
\equiv \left( a^{\ast },f\right)
\end{equation}%
defining the expectation on $\mathrm{M}_{\star }$ as $\mathbb{E}\left[ f%
\right] =\left( 1,f\right) $.

In quantum world, unfortunately, there are \emph{incompatible} properties
corresponding to inconsistent but not orthogonal (i.e. not mutually
excluding) \emph{questions} such that, if the infimum $P\wedge Q$ is zero,
it does not mean that $P\perp Q$. These questions cannot be \emph{surely}
answered simultaneously, i.e. tested with simultaneous events on any
universal measurable space $\Omega $, and they cannot be represented in any
Boolean algebra. Since the incompatibility is measured by noncommutativity
of orthoprojectors $P$ and $Q$ representing these questions as Hermitian
idempotents on a Hilbert space $\mathcal{H}$ of quantum \emph{vector-states}%
, the algebra $\mathcal{B}$ generated by all quantum properties must be
noncommutative. The set $\mathfrak{P}\left( \mathcal{B}\right) $ of all
orthoprojectors $P\in \mathcal{B}$, called \emph{property logic} of a
noncommutative algebra $\mathcal{B}$, clearly extends any \emph{eventum logic%
} of commuting orthoprojectors injectively representing the Boolean logic $%
\mathfrak{A}$ by a $\sigma $-homomorphism $E:\mathfrak{A}\rightarrow 
\mathfrak{P}\left( \mathcal{B}\right) $ such that $\sum E\left( \Delta
_{j}\right) =I$ for any measurable $\sigma $-partition $\Omega =\sum \Delta
_{j}$. Two normal quantum variables are said to be \emph{compatible} if
their orthoprojectors commute, and therefore can be represented classically
by measurable functions on their joint spectrum space $\Omega $, however
there is no such $\Omega $ if they do not commute. Since there are many
incompatible quantum variables, e.g. the position and momentum in quantum
mechanics, quantum properties cannot be identified with any commuting set $%
E\left( \mathfrak{A}\right) $ representing a Boolean logic $\mathfrak{A}$.

\subsection{Quantum causality and predictions}

Almost simultaneously with Kolmogorov's functional formulation of classical
probability theory von Neumann \cite{von32} gave another, more general
operator formulation, aiming to lay down the foundation of quantum
probability theory. It deals with not only commutative W*-algebras, called
von Neumann algebras when they are represented as algebras of operators on a
Hilbert space $\mathcal{H}$ with involution as Hermitian conjugation $\ast $
and unit as the identity operator $I$ on $\mathcal{H}$. In order to
understand the relation between these two formulations it is useful to
reformulate Kolmogorov's axioms in terms of von Neumann (wise versa is
impossible in the case of noncommutativity of the operator algebra). Any
random variable $a\in $ $\mathrm{M}$ can be represented by the diagonal
operator $\hat{a}$ of pointwise multiplication $\hat{a}g=ag$ in the Hilbert
space $\mathrm{H}=\mathrm{L}^{2}(\Omega ,\mathbb{P})$ such that the abelian
(commutative) operator algebra $\mathrm{\hat{M}}=\left\{ \hat{a}:a\in 
\mathrm{M}\right\} $ is maximal in the algebra $\mathcal{B}\left( \mathrm{H}%
\right) $ of all bounded operators on $\mathrm{H}$ in the sense that $%
\mathrm{\hat{M}}=\mathrm{\hat{M}}^{\prime }$. Here $\mathrm{\hat{M}}^{\prime
}=\left\{ B\in \mathcal{B}(\mathrm{H}):\left[ \mathrm{\hat{M}},B\right]
=0\right\} $ with $[\mathrm{\hat{M}},B]=\left\{ AB-BA:A\in \mathrm{\hat{M}}%
\right\} $ stands for the \emph{bounded commutant} of $\mathrm{\hat{M}}$,
which obviously coincides on $\mathrm{H}$ with the commutant%
\begin{equation}
\hat{1}_{\mathfrak{A}}^{\prime }=\left\{ B:\left[ \hat{1}_{\Delta },B\right]
=0,\Delta \in \mathfrak{A}\right\}  \label{a-algebra}
\end{equation}%
of the Boolean algebra $\hat{1}_{\mathfrak{A}}=\left\{ \hat{1}_{\Delta
}:\Delta \in \mathfrak{A}\right\} $ of all diagonal orthoprojectors (the
multiplications by $1_{\Delta }$) generating $\mathrm{\hat{M}}$. Note that
the commutant $\mathcal{B}=\mathcal{M}^{\prime }$ of any \emph{nonmaximal}
abelian subalgebra $\mathcal{M}\subseteq \mathcal{B}\left( \mathcal{H}%
\right) $ is a noncommutative W*-algebra with strict inclusion of $\mathcal{M%
}$ as the \emph{center} $\mathcal{B}\cap \mathcal{B}^{\prime }$ of $\mathcal{%
B}$. Thus the \emph{simple algebra} $\mathcal{B}=\mathcal{B}\left( \mathcal{H%
}\right) $ is the commutant of the abelian algebra of scalar multipliers $%
\mathcal{M}=\mathbb{C}I$ which is generated by the trivial Boolean algebra $%
\mathfrak{A}=\left\{ \emptyset ,\Omega \right\} $ represented by improper
orthoprojectors $P_{\emptyset }=O$, $P_{\Omega }=I$. The noncommutative
algebra $\mathcal{A}$ cannot be generated by any Boolean algebra of
orthoprojectors as the commuting Hermitian idempotents $P^{2}=P=P^{\ast }$
in $\mathcal{B}\left( \mathcal{H}\right) $.

\emph{Quantum causality}, assuming the existence of not only properties but
also observable events, requires that all quantum properties related to 
\emph{present and future} at each time-instant $t$ must be compatible with
all \emph{passed} events. This makes an allowance for simultaneous
predictability of incompatible properties at least in the statistical sense.
However the usual quantum mechanics, dealing only with irreducible
representations $\mathcal{B}=\mathcal{B}\left( \mathcal{H}\right) $ of
quantum properties but not with the events, is causal only for the trivial
eventum algebra of improper orthoprojectors $\left\{ O,I\right\} $ on $%
\mathcal{H}$. This is why any nontrivial causality requires an extension of
the orthodox framework of quantum mechanics to \emph{quantum stochastics}
unifying it with the framework of classical stochastics in a minimalistic
way allowing the distinction between the future quantum properties and past
classical events. This program was completed in \cite{Be85,Bel92b} on the
basis of quantum nondemolition (QND) principle \cite{Be80,Be87a,Bel94} as an
algebraic formulation quantum causality. The past events, corresponding to
the measurable histories $\Delta \in \mathfrak{A}_{t]}$ up to each $t\in 
\mathbb{R}_{+}$, should be represented in the commutant of a noncommutative
subalgebra $\mathcal{B}_{[t}\subseteq \mathcal{B}$ describing the present
and future on a universal Hilbert space $\mathcal{H}$. Thus, instead of a
single noncommutative algebra $\mathcal{B}$ extending the eventum W*-algebra 
$\mathcal{M}$ generated by $E\left( \mathfrak{A}\right) $ one should
consider a decreasing family $\left( \mathcal{B}_{[t}\right) $\ of reduced
subalgebras $\mathcal{B}_{[t}\subseteq \mathcal{B}_{[s}$ $\forall s<t$ in
the commutants $E\left( \mathfrak{A}_{t]}\right) ^{\prime }=\left\{ B\in 
\mathcal{B}:\left[ E\left( \mathfrak{A}_{t]}\right) ,B\right] =0\right\} $
of the past eventum logics $E\left( \mathfrak{A}_{t]}\right) $ representing
the consistent histories of increasing probability spaces $\left( \Omega
_{t]},\mathfrak{A}_{t]},\mathbb{P}_{t]}\right) $ in nonmaximal abelian
W*-algebras $\mathcal{M}_{t]}$ generated by $E\left( \mathfrak{A}%
_{t]}\right) $.

The nondemolition principle makes quantum causality irreversible by allowing
future observations represented by decreasing eventum algebras $E\left( 
\mathfrak{A}_{t}\right) \subset \mathcal{B}_{[t}$ to be incompatible with
some present-plus-future questions $Q\in \mathcal{B}_{[t}$. Although any
projectively increasing family of classical probability spaces can be
obtained by Kolmogorov construction from\ a single $\left( \Omega ,\mathfrak{%
A},\mathbb{P}\right) $ with projections $\kappa _{s]}:\Omega \rightarrow
\Omega _{s]}$ inverting the injections $\kappa _{s]}^{-1}\left( \mathfrak{A}%
_{s]}\right) \subseteq \kappa _{t]}^{-1}\left( \mathfrak{A}_{t]}\right)
\subseteq \mathfrak{A}$ for all $s\leq t$ such that $\mathbb{P}_{s]}=\mathbb{%
P}_{t]}\circ \kappa _{s]}^{-1}=\mathbb{P}\circ \kappa _{s]}^{-1}$, however
this projective limit may not be compatible with any noncommutative algebra $%
\mathcal{B}_{[t}$. Thus the maximal W*-algebra $\mathcal{B}=\mathcal{B}_{[0}$%
, satisfying the compatibility condition $\mathcal{B}^{\prime }=\mathcal{M}%
_{0]}$ with the initial central algebra $\mathcal{M}_{0]}=\mathcal{B}\cap 
\mathcal{B}^{\prime }$, coincides with the decomposable algebra $\mathcal{M}%
_{0]}^{\prime }$ which is not compatible with the total eventum algebra $%
\mathcal{M}=\vee \mathcal{M}_{t]}$ except the case $\mathcal{M}=\mathcal{M}%
_{0]}$ of absence of innovation $\mathcal{M}_{r]}=\mathcal{M}_{t]}$ for all $%
r$ and $t$. The latter with $\mathcal{M}=\mathbb{C}I$ is a standard
assumption in the orthodox quantum mechanics dealing in the absence of
observations with the constant $\mathcal{B}_{[t}$ equal to $\mathcal{B}%
\left( \mathcal{H}\right) $. We may assume that $\mathcal{M}_{0]}=\mathbb{C}%
I $ corresponding to trivial initial history, $\mathfrak{A}_{0]}=\left\{
\varnothing ,\Omega \right\} $ with $\mathbb{P}_{0]}=1$ on a single-point $%
\Omega _{0]}=\left\{ 0\right\} $, which allows $\mathcal{B}_{[0}=\mathcal{B}%
\left( \mathcal{H}_{[0}\right) $.

Note that since all operators $B\in E\left( \mathfrak{A}_{t]}\right)
^{\prime }$ commute with $\mathcal{M}_{t]}$, they are jointly decomposable,
given in the diagonal representation of $\mathcal{M}_{t]}$ by $\left( 
\mathfrak{A}_{t]},\mathbb{P}_{t]}\right) $-essentially bounded functions $%
B_{t}:\omega \mapsto B\left( \omega \right) $ on $\Omega _{t]}$ with
operator values $B\left( \omega \right) \in \mathcal{B}\left( \omega \right) 
$ on the Hilbert components $\mathcal{H}\left( \omega \right) $ of the
orthogonal decomposition $\int_{\Omega _{t]}}^{\oplus }\mathcal{H}\left(
\omega \right) \mathbb{P}_{t]}\left( \mathrm{d}\omega \right) \sim \mathcal{H%
}$ corresponding to the joint spectral representations 
\begin{equation}
E\left( \Delta \right) \simeq \int_{\Omega _{t]}}^{\oplus }1_{\Delta }\left(
\omega \right) I\left( \omega \right) \mathbb{P}_{t]}\left( \mathrm{d}\omega
\right) \mathbb{\equiv }I_{t]}\left( \Delta \right)  \label{decompos}
\end{equation}%
of commuting orthoprojectors $E\left( \Delta \right) $, $\Delta \in 
\mathfrak{A}_{t]}$.

\emph{Quantum state} (See Appendix 1.2) consistent with the trajectory
probability space $\left( \Omega ,\mathbb{P},\mathfrak{A}\right) $ is given
as the linear positive functional $\left\langle \varpi |Q\right\rangle
=\left( \varpi ,Q\right) $ by a Hermitian-positive $\varpi =\varpi ^{\ast }$
mass-one $\left\langle \varpi |I\right\rangle =1$ operator $\varpi \vdash 
\mathcal{B}$ ($\in \mathcal{B}$ in usual or a generalized sense as
affiliated to $\mathcal{B}$) defining the probability measure $\mathbb{P}$
as the projective limit of%
\begin{equation}
\mathbb{P}_{t]}\left( \Delta \right) =\left\langle \varpi |E\left( \Delta
\right) \right\rangle =\mathbb{P}\left( \Delta \right) ,\;\;\;\Delta \in 
\mathfrak{A}_{t]}  \label{induce}
\end{equation}%
\ (where $\left\langle \varpi ,B\right\rangle =\mathrm{tr}\left[ B\varpi %
\right] $\ for the semifinite $\mathcal{B}$). It is called the (probability)
density operator for $\mathcal{B}$ since it defines the probability $\Pr %
\left[ Q\right] =\left\langle \varpi |Q\right\rangle \in \left[ 0,1\right] $
of any quantum property described by an orthoprojector $Q\in \mathcal{A}$.
Since $Q\in \mathfrak{P}\left( \mathcal{B}_{[t}\right) $ is compatible with
each eventum projector $E\left( \Delta \right) $ for $\Delta \in \mathfrak{A}%
_{t]}$, the property $Q$ is statistically predictable with respect to all
past events due to the existence of \textit{a posteriori} conditional
probability%
\begin{equation}
\Pr \left[ Q|\Delta \right] =\frac{1}{\mathbb{P}\left( \Delta \right) }%
\left\langle \varpi |QE\left( \Delta \right) \right\rangle \;\;\forall
\Delta :\mathbb{P}\left( \Delta \right) \neq 0  \label{bayes}
\end{equation}%
such that $\Pr \left[ Q\right] =\mathbb{P}\left( \Delta \right) \Pr \left[
Q|\Delta \right] +\mathbb{P}\left( \Delta ^{\perp }\right) \Pr \left[
Q|\Delta ^{\perp }\right] $. Note that $\left\langle \varpi |QE\right\rangle 
$ is not positive and even not real without the compatibility of $Q$ and $E$%
. This leads to the existence of the \emph{posterior quantum states} on $%
\mathcal{B}_{[t}$ given by the conditional expectations%
\begin{equation}
\epsilon ^{t}\left[ B|\omega \right] =\left\langle \hat{\varpi}_{[t}\left(
\omega \right) |B\left( \omega \right) \right\rangle \;\;\forall B\in 
\mathcal{B}\cap \mathcal{M}_{t]}^{\prime }.  \label{post exp}
\end{equation}%
The posterior states are defined as classical stochastic adapted processes $%
\omega \mapsto \hat{\varpi}_{[t}\left( \omega \right) $ with density
operator values affiliated to the components $\mathcal{B}\left( \omega
\right) $ of $\mathcal{B}_{[t}$, and the corresponding conditional
expectations defined as positive normal projections $\mathcal{B}\cap 
\mathcal{M}_{t]}^{\prime }\rightarrow \mathcal{M}_{t]}$ will be denoted as $%
\epsilon ^{t}=\hat{\varpi}_{[t}^{\star }$.

\begin{theorem}
Let $\varpi $ be a normal state on $\mathcal{B}$. Then the induced state $%
\varpi _{\lbrack t}$ on the relative commutant $\mathcal{B}_{[t}\subseteq 
\mathcal{B}\cap \mathcal{M}_{t]}^{\prime }$ of the eventum algebra $\mathcal{%
M}_{t]}$ is given as classical expectation $\langle \varpi _{\lbrack
t},B\rangle =\mathbb{E}_{\Omega _{t]}}\left[ \hat{\varpi}_{[t}^{\star
}\left( B\right) \right] $ in terms of pairing 
\begin{equation}
\langle \varpi _{\lbrack t}|B\rangle :=\int_{\Omega _{t]}}\left\langle \hat{%
\varpi}_{[t}\left( \omega \right) |B\left( \omega \right) \right\rangle 
\mathbb{P}\left( \mathrm{d}\omega \right)  \label{b-pairing}
\end{equation}%
on $\mathcal{B}_{t}=\mathcal{B}\cap \mathcal{M}_{t]}^{\prime }$ with the
posterior density operators $\hat{\varpi}_{[t}\vdash \mathcal{B}_{t}$
uniquely defined by the conditional expectation $\epsilon ^{t}:\mathcal{B}%
_{t}\rightarrow \mathcal{M}_{t]}$ as positive integrable function for almost
all $\omega \in \Omega _{t]}$
\end{theorem}

\begin{proof}
Since $\varpi _{t}$ is normal state on $\mathcal{B}_{t}$, equivalent to the
space of essentially bounded functions on $\left( \Omega _{t]},\mathfrak{A}%
_{t]},\mathbb{P}_{t]}\right) $ with operator values in $\mathcal{B}\left(
\omega \right) \subseteq \mathcal{B}\left( \mathcal{H}\left( \omega \right)
\right) $, it is uniquely defined by an essentially integrable function $%
\varpi _{t}\left( \omega \right) $ in (\ref{b-pairing}) with operator values
in $\mathcal{B}_{\star }\left( \omega \right) $. Obviously it is the density
operator for the posterior states as the conditional expectations defined on 
$\mathcal{B}_{t}$ with respect to the central Abelian subalgebra $\mathcal{M}%
_{t]}\sim \mathrm{L}^{\infty }\left( \Omega _{t]},\mathbb{P}_{t]}\right) $
by the Radon-Nikodym derivatives 
\begin{equation}
\epsilon ^{t}\left[ B|\omega \right] :=\lim_{\mathfrak{A}_{t]}\ni \Delta
\searrow \left\{ \omega \right\} }\frac{\left\langle \varpi _{t}|BE\left(
\Delta \right) \right\rangle }{\mathbb{P}\left( \Delta \right) },\;\;B\in 
\mathcal{B}_{t}  \label{eq poststate}
\end{equation}%
where the limit is understood for and almost all $\omega \in \Omega _{t]}$
in the same way as in the classical case.
\end{proof}

Note that in the most important "white noise" cases considered in next
sections, all $\mathcal{H}\left( \omega \right) $ with $\omega \in \Omega
_{t]}$ are isomorphic to a single Hilbert space $\mathcal{H}_{[t}$ of a
decreasing family $\left( \mathcal{H}_{[t}\right) $ such that $B_{t}\left(
\omega \right) \in \mathcal{B}\left( \mathcal{H}_{[t}\right) $, uniquely
defined for almost all $\omega \in \Omega _{t]}$, represents $B\in \mathcal{B%
}_{t}$ as a bounded operator on $\mathcal{H}=\mathrm{H}_{t]}\otimes \mathcal{%
H}_{[t}$ commuting with $I_{t]}\left( \Delta \right) =\hat{1}_{\Delta
}\otimes I_{[t}$, where $\mathrm{H}_{t]}=\mathrm{L}^{2}\left( \Omega _{t]},%
\mathbb{P}_{t]}\right) $. Representing $\mathcal{H}$ as $\mathrm{H}%
_{t]}\otimes \mathcal{H}_{[t}$ in the case $\mathcal{H}\left( \omega \right)
=\mathcal{H}_{[t}$ $\forall \omega \in \Omega _{t]}$, the posterior states,
described by the positive mass one operators $\hat{\varpi}_{[t}\left( \omega
\right) =\hat{\varpi}_{[t}^{\omega }$ in $\mathcal{H}_{[t}$, can be
considered as the conditional states on the operator algebras $\mathcal{B}%
_{[t}=\mathcal{B}\left( \mathcal{H}_{[t}\right) $, controlled by the history
trajectory $\omega $. Thus the above quantum causality setting gives
immediately the posterior states $\hat{\varpi}_{[t}^{\omega }$ as the states
for quantum present and future conditioned by the classical past without any
reference to the projection or other phenomenological reduction postulate of
quantum measurement. This is main advantage of the extended, event enhanced
quantum mechanics, or \emph{eventum mechanics}, which allows treatment of \
the observable events on equal basis with other quantum properties of the
system. It can be shown, see (\ref{A.5}), that any reduction postulate of
the operational quantum mechanics can all be derived from QND causality, and
this principle is also applicable to the continuous measurements in both in
time and spectrum where projection postulates fails.

Thus posterior states provide the optimal in mean quadratic sense Bayesian
estimators for any number of unobservable quantum noncommuting variables $%
x\in \mathcal{B}\left( \mathcal{H}_{[t}\right) $ or properties from future,
given the observable history $\omega \in \Omega _{t]}$. For this reason, one
can consider quantum measurements in this "nondemolition" setup as a form of
quantum \emph{filtering}.

We now describe an appropriate Markovian model for the time-continuous
interactions between the open quantum system and the field.

\subsection{ Quantum open dynamics and input-output}

Quantum Markovian dynamics with observable trajectories, which entered into
physics in the 90's in terms of stochastic transfer-operators or stochastic
Master equations, define the phenomenological "instruments" of observation
without giving any microscopic dynamical model in terms of the fundamental
Hamiltonian interactions. In fact such approach is equivalent to the earlier
operational approach based on the instrumental transfer-measures (See
Appendix 1.3), and its starting point corresponds to already filtered Markov
dynamics in the classical case. Here we describe the general scheme for
underlying Hamiltonian interaction models with continuous observation for
open quantum dynamical objects in terms of quantum stochastic evolutions in
parallel to the classical stochastic models with partial observation,
following the original Belavkin approach suggested in \cite%
{Be85,Bel88,Bel92b}.

Let us fix a quantum probability space $\left( \mathcal{H},\mathcal{B}%
,\varpi \right) $ and an increasing family $\mathcal{B}_{s]}\subseteq 
\mathcal{B}_{t]}$, $\forall s<t$ of W*-subalgebras $\mathcal{B}%
_{t]}\subseteq \mathcal{B}$ containing the \emph{compatible histories} $%
E\left( \mathfrak{A}_{t]}\right) \subset \mathcal{B}_{[t}^{\prime }$ and 
\emph{nontrivial present} $\mathcal{B}_{[t]}=\mathcal{B}_{t]}\cap \mathcal{B}%
_{[t}$ for each $t$, assuming that each $\mathcal{B}_{t]}$ commutes with 
\emph{future} $\mathcal{A}_{t}\subset \mathcal{B}_{[t}$, $\mathcal{B}%
_{t]}\subseteq \mathcal{A}_{t}^{\prime }$, being generated by only \emph{%
nonanticipating questions} $Q\in \mathcal{A}_{t}^{\prime }\cap \mathcal{B}$.
The future is \emph{quantum noise} which is described by W*-subalgebras $%
\mathcal{A}_{t}\subseteq \mathcal{B}$, forming a decreasing family $\mathcal{%
A}_{t}\subseteq \mathcal{A}_{t+s}$ $\forall t,s>0$ with trivial intersection
such that we may assume that $\vee \mathcal{B}_{t]}=\mathcal{B}$. Moreover,
we shall assume that the family $\left( \mathcal{B}_{t]}\right) $ as well as 
$\left( \mathcal{M}_{t]}\right) $ form the\emph{\ W*-product systems} in the
sense of W*-isomorphisms%
\begin{equation}
\mathcal{B}_{t]}\bar{\otimes}\mathcal{A}_{t}^{s}\sim \mathcal{B}_{t+s]},\;\;%
\mathcal{M}_{t]}\bar{\otimes}\mathcal{M}_{t}^{s}\sim \mathcal{M}_{t+s]},
\label{eq Product}
\end{equation}%
where $\mathcal{A}_{t}^{s}=\mathcal{A}_{t}\cap \mathcal{B}_{t+s]}$, $%
\mathcal{M}_{t}^{s}=\mathcal{M}_{t}\cap \mathcal{M}_{t+s]}$ and $\left( 
\mathcal{M}_{t}\right) $ is decreasing family of W*-algebras $\mathcal{M}%
_{t}\subset \mathcal{B}_{[t}$ generated by future events $E\left( \mathfrak{A%
}_{t}\right) $. This implies that the family $\left( \mathcal{A}%
_{t}^{s}\right) $ satisfies the product condition such that $\mathcal{A}%
_{0}\sim \mathcal{A}^{t}\bar{\otimes}\mathcal{A}_{t}^{s}\bar{\otimes}%
\mathcal{A}_{t+s}$ for any $t$ and $s>0$, and similar for $\left( \mathcal{M}%
_{t}^{s}\right) $, corresponding to the split property%
\begin{equation}
\Omega =\Omega ^{t}\times \Omega _{t}^{r}\times \Omega _{t+r}^{s},\;%
\mathfrak{A}=\mathfrak{A}_{t}\otimes \mathfrak{A}_{t}^{r}\otimes \mathfrak{A}%
_{t+r}  \label{eq Split}
\end{equation}%
of the measurable trajectory space.

\emph{Quantum open object} under the observation is represented at each time 
$t$ by a past-future boundary W*-subalgebra $\mathfrak{b}_{t}\subseteq 
\mathcal{B}_{[t]}$ such that it is adapted with respect to the family $%
\left( \mathcal{B}_{t]}\right) $ quantum stochastic process (in the genral
sense \cite{Be85}), nonanticipating futures $\left( \mathcal{A}_{t}\right) $
and satisfying causality condition with respect to the histories $\left( 
\mathcal{M}_{t]}\right) $. We may assume that each $\mathfrak{b}_{t}$
represents a fixed $\mathfrak{b}$ or a variable boundary W*-algebra $%
\mathfrak{b}\left( t\right) $ by a W*-homomorphism $\pi _{t}$ of $\mathfrak{b%
}\left( t\right) $ onto $\mathfrak{b}_{t}$, with $\mathfrak{b}\left(
t\right) $ taken in the initial algebra $\mathcal{B}_{[0]}$, say. Due to the
causality condition the product%
\begin{equation*}
\Pi _{t}\left( \Delta ,\check{q}\right) =E\left( \Delta \right) \pi
_{t}\left( \check{q}\right) \;~\forall \Delta \in \mathfrak{A}_{t]}
\end{equation*}%
defines for each $t$ an adapted transfer-measure $\Pi _{t}\left( \Delta
\right) :$ $\mathfrak{b}\left( t\right) \rightarrow \mathcal{B}_{t]}$ (see
Appendix 1.4) with W*-homomorphic values normalized to the history eventum
projectors $E\left( \Delta \right) $. Obviously W*-algebras $\mathfrak{b}%
_{t} $ and $\mathcal{A}_{t}^{s}$ are in $\mathcal{B}_{[t}^{s}=\mathcal{B}%
_{t+s]}\cap \mathcal{B}_{[t}$, as well as $\mathfrak{b}_{t+s}$ and
W*-algebras $\mathcal{M}_{t}^{s}$.

Following \cite{Be85} we shall say that quantum open object $\mathfrak{b}%
\left( t\right) $ with eventum history $E\left( \mathfrak{A}_{t]}\right) $
is \emph{dynamical} with respect to $\left( \mathcal{A}_{t}\right) $ if%
\begin{equation*}
\mathcal{M}_{t}^{s}\vee \mathfrak{b}_{t+s}\subseteq \mathfrak{b}_{t}\vee 
\mathcal{A}_{t}^{s}\;\;\forall t,s>0.
\end{equation*}%
This is equivalent\emph{\ }\cite{Be85} to the existence of \emph{quantum
flow with observations} described as follows on the co-images $\mathfrak{b}%
\left( t\right) =\pi ^{t}\left( \mathfrak{b}_{t}\right) $ of the boundary
algebras $\mathfrak{b}_{t}$. Assuming that the W*-algebras $\mathcal{B}%
_{[t}^{s}$ are generated by $\mathfrak{b}_{t}$ and $\mathcal{A}_{t}^{s}$, we
can always consider the dynamical quantum open object with\ $\mathfrak{b}%
_{t}=\mathcal{B}_{[t]}$.

\begin{theorem}
Let $\pi ^{t}:\mathfrak{b}_{t}\rightarrow \mathfrak{b}\left( t\right) $ be
normal injections inverted by the dynamical representations $\pi _{t}$, and
let $\mathcal{B}_{t]},\mathcal{M}_{t]}$ form the product systems (\ref{eq
Product}). Then there exists a transitional spectral measure%
\begin{equation}
\Upsilon _{r}^{t}\left( \Delta ,\check{q}\right) =E^{t}\left( \Delta \right)
\gamma _{r}^{t}\left( \check{q}\right) ,\;\;\;\check{q}\in \mathfrak{b}%
\left( t+r\right)  \label{hom-measure}
\end{equation}%
on $\mathfrak{A}_{t}^{r}$ with values in $\mathcal{B}_{[t}^{r}=\mathfrak{b}%
\left( t\right) \bar{\otimes}\mathcal{A}_{t}^{r}$ given by adapted $\sigma $%
-homomorphisms $E^{t}:\mathfrak{A}_{t}\rightarrow \mathfrak{b}\left(
t\right) \bar{\otimes}\mathcal{A}_{t}$ and a Heisenberg flow $\left( \gamma
_{r}^{t}\right) $ of causal tensor-adapted W*-homomorphisms $\gamma _{r}^{t}:%
\mathfrak{b}\left( t+r\right) \rightarrow E^{t}\left( \mathfrak{A}%
_{t}^{r}\right) ^{\prime }\cap \mathcal{B}_{[t}^{r}$ such that%
\begin{eqnarray}
\gamma _{r}^{t}\circ \gamma _{s}^{t+r} &=&\gamma _{r+s}^{t}\;\;\forall r,s>0,
\label{hemflow} \\
\gamma _{r}^{t-r}\left( E^{t}\left( \Delta \right) \right) &=&E^{t-r}\left(
\Delta \right) \;\;\forall t>r  \label{hemout}
\end{eqnarray}%
under the trivial extensions onto $\mathcal{B}_{[t}=\mathfrak{b}\left(
t\right) \bar{\otimes}\mathcal{A}_{t}$.
\end{theorem}

\begin{proof}
The representations $\pi _{t}$ as well as $\pi ^{t-r}$ can be trivially
extended to the adapted W*-homomorphisms with respect to the identity maps $%
\mathrm{id}$ respectively on $\mathcal{A}_{t}$ and $\mathcal{A}_{t-r}$ by
virtue of commutativity $\mathcal{B}_{t]}\subseteq \mathcal{A}_{t}^{\prime }$
as $\pi _{t}\left( \check{q}\otimes A_{t}\right) =\pi _{t}\left( \check{q}%
\right) A_{t}$ and $\pi ^{t-r}\left( \check{q}_{t-r}A_{t-r}\right) =\pi
^{t-r}\left( \check{q}_{t-r}\right) \otimes A_{t-r}$ respectively for all $%
\check{q}\in \mathfrak{b}\left( t\right) $ and $\check{q}_{t}\in \mathfrak{b}%
_{t}$, $A_{t}\in \mathcal{A}_{t}$. This defines the compositions $\gamma
_{r}^{t-r}=\pi ^{t-r}\circ \pi _{t}$ of thus extended W*-representations as
tensor-adapted W*-homomorphisms $\gamma _{r}^{t-r}:\mathcal{B}%
_{[t}\rightarrow \mathcal{B}_{[t-r}$ on $\mathcal{B}_{[t}=\mathfrak{b}\left(
t\right) \otimes \mathcal{A}_{t}$ trivially extending the $\mathfrak{b}%
\left( t\right) \rightarrow \mathfrak{b}\left( t\right) \otimes \mathcal{A}%
_{t-r}^{r}$ and satisfying the hemigroup condition (\ref{hemflow}) such that 
$\gamma _{0}^{t}=\mathrm{id}\left( \mathcal{B}_{[t}\right) $ for each $t$.
Obviously these extensions satisfy causality condition%
\begin{equation*}
\pi _{t}\left( \mathcal{B}_{[t}\right) \subseteq E^{t}\left( \mathfrak{A}%
_{t]}\right) ^{\prime },\;\gamma _{r}^{t}\left( \mathcal{B}_{[t+r}\right)
\subseteq E^{t}\left( \mathfrak{A}_{t}^{r}\right) ^{\prime },
\end{equation*}%
where the eventum projectors $E^{t}\left( \Delta \right) \in \mathcal{B}%
_{[t} $ are defined for any $\Delta \in \mathfrak{A}_{t}$ as $\pi ^{t}\left(
E\left( \Delta \right) \right) $ by the extended injections $\pi ^{t}:%
\mathfrak{b}_{t}\vee \mathcal{A}_{t}\rightarrow \mathfrak{b}\left( t\right)
\otimes \mathcal{A}_{t}$ as right inverse of the extended $\pi _{t}$ on $%
E=\pi _{t}\left( E^{t}\right) \in E\left( \mathfrak{A}_{t}\right) $. The
second condition (\ref{hemout}) simply follows from $\pi _{t}\circ \gamma
_{r}^{t}=\pi _{t+r}$ due to $\pi _{t-r}\left( E^{t}\right) =E$ for any $r\in %
\left[ 0,t\right] $ and $E\in E\left( \mathfrak{A}_{t}\right) $. Thus the QS
flow with nondemolition observations can be described in terms of the
homomorphic transitional measures (\ref{hom-measure}) with (\ref{hemflow})
and (\ref{hemout}) satisfying the hemigroup composition low%
\begin{equation}
\Upsilon _{r}^{t-r}\left( \Delta _{t-r}^{r},\Upsilon _{s}^{t}\left( \Delta
_{t}^{s},\check{q}\right) \right) =\Upsilon _{r+s}^{t-r}{}\left( \Delta
_{t-r}^{r+s},\check{q}\right)  \label{hemihommeas}
\end{equation}%
where $\Delta _{t-r}^{r+s}=\Delta _{t-r}^{r}\times \Delta _{t}^{s}\in 
\mathfrak{A}_{t-r}^{r+s}$ and $\check{q}\in \mathfrak{b}\left( t+s\right) $.
\end{proof}

\begin{corollary}
The dynamical QS object is Markovian in the usual sense \cite{Be85} if the
initial state $\varpi =\varpi _{\lbrack 0}$\ on $\mathcal{B=B}_{[0}$ is
product state $\varpi \sim \varpi _{t]}\otimes \varrho _{t}$ for any $t$
such that%
\begin{equation*}
\left\langle \varpi |B_{t]}A_{t}^{s}\right\rangle =\left\langle \varpi
|B_{t]}\right\rangle \left\langle \varrho _{t}|A_{t}^{s}\right\rangle \;.
\end{equation*}%
It is operationally described is such state by the hemigroup of reduced
transitional measures 
\begin{equation}
\mathcal{T}_{r}^{t}\left( \Delta ,\check{q}\right) =\varrho _{t}^{\star }%
\left[ \Upsilon _{r}^{t}\left( \Delta ,\check{q}\right) \right] ,
\label{cpmeas}
\end{equation}%
where $\varrho _{t}^{\star }:\mathcal{B}_{[0}\rightarrow \mathcal{B}_{t]}$
is conditional expectation defined as%
\begin{equation}
\left\langle \varpi _{\lbrack 0}|B_{t]}\right\rangle =\left\langle \varpi
_{t]}|\varrho _{t}^{\star }\left[ B_{t]}\right] \right\rangle \;\;\forall
\varpi _{t]},B_{t]}\in \mathcal{B}_{t]}.  \label{eq NoiseCondExp}
\end{equation}%
They satisfy the operational Chapman-Kolmogorov equation%
\begin{equation}
\mathcal{T}_{r}^{t-r}\left( \Delta _{t-r}^{r},\mathcal{T}_{s}^{t}\left(
\Delta _{t}^{s},\check{q}\right) \right) =\mathcal{T}_{r+s}^{t-r}{}\left(
\Delta _{t-r}^{r+s},\check{q}\right)  \label{hemicpmeas}
\end{equation}%
as a normal completely positive map $\mathfrak{b}\left( t+s\right)
\rightarrow \mathfrak{b}\left( t-r\right) $ for each product $\Delta
_{t-r}^{r+s}$ of $\Delta _{t-r}^{r}\in \mathfrak{A}_{t-r}^{r}$ and $\Delta
_{t}^{s}\in \mathfrak{A}_{t}^{s}$.
\end{corollary}

\begin{remark}
The event representations $E^{t}=\pi ^{t}\left( E\right) $ are usually given
by \emph{input} $\sigma $-homomorphisms $I:\mathfrak{A}_{t}^{s}\rightarrow 
\mathcal{A}_{t}^{s}$, $I\left( \Delta \right) =\iota \left( 1_{\Delta
}\right) $ as $E^{t}\left( \Delta \right) =\upsilon ^{t}\left( I\left(
\Delta \right) \right) $ in terms of a two side adapted W*-representation $%
\iota :\mathrm{M}_{t}^{s}\rightarrow \mathcal{A}_{t}^{s}$ and a hemigroup $%
\left( \upsilon _{r}^{t}\right) $ of \emph{interaction isomorphisms} $%
\upsilon _{r}^{t}:\mathfrak{b}\left( t+r\right) \bar{\otimes}\mathcal{A}%
_{t}^{r}\rightarrow \mathfrak{b}\left( t\right) \bar{\otimes}\mathcal{A}%
_{t}^{r}$ such that $\upsilon _{r}^{t}|\mathfrak{b}\left( t+r\right) =\gamma
_{r}^{t}$. Here $\upsilon ^{t}$ defines the \emph{output representation}
induced on an \emph{input eventum algebra} $\iota \left( \mathrm{M}%
_{t}\right) \subset \mathcal{A}_{t}$ for $\mathrm{M}_{t}=\mathrm{L}^{\infty
}\left( \Omega _{t},\mathbb{P}_{t}\right) $ by the limit $\upsilon
^{t}=\lim_{r}\upsilon _{r}^{t}|\mathcal{A}_{t}$ of $\upsilon _{r}^{t}|%
\mathcal{A}_{t}^{r}$ which is well-defined on each $\mathcal{A}_{t}=\vee _{r}%
\mathcal{A}_{t}^{r}$ due to the \emph{localization property} $\upsilon
_{s+r}^{t}|\mathcal{A}_{t}^{r}=\upsilon _{r}^{t}|\mathcal{A}_{t}^{r}$ for
all $r,s>0$. Note that the localization property simply follows from the
hemigroup condition and the normalization $\upsilon ^{t}\left( I_{0}\right)
=I_{0}$ for these $\upsilon _{r}^{t}$, extended adaptively also on $%
A_{0}^{t}\in \mathcal{A}_{0}^{t}$ and $A_{t+r}\in \mathcal{A}_{t+r}$ such
that%
\begin{equation*}
\upsilon _{r}^{t}\left( A_{0}^{t}\otimes B\otimes A_{t+r}\right)
=A_{0}^{t}\otimes \upsilon _{r}^{t}\left( B\right) \otimes A_{t+r}.
\end{equation*}
\end{remark}

The quantum \emph{free evolution} is usually described by a semigroup $%
\left( \theta _{s}\right) $ of endomorphisms $\theta _{s}:\mathcal{B}%
_{[0}\rightarrow \mathcal{B}_{[0}$ shifting isomorphically any $\mathcal{A}%
_{t}$ onto $\mathcal{A}_{t+s}$ with trivial action on $\mathfrak{b}$. QS
Heisenberg flow $\left( \gamma _{s}^{t}\right) $ with observation over a
constant algebra $\mathfrak{b}\left( t\right) =\mathfrak{b}$ is called \emph{%
covariant} with respect to a shift semigroup $\left( \theta _{s}\right) $
acting also on $\mathfrak{A}_{0}$ by the shift of any $\mathfrak{A}_{t}$
onto $\mathfrak{A}_{t+s}=\theta _{s}\left( \mathfrak{A}_{t}\right) $ if$\;$ 
\begin{equation}
\gamma _{s}^{t}\circ \theta _{t+s}=\theta _{t}\circ \vartheta
_{s},\;\;E^{t}\circ \theta _{t}=\theta _{t}\circ E^{0},  \label{eq Cocycle}
\end{equation}%
where $\vartheta _{s}=\gamma _{s}^{0}\circ \theta _{s}$ and $\theta _{t}$ is
extended on the W*-algebra $\mathcal{B}_{[0}=\mathfrak{b}\otimes \mathcal{A}%
_{0}$ by $\theta _{t}\left( \check{q}\otimes A_{0}\right) =\check{q}\otimes
\theta _{t}\left( A_{0}\right) $. This defines a Heisenberg $\theta $%
-cocycle $\gamma _{s}=\gamma _{s}^{0}$ corresponding to the semigroup $%
\left( \vartheta _{s}\right) $ of W*-endomorphisms $\vartheta _{s}=\gamma
_{s}\circ \theta _{s}$ of the algebra $\mathcal{B}$, satisfying causality
condition $\vartheta _{s}\left( \mathcal{B}\right) \subseteq E\left( 
\mathfrak{A}_{s]}\right) ^{\prime }$. Note that the shift semigroup can be
extended to a group $\left\{ \theta _{t}:t\in \mathbb{R}\right\} $ on $%
\mathcal{B}=\widetilde{\mathcal{A}}_{0}\bar{\otimes}\mathfrak{b}\bar{\otimes}%
\mathcal{A}_{0}$, where $\widetilde{\mathcal{A}}_{0}$ is an independent copy
of the algebra $\mathcal{A}_{0}$, with $\theta _{t}$ transforming each
segment $\widetilde{\mathcal{A}}_{r}^{\tau }$ onto $\mathcal{A}_{s}^{\tau }$
for any positive $\tau ,r,s$ and $t=\tau +r+s$, shifting $\widetilde{%
\mathcal{A}}_{t}$ onto $\widetilde{\mathcal{A}}_{0}$ similar to the inverted
shift $\theta _{-t}:\mathcal{A}_{t}\rightarrow \mathcal{A}_{0}$, with
backward transformation of each $\mathcal{A}_{s}^{\tau }$ onto $\widetilde{%
\mathcal{A}}_{r}^{\tau }$ for $\tau +r+s=-t$. This free group dynamics $%
\theta _{t}$ defines the reversible quantum dynamics interaction on such $%
\mathcal{B}$ by one-parametric group of $\vartheta _{t}=\upsilon _{t}\circ
\theta _{t+s}$ extending $\vartheta _{t}$ onto $\mathcal{B}$ by interaction
W*-automorphisms $\upsilon _{t}=\upsilon _{t}^{0}$ on $\mathcal{B}_{[0}=%
\mathfrak{b}\otimes \mathcal{A}_{0}$, acting identically on $\widetilde{%
\mathcal{A}}_{0}$ for any $t>0$, with the identical action on $\mathcal{A}%
_{0}$ and the reflected cocycle action $\upsilon _{t}=\theta ^{-t}\circ 
\widetilde{\upsilon }_{-t}\circ \theta _{t+s}$\ on $\widetilde{\mathcal{B}}%
_{0]}=\widetilde{\mathcal{A}}_{0}\otimes \mathfrak{b}$ for $t<0$, where $%
\widetilde{\upsilon }_{s}$ is defined\ on $\widetilde{\mathcal{B}}_{0]}$
exactly as $\upsilon _{s}^{-1}$ on $\mathcal{B}_{[0}$ for any $s>0$. However
the reversible quantum dynamics on such noncommutative $\mathcal{B}$ cannot
satisfy the causality in both directions of time with respect to a
nontrivial eventum algebra $E\left( \mathfrak{A}\right) $, except the case
of absence innovation as in the conservative quantum mechanics without
observation. To keep the causality in the positive direction of time one
must replace the nonabelian $\widetilde{\mathcal{A}}_{0}$ by the smaller,
abelian subalgebra $\mathrm{\check{M}}_{0}$, a copy of the eventum algebra $%
\mathrm{\hat{M}}_{0}=I\left( \mathrm{M}_{0}\right) $, which makes $\theta
_{s}$ and $\vartheta _{s}$ irreversible on $\mathcal{B}=\mathrm{\check{M}}%
_{0}\bar{\otimes}\mathfrak{b}\bar{\otimes}\mathcal{A}_{0}$ with
noncommutative $\mathcal{A}_{0}$.

\section{Quantum Stochastics of Eventum Mechanics}

In this section we consider quantum noise models defining quantum Markovian
dynamics with continuous nondemolition observation and show the quantum
filtering equations derived from this models. Such observation can be based
only on indirect measurement of a quantum open object via a coupled channel
representing a classical measured process $m_{0}^{t}$ in a bath $\mathcal{A}%
_{0}$ which is usually assumed to be initially independent of the quantum
object $\mathfrak{b}$. We shall consider the measurement processes having
initially independent increments $m_{t}^{s}=m_{0}^{t+s}-m_{0}^{t}\equiv
m\left( \mathrm{I}_{t}^{s}\right) $ as random measure on the intervals $%
\mathrm{I}_{t}^{s}=[t,t+s)$. They generate independent W*-algebras $\mathrm{M%
}_{t}^{s}=\mathrm{L}_{\mathfrak{A}}^{\infty }\left( \Omega _{t}^{s},\mathbb{P%
}_{t}^{s}\right) \equiv \mathrm{M}\left( \mathrm{I}_{t}^{s}\right) $, and an
input quantum process with the increments $M_{t}^{s}=\iota \left(
m_{t}^{s}\right) \equiv M\left( \mathrm{I}_{t}^{s}\right) $ generates
independent eventum algebras $\iota \left( \mathrm{M}_{t}^{s}\right) \subset 
\mathcal{B}\left( \mathcal{F}_{t}^{s}\right) $ on a Hilbert space $\mathcal{F%
}_{0}$ with respect to an initial unit state vector $\chi _{0}$ satisfying
the divisibility condition%
\begin{equation}
\mathcal{F}_{0}\sim \mathcal{F}_{0}^{t}\bar{\otimes}\mathcal{F}_{t}^{s}\bar{%
\otimes}\mathcal{F}_{t+s},\ \chi _{0}\sim \chi _{0}^{t}\otimes \chi
_{t}^{s}\otimes \chi _{t+s},  \label{eq Fsplit}
\end{equation}%
such that it induces the \emph{initial product probability measure} $\mathbb{%
P}_{0}=\mathbb{P}_{0}^{t}\otimes \mathbb{P}_{t}^{s}\otimes \mathbb{P}_{t+s}$
on the measurable space of observable trajectories under the split condition
(\ref{eq Split}).

An appropriate candidate for such Hilbert space suitable to accommodate any
kind of classical independent increment process is Guichardet-Fock space $%
\mathcal{F}_{0}=\Gamma \left( \mathcal{E}_{0}\right) $ over the Hilbert
space $\mathcal{E}_{0}$ of $\mathrm{L}^{2}$-functions $\xi :t\mapsto 
\mathfrak{k}$ on $\mathbb{R}_{0}^{+}=\left\{ t\geq 0\right\} $ with values
in a Hilbert space $\mathfrak{k}$ such that $\mathcal{F}_{t}^{s}=\Gamma
\left( \mathcal{E}_{t}^{s}\right) $ with $\mathcal{E}_{t}^{s}=\mathrm{L}%
^{2}\left( \mathrm{I}_{t}^{s}\rightarrow \mathfrak{k}\right) $ (see the
definitions in \cite{Bel92b},\cite{Par92} summarized in the Appendix2).
There are sufficiently many product vectors in $\mathcal{F}_{0}$, called 
\emph{exponential vectors} $\left\{ \xi ^{\otimes }:\xi \in \mathcal{E}%
_{0}\right\} $, generating Fock space $\mathcal{F}_{0}$ such that any \emph{%
coherent state vector}%
\begin{equation}
\chi _{\xi }=\mathrm{e}^{-\frac{1}{2}\left\Vert \xi \right\Vert _{\mathcal{E}%
}^{2}}\xi ^{\otimes },\;\left\Vert \xi \right\Vert _{\mathcal{E}%
}^{2}=\int_{0}^{\infty }\left\Vert \xi \left( t\right) \right\Vert _{%
\mathfrak{k}}^{2}\mathrm{d}t  \label{eq Coherent}
\end{equation}%
defines a product state $\left\langle \varphi _{\xi }|A\right\rangle
=\left\langle \chi _{\xi }|A\chi _{\xi }\right\rangle $ on any subalgebra $%
\mathcal{A}_{0}\subseteq \mathcal{B}\left( \mathcal{F}_{0}\right) $
satisfying the divisibility condition $\mathcal{A}_{0}\sim \mathcal{A}%
_{0}^{t}\otimes \mathcal{A}_{t}^{s}\otimes \mathcal{A}_{t+s}~$ for any $%
t,s>0 $ such that each $\mathcal{A}_{t}^{s}=\mathcal{A}\left( \mathrm{I}%
_{t}^{s}\right) $ is represented in $\mathcal{B}\left( \mathcal{F}%
_{t}^{s}\right) $. However there is only one shift-invariant such state
which is given by the \emph{vacuum vector} $\chi _{0}=\delta _{\emptyset }$
corresponding to $\xi =0$. In fact, not only shift invariant but any
infinitely divisible normal state on $\mathcal{A}_{0}$ can be induced by the
vacuum state $\varphi _{0}$ on $\mathcal{B}\left( \mathcal{F}_{0}\right) $
by choosing in general time dependent Hilbert space $\mathfrak{k}\left(
t\right) $ in a canonical way \cite{Be92a},\cite{Be92c}. In particular,
since the state on the abelian algebra $\mathrm{M}_{0}$ defined by the
probability $\mathbb{P}_{0}$ of any classical process with independent
increments is infinitely divisible, any such state can be induced from the
quantum vacuum state by restricting it to the abelian part $\mathcal{\iota }%
\left( \mathrm{M}_{0}\right) \subset \mathcal{A}_{0}$ given by an adapted
input W*-representation $\iota :\mathrm{M}_{0}\rightarrow \mathcal{A}_{0}$
such that $\mathcal{\iota }\left( \mathrm{M}_{t}^{s}\right) \subset \mathcal{%
A}_{t}^{s}$ for any $t,s>0$. It defines the infinitely divisible probability
measure $\mathbb{P}_{0}$ as%
\begin{equation}
\mathbb{P}_{t}^{s}\left( \Delta \right) :=\left\langle \delta _{\emptyset
}|I\left( \Delta \right) \delta _{\emptyset }\right\rangle \equiv
\left\langle \varphi _{0}|I\left( \Delta \right) \right\rangle \;\forall
\Delta \in \mathfrak{A}_{t}^{s},  \label{eq InputP}
\end{equation}%
where $I\left( \Delta \right) =\iota \left( 1_{\Delta }\right) $, by the
vacuum vectors $\chi _{t}^{s}=\delta _{\emptyset }$ of $\mathcal{F}_{t}^{s}$.

There are two basic processes with additive independent increments which can
be realized in Fock space with finite-dimensional $\mathfrak{k}=\mathbb{C}%
^{d}$: Wiener vector-valued process $\boldsymbol{w}^{t}=\left(
w_{i}^{t}\right) _{i\in I_{W}}$ which we index by a subset $I_{W}\subseteq
\left\{ 1,\ldots ,d\right\} $, and Poisson compound process $\boldsymbol{n}%
\left( \mathrm{I}_{0}^{t}\right) =\left( n_{i}^{t}\right) _{i\in I_{N}}$
which we index by another subset $I_{N}\subseteq \left\{ 1,\ldots ,d\right\} 
$ (It should be thought as diagonal matrix-valued rather than vector). Their
differential increments $\mathrm{d}m=m\left( \mathrm{d}t\right) $ satisfy
quite different It\^{o} multiplication tables%
\begin{equation}
\mathrm{d}w_{i}\mathrm{d}w_{k}=\delta _{ik}\mathrm{d}t,\;\mathrm{d}n_{i}%
\mathrm{d}n_{k}=\delta _{i}^{j}\delta _{k}^{j}\mathrm{d}n_{j},
\label{eq w&n ItoTable}
\end{equation}%
where the summation rule is applied over $j$. Their canonical input
representations $M_{i}^{t}=\iota \left( m_{i}^{t}\right) $ in Fock space
over $\mathcal{E}_{0}=\mathbb{C}^{d}\otimes \mathrm{L}^{2}\left( \mathbb{R}%
_{0}^{+}\right) $ are defined as%
\begin{eqnarray}
W_{i}^{t} &=&A_{i}^{+}\left( t\right) +A_{-}^{i}\left( t\right) \equiv \Re 
\left[ A_{i}^{+}\left( t\right) \right] ,  \label{eq Winput} \\
N_{i}^{t} &=&A_{i}^{i}\left( t\right) +A_{i}^{+}\left( t\right)
+A_{-}^{i}\left( t\right) +A_{-}^{+}\left( t\right)  \label{eq Ninput}
\end{eqnarray}%
in terms of four basic operator-processes in $\mathcal{F}_{0}$ defined in
Appendix B. These are creation $A_{\circ }^{+}\left( t\right) :=A_{\circ
}^{+}\left( \mathrm{I}_{0}^{t}\right) $ (row-valued, $A_{\circ }^{+}=\left(
A_{i}^{+}\right) $), annihilation $A_{-}^{\circ }\left( t\right)
:=A_{-}^{\circ }\left( \mathrm{I}_{0}^{t}\right) $ (column-valued, $%
A_{-}^{\circ }:=\left( A_{-}^{k}\right) $), exchange $A_{\circ }^{\circ
}\left( t\right) :=A_{\circ }^{\circ }\left( \mathrm{I}_{0}^{t}\right) $
(matrix-valued, $A_{\circ }^{\circ }=\left( A_{i}^{k}\right) $ ) and
preservation $A_{+}^{-}\left( t\right) :=A_{+}^{-}\left( \mathrm{I}%
_{0}^{t}\right) $ (scalar-valued, $A_{-}^{+}\left( t\right) =tI$).\ These
canonical processes forming a pseudo-Hermitian matrix $\boldsymbol{A}=\left(
A_{\iota }^{\kappa }\right) _{\iota =-,\circ }^{\kappa =\circ ,+}=%
\boldsymbol{A}^{\star }$ under the involution $\left( A_{-\iota }^{\kappa
}\right) ^{\star }=\left( A_{-\kappa }^{\iota \ast }\right) $ with respect
to the index reflection $-\left( -,\circ ,+\right) =\left( +,\circ ,-\right) 
$, satisfy pseudo-Poisson multiplication table%
\begin{equation}
\mathrm{d}A_{\mu }^{\iota }\mathrm{d}A_{\kappa }^{\nu }=\delta _{\kappa
}^{\iota }\mathrm{d}A_{\mu }^{\nu }\;\;\ \forall \iota ,\nu \in \left\{
i,+\right\} ,\mu ,\kappa \in \left\{ -,k\right\}  \label{eq qItoTable}
\end{equation}%
of \emph{quantum stochastic calculus} discovered as a noncommutative
generalization of the classical It\^{o}-Poisson table by Belavkin in \cite%
{Bel92b}. Note that from (\ref{eq qItoTable}) it follows that%
\begin{equation*}
\mathrm{d}W_{i}\mathrm{d}N_{k}=\delta _{k}^{i}\mathrm{d}A_{-}^{i}+\delta
_{ik}\mathrm{d}t,\;\mathrm{d}N_{i}\mathrm{d}W_{k}=\delta _{k}^{i}\mathrm{d}%
A_{k}^{+}+\delta _{ik}\mathrm{d}t\;
\end{equation*}%
which cannot be realized in the classical category of commutative processes
in which it always $\mathrm{d}w_{i}\mathrm{d}n_{k}=0=\mathrm{d}n_{k}\mathrm{d%
}w_{i}$. Thus the \emph{joint} operator representation of two types of basic
classical processes with independent increments is possible only in splitted
Fock spaces such that $I_{W}\cap I_{N}=\emptyset $ corresponding to
orthogonal subspaces $\mathfrak{k}_{W}\perp \mathfrak{k}_{N}$, which
reflects the classical split%
\begin{equation*}
\Omega =\Omega _{W}\times \Omega _{N},\;\mathfrak{A}=\mathfrak{A}_{W}\otimes 
\mathfrak{A}_{N},\ \mathbb{P}=\mathbb{P}_{W}\otimes \mathbb{P}_{N}.
\end{equation*}

The four basic processes $A_{\mu }^{\nu }$ form a linear basis of quantum It%
\^{o} $\star $-algebra as noncommutative integrators for the increments 
\begin{equation}
M_{t}^{s}=\sum_{\mu ,\nu }\int_{t}^{t+s}K_{\nu }^{\mu }\left( r\right) 
\mathrm{d}A_{\mu }^{\nu }\equiv \mathfrak{i}_{t}^{s}\left( \boldsymbol{K}%
\right)
\end{equation}%
defined by four integrants $\boldsymbol{K}=\left( K_{\nu }^{\mu }\right)
_{\nu =\circ ,+}^{\mu =-,\circ }$ as operator-valued functions integrable in
a quantum-stochastic \cite{Bel92b}. These quantum stochastic integrals
satisfy $\star $-property such that $\mathfrak{i}_{t}^{s}\left( \boldsymbol{K%
}\right) ^{\ast }=\mathfrak{i}_{t}^{s}\left( \boldsymbol{K}^{\star }\right) $
under the involution $\left( K_{-\nu }^{\mu }\right) ^{\star }=\left(
K_{-\mu }^{\nu \ast }\right) $, and the It\^{o} product rule 
\begin{equation}
\mathrm{d}(M^{\ast }M)=\mathrm{d}\left( M^{\ast }\right) M+M^{\ast }\left( 
\mathrm{d}M\right) +\left( \mathrm{d}M^{\ast }\right) \mathrm{d}(M),
\label{eq ItoRule}
\end{equation}%
where $\mathrm{d}M=K_{\nu }^{\mu }\mathrm{d}A_{\mu }^{\nu }$ (the usual
summation convention is assumed), with the It\^{o} correction calculated as%
\begin{equation}
\mathrm{d}\mathfrak{i}_{t}^{s}\left( \boldsymbol{K}\right) ^{\ast }\mathrm{d}%
\mathfrak{i}_{t}^{s}\left( \boldsymbol{K}\right) =\mathrm{d}\mathfrak{i}%
_{t}^{s}\left( \boldsymbol{K}^{\star }\boldsymbol{K}\right)
\label{eq QSItotable}
\end{equation}%
for adapted quantum stochastic integrators $K_{\nu }^{\mu }\left( t\right) $.

Note that in the case of a single degree of freedom $d=1$ this quantum It%
\^{o} table reads in the Hudson-Parthasarathy (HP) form \cite{HuP84} as%
\begin{eqnarray*}
\mathrm{d}A_{-}^{\circ }\mathrm{d}A_{\circ }^{+} &=&\mathrm{d}tI,\;\;\;%
\mathrm{d}A_{-}^{\circ }\mathrm{d}A_{\circ }^{\circ }=\mathrm{d}A_{-}^{\circ
},\; \\
\mathrm{d}A_{\circ }^{\circ }\mathrm{d}A_{\circ }^{\circ } &=&\mathrm{d}%
A_{\circ }^{\circ },~\ \;\mathrm{d}A_{\circ }^{\circ }\mathrm{d}A_{\circ
}^{+}=\mathrm{d}A_{\circ }^{+},
\end{eqnarray*}%
with all other increment multiplications vanishing, in terms of four
scalar-operator processes $A_{\nu }^{\mu }$ with pseudo-Hermitian property
written as%
\begin{equation*}
A_{-}^{+}\left( t\right) =A_{-}^{+}\left( t\right) ^{\ast },\;A_{\circ
}^{+}\left( t\right) =A_{-}^{\circ }\left( t\right) ^{\ast },\;A_{\circ
}^{\circ }\left( t\right) =A_{\circ }^{\circ }\left( t\right) ^{\ast }.
\end{equation*}

\subsection{Quantum mechanics with observations}

Interaction automorphic evolutions $\left( \upsilon _{r}\right) _{r>0}$ on
the the tensor product $\mathcal{B}_{[0}=\mathfrak{b}\otimes \mathcal{A}_{0}$
of simple algebras $\mathfrak{b}=\mathcal{B}\left( \mathfrak{h}\right) $ and 
$\mathcal{A}_{0}=\mathcal{B}\left( \mathcal{F}_{0}\right) $ are usually
described by right unitary cocycles $\left\{ U_{r}:r>0\right\} $ as $%
\upsilon _{t}\left( B\right) =U_{t}BU_{t}^{\ast }$. The cocycles satisfy the
operator hemigroup condition $U_{r}^{t}U_{s}^{t+r}=U_{r+s}^{t}$ similar to (%
\ref{hemflow}) in terms of the shifted operators $U_{r}^{s}=\theta
_{s}\left( U_{r}\right) $, where $\left( \theta _{s}\right) $ is the
semigroup of right shift W*-endomorphisms $\theta _{s}:\mathcal{A}%
_{0}\rightarrow \mathcal{A}_{s}\subseteq \mathcal{A}_{0}$ describing free
evolution of the bath by trivial placing of each subalgebra $\mathcal{A}%
\left( \mathrm{I}_{t-s}^{s}\right) $ onto $\mathcal{A}\left( \mathrm{I}%
_{t}^{s}\right) $ as $\mathcal{A}_{t}^{s}=\mathcal{B}\left( \mathcal{F}%
_{t}^{s}\right) $. Hudson and Parthasarathy \cite{HuP84},\cite{Par92}
derived a QS \emph{forward equation}%
\begin{equation}
\mathrm{d}U_{t}=U_{t}\left( R\mathrm{d}A_{\circ }^{\circ }+R_{+}\mathrm{d}%
A_{\circ }^{+}+R^{-}\mathrm{d}A_{-}^{\circ }+R_{+}^{-}\mathrm{d}t\right)
\label{eq HP}
\end{equation}%
for the unitary cocycles in Fock space $\mathcal{F}_{0}=\Gamma \left( 
\mathrm{L}^{2}\left( \mathbb{R}_{0}^{+}\right) \right) $ defined as the
solution of QS integral equation $U_{t}=U_{0}+\mathfrak{i}_{0}^{t}\left( U%
\boldsymbol{R}\right) $ with four adapted operator-valued coefficients $%
R_{\nu }^{\mu }$ and $U_{0}=I$. They gave the necessary conditions for
unitarity of this solution written in terms $S=I+R$ as%
\begin{equation}
S^{\ast }=S^{-1},\;R^{-}=-R_{+}^{\ast }S,\;2\Re \left( R_{+}^{-}\right)
=-R_{+}^{\ast }R_{+},  \label{eq HPunicond}
\end{equation}%
where $\Re \left( A\right) $ denotes the Hermitian part $\left( A+A^{\ast
}\right) /2$ of an operator $A$ (the sufficiency was shown only for the
constant bounded initial-valued coefficients $R_{\nu }^{\mu }$).

It is important for control to have also the sufficient unitarity conditions
in the case of time-dependent and multi-dimensional noise $A$. As it was
proved in \cite{Be91} under the natural QS-integrability conditions, the
relations (\ref{eq HPunicond}) are also sufficient for the uniqueness of
unitary solution even in the case of time-dependent infinite dimensional
coefficients with values $R_{\nu }^{\mu }\left( t\right) =\left( \check{s}%
_{\nu }^{\mu }\left( t\right) -\delta _{\nu }^{\mu }\check{1}\right) \otimes
I$ in the initial operator algebra $\mathfrak{b}$. In terms of these $\check{%
s}_{\nu }^{\mu }\left( t\right) $ the multidimensional version of HP
equation can be simply written in Belavkin's $\star $-algebraic notations 
\cite{Be88,Be92} as 
\begin{equation}
\mathrm{d}U_{t}=U_{t}\left( \check{s}_{\nu }^{\mu }\left( t\right) -\delta
_{\nu }^{\mu }\check{1}\right) A_{\mu }^{\nu }\left( \mathrm{d}t\right)
\label{eq MultiHP}
\end{equation}%
where usual summation convention over all $\mu ,\nu \in \left\{ -,\circ
,+\right\} $ can be restricted to the domain $\mu \leq \nu $ under the order 
$-<\circ <+$ of the the triangular matrix $\boldsymbol{\check{s}}=\left( 
\check{s}_{\nu }^{\mu }\right) _{\nu =-,\circ ,+}^{\mu =-,\circ ,+}$ with
zero operator entries for $\mu >\nu $ and $\check{s}_{-}^{-}=\check{1}=%
\check{s}_{+}^{+}$ (the usual summation convention then can be applied). In
this notations the algebraic relations between nonzero matrix elements $%
\check{s}_{\nu }^{\mu }$ generalizing the unitarity conditions (\ref{eq
HPunicond}) are simply expressed as the pseudo-unitarity $\boldsymbol{\check{%
s}}^{\star }=\boldsymbol{\check{s}}^{-1}$ of the operator matrix $%
\boldsymbol{\check{s}}$ in terms of the pseudo-Hermitian adjoint matrix $%
\boldsymbol{\check{s}}^{\star }=\left( \check{s}_{-\mu }^{-\nu \ast }\right) 
$.

From the quantum It\^{o} rule (\ref{eq ItoRule}) applied to Heisenberg QS
flow $X\left( t\right) =\gamma _{t}\left( \check{x}\right) $ given by
interaction dynamics as $\gamma _{t}\left( \check{x}\right) =U_{t}(\check{x}%
\otimes I)U_{t}^{\ast }\equiv \upsilon _{t}(\check{x}\otimes I)$, and the
quantum It\^{o} multiplication table (\ref{eq qItoTable}), we obtain the
general \emph{QS Langevin equation} 
\begin{equation}
\mathrm{d}X=\left( \Sigma _{\nu }^{\mu }\left( X\right) -X\delta _{\nu
}^{\mu }\right) \mathrm{d}A_{\mu }^{\nu }\equiv \left( \boldsymbol{\Sigma }%
\left( X\right) -X\boldsymbol{1}\right) \cdot \mathrm{d}\mathbf{A},
\label{eq FlowX}
\end{equation}%
Here $\boldsymbol{\Sigma }\left( t,X\right) =\boldsymbol{S}\left( t\right)
\left( X\otimes \boldsymbol{1}\right) \boldsymbol{S}\left( t\right) ^{\star
} $, called \emph{QS germ}, is given on $X=X\left( t\right) $ by
matrix-function $\boldsymbol{\Sigma }\left( t\right) $ which is defined as a
triangular matrix $\left( \Sigma _{\nu }^{\mu }\right) _{\nu =-,\circ
,+}^{\mu =-,\circ ,+}$ of\ six time evolved maps%
\begin{equation*}
\Sigma _{\nu }^{\mu }\left( t,U_{t}\left( \check{x}\otimes I\right)
U_{t}^{\ast }\right) =U_{t}\left( \check{\sigma}_{\nu }^{\mu }\left( t,%
\check{x}\right) \otimes I\right) U_{t}^{\ast }
\end{equation*}%
given on $\mathfrak{b}$ by $\boldsymbol{\check{\sigma}}\left( \check{x}%
\right) =\boldsymbol{\check{s}}\left( \check{x}\otimes \boldsymbol{1}\right) 
\boldsymbol{\check{s}}^{\star }$, where $\boldsymbol{1}=\left( \delta _{\nu
}^{\mu }\right) _{\nu =-,\circ ,+}^{\mu =-,\circ ,+}$, with trivial $\check{%
\sigma}_{-}^{-}=\mathrm{id}\left( \mathfrak{b}\right) =\check{\sigma}%
_{+}^{+} $ and $\check{\sigma}_{\nu }^{\mu }\left( \check{x}\right) =0$ for $%
\mu >\nu $ such that the summation convention can be applied only for $\mu
\leq \nu $. It was proved in \cite{Be91} under the natural QS-integrability
conditions that the unitality and $\star $-multiplicativity of the germ%
\begin{equation}
\boldsymbol{\check{\sigma}}\left( t,\check{x}^{\ast }\check{x}\right) =%
\boldsymbol{\check{\sigma}}\left( t,\check{x}\right) ^{\star }\boldsymbol{%
\check{\sigma}}\left( t,\check{x}\right) ,\;\;\boldsymbol{\check{\sigma}}%
\left( t,\check{1}\right) =\boldsymbol{\check{1},}  \label{eq Homcond}
\end{equation}%
are the necessary and sufficient conditions for the existence and uniqueness
of the unital $\ast $-homomorphic solutions $X\left( r+s\right) =\gamma
_{s}^{r}\left( \check{x}\right) $ to the Langevin equation (\ref{eq FlowX})
with $X\left( t\right) =\check{x}$.

The composition $\check{\tau}_{t-r}^{r}\left( \check{x}\right) =\varrho
_{r}^{\star }\left[ \gamma _{t-r}^{r}\left( \check{x}\right) \right] $ with
noise conditional expectation (\ref{eq NoiseCondExp}), defined by the vacuum
state $\varrho _{r}=\varphi _{0}$, describes a \emph{dynamical hemigroup} $%
\left( \check{\tau}_{s}^{r}\right) $ (or semigroup $\left( \check{\tau}%
_{s}\right) $ in the stationary case) of unital completely positive maps $%
\check{\tau}_{s}^{r}:\mathfrak{b}\rightarrow \mathfrak{b}$ on operator
algebra $\mathfrak{b}\subseteq \mathcal{B}\left( \mathfrak{h}\right) $. This
bath expectation, given by solutions $X\left( r+s\right) =\gamma
_{s}^{r}\left( \check{x}\right) $ of QS flow equation (\ref{eq FlowX}) with $%
X\left( r\right) =\check{x}\otimes I$ for the evolved on time interval $%
\mathrm{I}_{r}^{s}$ operators $\check{x}\in \mathfrak{b}$, satisfies the%
\emph{\ master equation} \cite{GKS76,Lin76} $\frac{\mathrm{d}}{\mathrm{d}t}%
\check{\tau}_{t}\left( \check{x}\right) =\check{\tau}_{t}\left( \check{%
\lambda}\left( \check{x}\right) \right) $ with a \emph{Lindblad generator} $%
\check{\lambda}$ which can be written as a linear conditionally positive map
in $\check{x}\in \mathfrak{b}$ in the form 
\begin{equation}
\check{\lambda}\left( \check{x}\right) =\sum_{i,k}K^{i\ast }\check{\sigma}%
_{k}^{i}\left( \check{x}\right) K^{k}+L\check{x}+\check{x}L^{\ast }.
\label{eq Lindblad}
\end{equation}%
Here $K^{i}=-R_{+}^{i}$, $L=R_{+}^{-}$ satisfies the condition $L+L^{\ast
}=-\sum_{i}K^{i\ast }K^{i}$ and $\check{\sigma}_{k}^{i}\left( \check{x}%
\right) =\sum_{j}S_{j}^{i}\check{x}S_{j}^{k\ast }$ ($R_{+}^{i}=\check{s}%
_{+}^{i}$, $R_{+}^{-}=\check{s}_{+}^{-}$, $S_{j}^{i}=\check{s}_{j}^{i}$ to
denote that these operators belong or affiliated to the algebra $\mathfrak{b}
$). In fact, for quantum coherent control we need a time dependent version
of this equation in the following decomposed form.

\begin{theorem}
Let $\varphi _{r,\xi }^{\star }$ be coherent conditional expectation $%
\mathcal{B}_{[0}\rightarrow \mathcal{B}_{r]}$ defined on $\mathcal{B}_{[0}=%
\mathfrak{b}\otimes \mathcal{A}_{0}$ as normal positive projection such that%
\begin{equation*}
\left\langle \varpi _{r]}|\varphi _{r,\xi }^{\star }\left[ B\right]
\right\rangle =\left\langle \psi _{r]}\otimes \chi _{\xi _{r}}|B|\psi
_{r]}\otimes \chi _{\xi _{r}}\right\rangle
\end{equation*}%
for any vector product-state $\varpi _{r]}=\varsigma _{\eta }\otimes \varphi
_{\xi _{0}^{r}}$ with $\eta \in \mathfrak{h}$ and coherent vector (\ref{eq
Coherent}) in $\mathcal{F}_{0}^{r}$. Then the dynamical map $\check{\tau}%
\left( t,\check{x}\right) =\varphi _{r,\xi }^{\star }\left[ \gamma
_{t-r}^{r}\left( \check{x}\right) \right] \equiv \check{\tau}%
_{t-r}^{r}\left( \xi ,\check{x}\right) $ satisfies forward evolution
equation 
\begin{equation}
\frac{\mathrm{d}}{\mathrm{d}t}\check{\tau}\left( t,\check{x}\right) =\check{%
\tau}\left( t,\check{\lambda}\left( t,\check{x}\right) \right) ,\;\;\check{%
\tau}\left( r\right) =\mathrm{Id}\left( \mathfrak{b}\right)
\label{eq LindbX}
\end{equation}%
with Lindblad type generator decomposed as%
\begin{equation}
\check{\lambda}\left( \check{x}\right) =\frac{\mathrm{i}}{\hbar }\left[ H,%
\check{x}\right] +\check{\lambda}_{R}\left( \check{x}\right) +K^{\circ \ast
}\left( \check{\sigma}_{\circ }^{\circ }\left( \check{x}\right) -\check{x}%
\delta _{\circ }^{\circ }\right) K^{\circ }  \label{eq Lindbdecomp}
\end{equation}%
where $H\left( t\right) =\hbar \Im \left[ \left( 2\xi \left( t\right) ^{\ast
}R_{+}^{\circ }+R_{+}^{-}\right) \right] $, $K^{\circ }\left( t\right) =\xi
\left( t\right) -R_{+}^{\circ }$, $\check{\sigma}_{\circ }^{\circ }\left( 
\check{x}\right) =S_{\circ }^{\circ }\check{x}S_{\circ }^{\circ \ast }$ and%
\begin{equation*}
\check{\lambda}_{R}\left( x\right) =\frac{1}{2}\sum_{i}\left( R_{+}^{i\ast }%
\left[ \check{x},R_{+}^{i}\right] +[R_{+}^{i\ast },\check{x}%
]R_{+}^{i}\right) .
\end{equation*}
\end{theorem}

\begin{proof}
Indeed, it can be shown \cite{Be97} that conditional coherent expectations $%
\check{\tau}_{t-r}^{r}\left( \xi ,\check{x}\right) $, evaluated from the QS
flow equation (\ref{eq FlowX}), satisfy Lindblad type equation with\emph{\ }%
generator%
\begin{equation*}
\check{\lambda}\left( t\right) =\check{\sigma}_{+}^{-}+\check{\sigma}_{\circ
}^{-}\xi \left( t\right) +\xi \left( t\right) ^{\ast }\check{\sigma}%
_{+}^{\circ }+\xi \left( t\right) ^{\ast }\check{\rho}_{\circ }^{\circ }\xi
\left( t\right)
\end{equation*}%
where $\check{\rho}_{k}^{i}\left( \check{x}\right) =\check{\sigma}%
_{k}^{i}\left( \check{x}\right) -\check{x}\delta _{k}^{i}$. Using HP
conditions (\ref{eq HPunicond}) in multidimensional form 
\begin{equation*}
S_{\circ }^{\circ \ast }S_{\circ }^{\circ }=\delta _{\circ }^{\circ
}1,\;R_{+}^{\circ \ast }S_{\circ }^{\circ }=-R_{\circ }^{-},R_{+}^{\circ
\ast }R_{+}^{\circ }=-R_{+}^{-\ast }-R_{+}^{-}
\end{equation*}%
as pseudo-unitarity conditions $T_{-\nu }^{\mu }=S_{-\mu }^{\nu \ast }$ in
terms of operator matrix elements of inverse $\boldsymbol{T}=\boldsymbol{S}%
^{-1}$ to the triangular operator matrix $\boldsymbol{S}$ for $\check{\sigma}%
_{-\nu }^{\mu }\left( \check{x}\right) =\sum_{\iota }S_{\iota }^{\mu }\check{%
x}S_{-\iota }^{\nu \ast }$, this generator can be written as%
\begin{equation*}
\check{\lambda}\left[ \check{x}\right] =\sum_{\iota =-,\circ ,+}\left(
S_{-\iota }^{-}+\xi ^{\ast }S_{-\iota }^{\circ }\right) \check{x}\left(
S_{\iota }^{-}+\xi ^{\ast }S_{\iota }^{\circ }\right) ^{\ast }-\xi ^{\ast
}\xi \check{x}
\end{equation*}%
where $\xi \left( t\right) ^{\ast }\xi \left( t\right) =\left\Vert \xi
\left( t\right) \right\Vert _{\mathfrak{k}}^{2}$. This gives (\ref{eq
Lindbdecomp}) after taking into account again the unitarity conditions.
\end{proof}

Given a normal quantum state $\varsigma \left( r\right) $ on $\mathfrak{b}$
at time $r$, the hemigroup $\left( \check{\tau}_{s}^{r}\right) $ defines an
averaged coherent controlled non-Hamiltonian state evolution $\left[
\varsigma \left( t\right) :t\geq r\right] $ of the quantum dynamical object
by composing it with the adjoint CP maps $\tau _{r}^{t}=\check{\tau}%
_{t-r}^{r\star }$ as $\varsigma \left( t\right) =\tau _{r}^{t}\left(
\varsigma \left( t-r\right) \right) $. It satisfies the \emph{hemigroup
master equation} 
\begin{equation}
\frac{\mathrm{d}}{\mathrm{d}t}\varsigma +K\varsigma +\varsigma K^{\ast
}=\sum_{i}L^{i}\varsigma L^{i\ast },  \label{eq mastereq}
\end{equation}%
where the operators $K=-L_{\star }$ and $L^{i}=L_{i\star }$ are defined by
left adjoints $L_{\star }=L^{\sharp },L_{i\star }=L_{i}^{\sharp }$ to%
\begin{equation*}
L=R_{+}^{-}+\xi ^{\ast }R_{+}^{\circ }-\frac{1}{2}\xi ^{\ast }\xi
,\;L_{i}=R_{i}^{-}+\xi ^{\ast }S_{i}^{\circ }
\end{equation*}%
with respect to the $\left\langle \mathfrak{b}^{\star }|\mathfrak{b}%
\right\rangle $-pairing: $\left\langle L^{\sharp }\varsigma |\check{x}%
\right\rangle =\left\langle \varsigma |L\check{x}\right\rangle $ (which are
usual Hilbert space adjoints, $L_{\star }=L^{\ast }$, $L_{i\star
}=L_{i}^{\ast }$ in the case of the trace pairing $\left\langle \varsigma |%
\check{x}\right\rangle =\mathrm{tr}\left[ \varsigma ^{\ast }\check{x}\right] 
$). This master equation is usually written in the Lindbladian form as $%
\frac{\mathrm{d}}{\mathrm{d}t}\varsigma =\lambda \left( t,\varsigma \right) $%
, and it is a particular case of the general QS Master equation (\ref{eq
QSmaster}) derived in \cite{Be95a},\cite{Be96a}.

\subsection{Quantum nonlinear filtering equations}

A time continuous measurement of each Wiener process $w_{i}^{t}$ represented
by the field quadratures (\ref{eq Winput}) after interaction with quantum
object as $Y_{i}^{t}=U_{t^{\prime }}W_{i}^{t}U_{t^{\prime }}^{\ast }$ due to
locality for any $t^{\prime }\geq t$, realizes an indirect measurement of
the evolved generalized coordinate $Q_{i}\left( t\right) =2\Re \left[
L_{i}\left( t\right) \right] $, where $L_{i}\left( t\right) =U_{t}\left( 
\check{s}_{i}^{-}\otimes I\right) U_{t}^{\ast }$. This can be seen from the
quantum It\^{o} formula (\ref{eq ItoRule}), (\ref{eq QSItotable}) applied to
the output operators $\upsilon _{t}\left( W_{i}^{t}\right)
=U_{t}W_{i}^{t}U_{t}^{\ast }$: 
\begin{equation}
\mathrm{d}\upsilon _{t}\left( W_{i}^{t}\right) =2\Re \left( L_{i}^{\ast
}\left( t\right) \mathrm{d}t+\mathrm{d}A_{i}^{+}\right) =Q_{i}\left(
t\right) \mathrm{d}t+\mathrm{d}W_{i}^{t}  \label{eq flowW}
\end{equation}%
Similarly, the output process corresponding to the field counting process (%
\ref{eq Ninput}) as $Y_{i}^{t}=U_{t^{\prime }}N_{i}^{t}U_{t^{\prime }}^{\ast
}$ for any $t^{\prime }\geq t$ is given by $\upsilon _{t}\left(
N_{i}^{t}\right) =U_{t}N_{i}^{t}U_{t}^{\ast }$ as the QS integral of 
\begin{equation}
\mathrm{d}\upsilon _{t}\left( N_{i}^{t}\right) =L_{i}\left( t\right)
L_{i}^{\ast }\left( t\right) \mathrm{d}t+2\Re \left( L_{i}^{\ast }\left(
t\right) \mathrm{d}A_{i}^{+}\right) +\mathrm{d}A_{i}^{i}.  \label{eq flowN}
\end{equation}

Classically, filtering equations are used when we need to estimate the value
of dynamical variables about which we have incomplete knowledge due to an
indirect observation. For example, the Kalman-Bucy filter \cite{Kal60},\cite%
{KaB61} gives a continuous least-squares estimator for a Gaussian classical
random variable with linear dynamics when we only have access to a
correlated, noisy output signal. Since closed quantum systems are
fundamentally unobservable (hidden) unless they are open, e.g.\ disturbed by
quantum noise processes (c.f. (\ref{eq FlowX}) such that equations (\ref{eq
flowW})) and (\ref{eq flowN}) have nontrivial input from the quantum object
in terms of the non-Hamiltonian part $L^{i}=\check{s}_{i}^{\ast }$ of the
Lindblad generator, filtering of quantum noise plays an important role in
quantum measurement. As it follows immediately from the localization
property of quantum interaction evolution due to the hemigroup property the
output operators $Y_{i}^{t}$ are self non-demolition (i.e. mutually
compatible at all times) and satisfy the quantum non-demolition (QND)
condition 
\begin{equation}
\lbrack X\left( s\right) ,Y_{i}^{t}]=0\quad \forall s\leq t,i\in \left\{
1,\ldots ,d\right\}  \label{eq QND}
\end{equation}%
with respect to any evolved quantum object process $X\left( s\right)
=\upsilon _{s}\left( \check{x}\otimes I\right) \equiv \gamma _{s}\left( 
\check{x}\right) $. Belavkin was the first to realize that an optimal
estimation without further disturbance is possible in the general quantum
open dynamical models when based on any output QND measurements \cite{Bel79},%
\cite{Bel80},\cite{Be87a},\cite{Bel88}. He constructed the quantum filtering
equation which describes the evolution of the optimal estimate given by the
density matrix conditioned on a classical output of the noisy quantum
channel. This is used to estimate arbitrary object variable $X\left(
t\right) \vdash \mathfrak{b}_{t}$ which are driven by environmental quantum
noises. The QND condition insists that the expectation of $X\left( t\right) $
is not disturbed when we measure $Y_{i}^{s}$ for $s\leq t$. As it was
already pointed out by the Section 1 Theorem, this is necessary and
sufficient for the existence of a well defined conditional expectation of $%
X\left( t\right) $ with respect to past measurement results of $Y^{t]}$.

Let $\mathcal{M}_{t]}$ be the the history abelian W*-algebra $\mathcal{W}%
_{t]}\subset \mathcal{B}\left( \mathcal{F}_{t]}\right) $ generated by the
output operators $\left\{ \upsilon _{r}\left( W_{i}^{r}\right) :r\in \mathrm{%
I}_{0}^{t},i\in I_{W}\right\} $ for an index subset $I_{W}\subseteq \left\{
1,\ldots ,d\right\} $, or another abelian W*-algebra $\mathcal{N}%
_{t]}\subset \mathcal{B}\left( \mathcal{F}_{t]}\right) $ generated by $%
\{\upsilon _{r}\left( N_{i}^{r}\right) :r\in \mathrm{I}_{0}^{t},i\in I_{N}\}$
for the same or another index subset $I_{N}$, or the product algebra $%
\mathcal{M}_{t]}\sim \mathcal{W}_{t]}\bar{\otimes}\mathcal{N}_{t]}$
generated by $Y_{e}^{t]}=\left\{ Y_{i}^{r}:t\in \mathrm{I}_{0}^{t},i\in
I_{e}\right\} $ corresponding to the union $I_{e}\subseteq \left\{ 1,\ldots
,d\right\} $ of disjoint index subsets $I_{W}$ and $I_{N}$. Also let $%
\mathcal{B}_{[t}$ denote the future nonabelian W*-algebra generated by the
system operators $X\left( s\right) \vdash \gamma _{s}\left( \mathfrak{b}%
\right) $ for $s\geq t$. From the compatibility of the output operators, we
have quantum causality condition $\mathcal{M}_{t]}\subset \mathcal{B}%
_{[t}^{\prime }$, so again we have a unique well defined conditional
expectation $\epsilon ^{t}=\hat{\varpi}_{[t}^{\star }$ given by the
posterior states $\hat{\varpi}_{[t}$ on $\mathcal{B}_{[t}$ onto $\mathcal{M}%
_{t]}$.

The conditional expectation $\hat{x}_{r}^{t}=\epsilon ^{t}[X\left(
t+r\right) ]$ gives for any $r\geq 0$ the best prediction of $X\left(
t+r\right) =\upsilon _{r}^{t}\left( X\left( t\right) \right) $ as least
squares estimator of any operator $X\left( t\right) =\upsilon _{t}\left( 
\check{x}\otimes I\right) $ evolved to the time $t^{\prime }=t+r$
conditional on the output operators $Y_{e}^{t]}$ and so is equivalent to a
classical random variable on the space of measurement trajectories $\Omega
_{t]}=\left\{ \omega _{i}^{r}:r\in \mathrm{I}_{0}^{t},i\in I_{e}\right\} $
such that $\omega _{i}^{r}$ is an eigenvalue of $Y_{i}^{r}$. This
conditional expectation for is most conveniently written in the Schr\"{o}%
dinger picture as $\hat{x}_{r}^{t}=\left\langle \hat{\varsigma}^{t}|\check{%
\tau}_{r}^{t}\left( \check{x}\right) \right\rangle $ in terms of the
expected CP hemigroup $\check{\tau}_{r}^{t}$ and posterior states $\hat{%
\varsigma}^{t}$ is defined by the relation%
\begin{equation*}
\left\langle \hat{\varsigma}^{t}|\check{x}\right\rangle :=\epsilon
^{t}[\upsilon _{t}\left( \check{x}\otimes I\right) ]=\left\langle \hat{\varpi%
}_{[t}|\upsilon _{t}\left( \check{x}\otimes I\right) \right\rangle .
\end{equation*}%
In the case of product state $\varpi _{\lbrack r}=\varsigma \otimes \varrho
_{r}$ with infinitely divisible $\varrho _{r}$ realized on $\mathcal{A}_{r}=%
\mathcal{B}\left( \mathcal{F}_{r}\right) $ by the vacuum state, the
posterior state is given for any $t>r$ as $\hat{\varsigma}\left( t\right)
=\phi _{r}^{t-r}\left( \varsigma \right) $ by a hemigroup $\left\{ \phi
_{r}^{s}\right\} $ of nonlinear transformations of a starting state $%
\varsigma =\hat{\varsigma}\left( r\right) $ on the object algebra $\mathfrak{%
b}$, resolving the quantum nonlinear filtering (\emph{Belavkin}) equation 
\begin{equation}
\mathrm{d}\hat{\varsigma}\left( t\right) =\lambda \left[ \hat{\varsigma}%
\right] \left( t\right) \mathrm{d}t+\sum_{i\in I_{e}}\delta ^{i}\left( \hat{%
\varsigma}\right) \left( t\right) \hat{M}_{i}\left( \mathrm{d}t\right) .\;\;
\label{eq GenBeleq}
\end{equation}%
Here $\lambda $ is Lindblad generator of the adjoint equation (\ref{eq
mastereq}), and quantum filtering coefficients $\delta ^{i}$ against the
innovation martingales%
\begin{equation*}
\hat{M}_{i}\left( \mathrm{I}_{r}^{s}\right) =Y_{i}\left( \mathrm{I}%
_{r}^{s}\right) -\int_{r}^{r+s}\epsilon _{r}^{t}\left[ Y_{i}\left( \mathrm{d}%
t\right) \right]
\end{equation*}%
on the time intervals $\mathrm{I}_{t}^{s}$ where first specified in \cite%
{Be88a} as the functionals of the posterior quantum states $\hat{\varsigma}%
_{r}^{s}=\hat{\varsigma}_{r}\left( r+s\right) =\phi _{r}^{s}\left( \varsigma
\right) $ resolving It\^{o} equation (\ref{eq GenBeleq}) for $\hat{\varsigma}%
\left( r\right) =\varsigma $.

We present here two separate cases of Belavkin quantum filtering equation
corresponding to the diffusive and counting measurements. For rigorous
derivation see \cite{Be90e}, and for the most general mixed case we refer to 
\cite{Bel92b}.

\subsubsection{Diffusive Belavkin equation}

The diffusive version \cite{Be88a},\cite{Bel89},\cite{Be90e} of Belavkin
quantum filtering equation corresponding to continuous observation of the
diffusive row-vector $\boldsymbol{Y}_{W}^{t}=\upsilon _{t}\left(
W_{I_{e}}^{t}\right) =\left( Y_{i}^{t}\right) _{i\in I_{W}}$ indexed by a
fixed set $I_{e}=I_{W}$ specifying the \emph{estimation channels} is a
classical non-linear stochastic differential equation given by 
\begin{equation}
\mathrm{d}\hat{\varsigma}=\lambda \left[ \hat{\varsigma}\right] \mathrm{d}%
t+\sum_{i\in I_{W}}\delta ^{i}(\hat{\varsigma})(Y_{i}\left( \mathrm{d}%
t\right) -\left\langle \hat{\varsigma}|L_{i}+L_{i}^{\ast }\right\rangle 
\mathrm{d}t).  \label{eq filt1}
\end{equation}%
Here $\lambda \left( \varsigma \right) =\sum_{i}L^{i}\varsigma L^{i\ast
}-K\varsigma -\varsigma K^{\ast }$ is defined by the left adjoints $%
L^{i}=L_{i}^{\sharp }$, $K=-L^{\sharp }$ of the Lindblad operators in (\ref%
{eq Lindblad}), and 
\begin{equation*}
\delta ^{i}\left( \varsigma \right) =\varsigma L^{i\ast }+L^{i}\varsigma
-\left\langle \varsigma |L_{i}+L_{i}^{\ast }\right\rangle \varsigma
\end{equation*}%
is the nonlinear \emph{fluctuation coefficient} such that $\left\langle
\delta ^{i}\left( \varsigma \right) |\check{1}\right\rangle =0$ for any $%
\varsigma \in \mathfrak{b}_{\star }$ with respect to the pairing of $%
\mathfrak{b}$ and $\mathfrak{b}_{\star }$.

\subsubsection{Counting Belavkin equation}

The counting, or quantum jump version \cite{Be88a, Be89a}, \cite{Be90e} of
Belavkin quantum filtering equation corresponding to counting observation of
the number processes $\boldsymbol{Y}_{N}^{t}=\upsilon _{t}\left(
N_{I_{e}}^{t}\right) =\left( Y_{i}^{t}\right) _{i\in I_{N}}$ indexed by $%
I_{e}=I_{N}$ is given as a stochastic differential equation in the classical
It\^{o} form by 
\begin{equation}
\mathrm{d}\hat{\varsigma}=\lambda \left[ \hat{\varsigma}\right] \mathrm{d}%
t+\sum_{i\in I_{N}}\delta ^{i}\left( \hat{\varsigma}\right) (Y_{i}\left( 
\mathrm{d}t\right) -\langle \hat{\varsigma}|L_{i}L_{i}^{\ast }\rangle 
\mathrm{d}t)  \label{eq filt2}
\end{equation}%
for a counting measurement in the field. Here $\lambda \left( \varsigma
\right) =\sum_{i}L^{i}\varsigma L^{i\ast }-K\varsigma -\varsigma K^{\ast }$
such that $\left\langle \lambda \left( \varsigma \right) |\check{1}%
\right\rangle =0$ for any $\varsigma \in \mathfrak{b}_{\star }$ and 
\begin{equation}
\delta ^{i}(\varsigma )=\alpha ^{i}\left( \varsigma \right) -\varsigma
,\;\;\alpha ^{i}\left( \varsigma \right) =\frac{L^{i}\varsigma L^{i\ast }}{%
\langle \varsigma |L_{i}L_{i}^{\ast }\rangle }  \label{eq difference}
\end{equation}%
is the non-linear \emph{normalized difference coefficient} defined by $%
L^{i}=L_{i}^{\sharp }$ (=$L_{i}^{\ast }$ for the trace pairing of $\mathfrak{%
b}$ with $\mathfrak{b}_{\star }$) such that $\left\langle \delta ^{i}\left(
\varsigma \right) |\check{1}\right\rangle =0$ for any $\varsigma \in 
\mathfrak{b}_{\star }$ with respect to the standard pairing of $\mathfrak{b}%
_{\star }$ and $\mathfrak{b}$.

\section{Optimal Quantum Feedback Control}

We now couple the system to a control force (row-vector) $\boldsymbol{u}%
=\left( u_{i}\right) _{i\in I_{f}}$ via \emph{forward} (feedback) channels
indexed by a finite set $I_{f}$, $\left\vert I_{f}\right\vert =d_{f}$. The
force perturbs open quantum dynamics described by unitary cocycle $\left\{
U_{r}\right\} $ by making it causally dependent on each time interval $%
[r,r+s)$ on the control segment $\boldsymbol{u}_{r}^{s}=u\left( t\right)
:t\in \lbrack r,r+s))$ such that $U_{\boldsymbol{u}}^{r}\left( r+s\right)
=U_{s}^{r}\left( \boldsymbol{u}_{r}^{s}\right) $. The family $\left\{
U_{s}^{r}\left( \boldsymbol{u}_{r}^{s}\right) \right\} $ is assumed to
satisfy QS equation (\ref{eq MultiHP}) with controlled pseudo-unitary germ $%
\boldsymbol{\check{s}}\left( t,\boldsymbol{u}\right) =\boldsymbol{\check{s}}%
\left( \boldsymbol{u}\left( t\right) \right) $ such that its solution forms
a hemigroup%
\begin{equation*}
U_{r}^{t}\left( \boldsymbol{u}_{t}^{r}\right) U_{s}^{t+r}\left( \boldsymbol{u%
}_{t+r}^{s}\right) =U_{r+s}^{t}\left( \boldsymbol{u}_{t}^{r+s}\right) .
\end{equation*}%
for any composed segment $\boldsymbol{u}_{t}^{r+s}=\left( \boldsymbol{u}%
_{t}^{r},\boldsymbol{u}_{t}^{r+s}\right) $. This defines a hemigroup\ $%
\left( \gamma _{s}^{r}\right) $ of time-dependent interaction dynamics 
\begin{equation*}
\gamma _{\boldsymbol{u}}^{r}\left( t,\check{x}\right) =U_{\boldsymbol{u}%
}^{r}(t)(\check{x}\otimes I)U_{\boldsymbol{u}}^{r\ast }(t)=\gamma
_{t-r}^{r}\left( \boldsymbol{u}_{r}^{t-r},\check{x}\right)
\end{equation*}%
satisfying the controlled Langevin equation (\ref{eq FlowX}) with $%
\boldsymbol{\Sigma }_{\boldsymbol{u}}\left( t\right) $ given by time evolved
germ $\boldsymbol{\check{\sigma}}_{\boldsymbol{u}}\left( t\right) =%
\boldsymbol{\check{\sigma}}\left( \boldsymbol{u}\left( t\right) \right) $.

Following the original Belavkin's formulation \cite{Bel79},\cite{Bel83},\cite%
{Bel99} of quantum optimal control theory, we assume that the quality of a
control process on a quantum object over a finite period $[r,T)$ with
starting product state $\varpi _{\lbrack r}=\varsigma \otimes \varrho _{r}$
on the object plus noise algebra $\mathfrak{b}\otimes \mathcal{A}_{r}$ is
judged by the integral expectation 
\begin{equation}
\mathsf{J}_{r}(\varsigma ,\boldsymbol{u}_{r})=\int_{r}^{T}\left\langle
\varpi _{\lbrack r}|C\left( \boldsymbol{u}_{r},t\right) \right\rangle 
\mathrm{d}t+\left\langle \varpi _{\lbrack r}|S\left( \boldsymbol{u}%
_{r},T\right) \right\rangle  \label{eq QExpCost}
\end{equation}%
of the \emph{operator-valued cost functionals} of $\boldsymbol{u}_{r}$ given
by the evolved object operator-valued measurable\ positive cost function $%
\check{c}\left( t\right) :\mathbb{U}\rightarrow \mathfrak{b}$ and a terminal
positive cost operator $\check{s}\in \mathfrak{b}$ in%
\begin{equation}
C\left( \boldsymbol{u}_{r},t\right) =\gamma _{r,\boldsymbol{u}%
_{r}}^{t-r}\left( \check{c}(\boldsymbol{u})\right) ,\;S\left( \boldsymbol{u}%
_{r},T\right) =\gamma _{r,\boldsymbol{u}_{r}}^{T-r}\left( \check{s}\right)
\label{eq OpValuedCost2}
\end{equation}%
for self-adjoint positive system operators $\check{c}(\boldsymbol{u}\left(
t\right) ),\check{s}\vdash \mathcal{A}$. An alternative problem of \emph{%
risk-sensitive} control has also been studied by James \cite{Jam04,Jam05}
where the cost is exponentiated to enforce higher penalties for undesirable
behavior.

\emph{Coherent control} of open quantum dynamics uses field channels indexed
by a subset $I_{f}\subseteq \left\{ 1,\ldots ,d\right\} $. It is realized by
controlling quantum noise in these channels by $\boldsymbol{u}\in \mathbb{R}%
^{d_{f}}$ via their coherent states. One can start with uncontrolled
dynamics described by Hamiltonian and Lindbladian operators $H_{0}=\hbar \Im
\left( R_{+}^{-}\right) $ and $K_{0}^{i}=-R_{+}^{i}$ for $R_{+}^{\iota }=%
\check{s}_{+}^{\iota }$, $\iota =-,i$, defining QS unitary evolution by (\ref%
{eq MultiHP}) with arbitrary unitary scattering $S_{\circ }^{\circ }=\check{s%
}_{\circ }^{\circ }$, and apply coherent conditional expectation $\varphi
_{r,\xi }^{\star }$ to corresponding QS flow $\gamma _{s}^{r}\left( \check{x}%
\right) $ with $\xi \left( t\right) =\frac{\mathrm{i}}{\hbar }\boldsymbol{u}%
^{\intercal }\left( t\right) $ defining controlling field expectations%
\begin{equation*}
\left\langle \varphi _{t,\xi }|\hbar \Im \left[ A_{+}^{i}\left( \mathrm{I}%
_{t}^{s}\right) \right] \right\rangle =\int_{0}^{s}u_{i}\left( t+s\right) 
\mathrm{d}s\;\;\forall i\in I_{f}.
\end{equation*}%
This effectively results in change of the Hamiltonian $H_{0}$ and all
operators $K_{0}^{i}$ with $i\in I_{f}$ to 
\begin{equation}
H_{\boldsymbol{u}}=H_{0}+u_{i}\Re \left( K_{0}^{i}\right) ,\;K_{\boldsymbol{u%
}}^{i}=K_{0}^{i}+\frac{\mathrm{i}}{\hbar }u_{i},  \label{eq H&Lcohcontr}
\end{equation}%
and no change for other $K_{0}^{i}$ with $i\notin I_{f}$ (The summation is
taken only over $i\in I_{f}$.) The resulting conditioned dynamics $\check{%
\tau}_{\boldsymbol{u}}^{r}\left( t,\check{x}\right) $ satisfies time
dependent Lindblad equation which can be written in the form (\ref{eq
Lindbdecomp}) as%
\begin{equation}
\check{\lambda}_{\boldsymbol{u}}\left( \check{x}\right) =\frac{\mathrm{i}}{%
\hbar }\left[ H_{2\boldsymbol{u}},\check{x}\right] +\check{\lambda}%
_{R}\left( \check{x}\right) +\sum_{i,k\in I_{f}}K_{\boldsymbol{u}}^{i\ast }%
\check{\rho}_{k}^{i}\left( \check{x}\right) K_{\boldsymbol{u}}^{k}
\label{eq Lindbcohcontr}
\end{equation}%
with $\check{\rho}_{k}^{i}\left( \check{x}\right) =\check{\sigma}%
_{k}^{i}\left( \check{x}\right) -\delta _{k}^{i}\check{x}$, without change
of the part $\check{\lambda}_{R}$ (but with doubled $\boldsymbol{u}$ in the
Hamiltonian $H_{\boldsymbol{u}}$).

We are going to consider quantum feedback control problem in which it is
natural to assume that the forward feedback control channels are disjoint to
the set $I_{e}=I_{W}\cup I_{N}$ of \emph{estimation}. This is achieved by
considering coherent controls in the channels $I_{f}$ such that $I_{f}\cap
I_{e}=\emptyset $. In this case the controlling amplitude $\xi \left(
t\right) \in \mathfrak{k}_{f}$ is orthogonal to the subspace $\mathfrak{k}%
_{e}=\mathfrak{k}_{W}\oplus \mathfrak{k}_{N}$ of observation channels, so
the output equations (\ref{eq flowW}), (\ref{eq flowN}) are not affected by
the coherent control which will simplify optimal feedback control problem
which we solve by applying dynamical programming to coherent controlled
quantum states. The controlled posterior density operator $\hat{\varsigma}%
\left( t\right) =\hat{\varsigma}_{r,\boldsymbol{u}}^{t-r}$ can then be
obtained from the relevant uncontrolled filtering equation by replacing
Lindblad generator $\lambda =\check{\lambda}_{\star }$ in (\ref{eq GenBeleq}%
) by time dependent $\lambda _{\boldsymbol{u}}\left( t\right) =\lambda
\left( \boldsymbol{u}\left( t\right) \right) $, and so we now have a
controlled time dependent nonlinear filtering dynamics $\hat{\varsigma}_{r,%
\boldsymbol{u}}^{s}=\hat{\varsigma}_{r}^{s}\left( \boldsymbol{u}%
_{r}^{s}\right) $ satisfying Belavkin equation of the form, say (\ref{eq
filt1}) or\ (\ref{eq filt2}), in which the fluctuating part under the above
coherent control assumption is independent of $\boldsymbol{u}\left( t\right) 
$.

\subsection{Dynamical programming of quantum states}

In the search for optimal control inputs, it is desirable to allow the
control to be determined in terms of measurement results on the system,
particularly in the quantum setting where quantum noises introduce an
inevitable stochastic nature. A feedback strategy $\boldsymbol{\varkappa }%
_{[0}$ consists of measurable maps $\boldsymbol{\varkappa }\left( t\right) $
which give for each $0\leq t<T$ an operator-valued control law $\boldsymbol{u%
}\left( t\right) =\boldsymbol{\varkappa }(t,\boldsymbol{Y}_{e}^{t]})$ as a
function of the current and previous commuting output operators $\boldsymbol{%
Y}_{e}^{t]}=\left\{ \boldsymbol{Y}_{e}^{r}:r\in (0,t]\right\} $. Thus the
control law $\boldsymbol{\varkappa }\left( t\right) $ is realized by an
adapted random vector variable $\boldsymbol{\varkappa }(t,\omega _{t]})$ on
the probability space $(\Omega _{t]},\mathfrak{A}_{t]},\mathbb{P}_{t]})$ of
output measurement results in the value space $\mathbb{U}$ of admissible
control inputs in the spectral representation%
\begin{equation*}
\boldsymbol{u}\left( t\right) =\int_{\Omega _{t]}}\boldsymbol{\varkappa }%
(t,\omega _{t]})E_{0}\left( \mathrm{d}\omega _{t]}\right) \equiv \boldsymbol{%
\varkappa }(t,\boldsymbol{Y}_{e}^{t]}).
\end{equation*}%
We denote the space of admissible operator-valued feedback controls on the
interval $[r,r+s)$ by $\mathbb{U}_{r}^{s}\left( \boldsymbol{Y}_{e}\right) $.
Note that no measurement results are available initially, so the initial
control $\boldsymbol{u}\left( 0\right) $ is deterministic and also no
controls are applied at the termination time $T$. It is too restrictive to
consider only continuous sample paths $\left\{ \boldsymbol{\varkappa }%
(t,\omega _{e}^{t}):t>0\right\} $, since for example the Poisson process $\{%
\boldsymbol{Y}_{N}^{t}\}$ certainly does not have continuous sample paths.
Instead we give the following definition of an admissible strategy.

\begin{definition}
An admissible feedback control strategy $\boldsymbol{\varkappa }_{r}=\{%
\boldsymbol{\varkappa }\left( t\right) :r\leq t<T\}$ determines randomized
control laws $\boldsymbol{u}\left( t\right) =\boldsymbol{\varkappa }(t,%
\boldsymbol{Y}_{e}^{t]})$ at each time $t>0$ which realize values in $%
\mathbb{U}$ and form c\`{a}dl\`{a}g sample paths $\left\{ \boldsymbol{%
\varkappa }\left( t,\omega \right) :t\geq r\right\} $ \footnote{%
Right-continuous paths ($\lim_{h\rightarrow 0}\mu _{t+h}(\omega ^{t+h})=\mu
_{t}(\omega ^{t})$) having well defined left limits $\mu _{t}(\omega
^{t})_{-}=\lim_{h\rightarrow 0}\mu _{t-h}(\omega ^{t-h})$}. Moreover, an
admissible strategy $\boldsymbol{\varkappa }_{t}^{o}$ shall be called \emph{%
optimal} if it realizes the infimum 
\begin{equation}
\mathsf{S}(r,\varsigma )=\inf_{\boldsymbol{\varkappa }_{r}\in \mathcal{K}%
_{r}}\mathsf{J}_{r}(\varsigma ,\boldsymbol{\varkappa }_{r})=\mathsf{J}%
_{r}(\varsigma ,\boldsymbol{\varkappa }_{r}^{o})  \label{eq ActionS}
\end{equation}%
over the space $\mathcal{K}_{r}$ of admissible feedback control strategies,
where $\mathsf{J}_{r}(\varsigma ,\boldsymbol{\varkappa }_{r})$ is the
expected cost for the control process determined by the feedback strategy $%
\boldsymbol{\varkappa }_{r}$.
\end{definition}

It is a simple exercise to show that under a feedback strategy, the output
operators once again form a QND measurement with respect to the controlled
system operators which justifies the existence of the conditional
expectation 
\begin{equation}
\epsilon ^{t}\big[\gamma _{r,\boldsymbol{u}}^{t-r}(\check{x}]\big]%
=\left\langle \hat{\varsigma}_{r,\boldsymbol{u}}^{t-r}|\check{x}\right\rangle
\label{eq Contcondexp}
\end{equation}%
with respect to the output operators $\boldsymbol{Y}_{e}^{t]}$ for $t>r$ .
It is given by the posterior state $\hat{\varsigma}_{r,\boldsymbol{u}}^{t-r}=%
\hat{\varsigma}_{r,\boldsymbol{u}}\left( t\right) $ as solution of Belavkin
equation (\ref{eq GenBeleq}) which now has dependence on the chosen control $%
\boldsymbol{u}_{s}^{r}$ inputs through the dynamics $\gamma _{r}^{s}\left( 
\boldsymbol{u}_{s}^{r},\check{x}\right) $, and, given an initial condition $%
\hat{\varsigma}_{r,\boldsymbol{u}}\left( r\right) =\varsigma $, it does not
really depend on $\boldsymbol{Y}_{e}^{r]}$ due to Markovianity of the
process $\hat{\varsigma}_{\boldsymbol{u}}^{t}$ proved in \cite{Bel83}. The
existence of this conditional expectation permits the following theorem
which lies at the heart of quantum feedback control.

\begin{theorem}
The expectation (\ref{eq QExpCost}) of the operator valued cost (\ref{eq
OpValuedCost2}) when a feedback control strategy $\boldsymbol{\varkappa }%
_{r} $ is in operation can be written as a classical expectation 
\begin{equation}
\mathsf{J}_{r}(\varsigma ,\boldsymbol{\varkappa })=\int_{\Omega }\mathsf{J}%
_{r}^{\omega }(\varsigma ,\boldsymbol{\varkappa })\mathbb{P}\left( \mathrm{d}%
\omega \right) \equiv \mathbb{E}_{\Omega }[\mathsf{J}_{r}^{\bullet
}(\varsigma ,\boldsymbol{\varkappa })]  \label{eq CosttogoEDB}
\end{equation}%
of the random cost-to-go function 
\begin{equation}
\mathsf{J}_{r}^{\omega }(\varsigma ,\boldsymbol{\varkappa }%
)=\int_{r}^{T}\langle \hat{\varsigma}_{r,\boldsymbol{\varkappa }%
}^{t-r}\left( \omega \right) |\check{c}(\boldsymbol{\varkappa }(t,\omega
_{t]})\rangle \mathrm{d}t+\langle \hat{\varsigma}_{r,\boldsymbol{\varkappa }%
}^{T-r}\left( \omega \right) |\check{s}\rangle
\end{equation}%
where $\hat{\varsigma}_{r,\varkappa }^{t-r}\left( \omega \right) $ is the
solution $\hat{\varsigma}\left( t,\omega \right) $ to the controlled
filtering equation corresponding to the chosen measurement process $%
\boldsymbol{Y}_{e}^{t}$ classically represented as $\boldsymbol{\omega }%
_{e}^{t}$ for the feedback strategy $\boldsymbol{\varkappa }$ with the
initial condition $\hat{\varsigma}\left( r\right) =\varsigma $.
\end{theorem}

\begin{proof}
Using the existence and state invariance of the conditional expectation and
the classical isomorphism proved in the first Section, it is straight
forward application of the formula (\ref{eq Contcondexp}) to $\check{x}%
\left( t\right) =\check{c}\left( \boldsymbol{u}\left( t\right) \right) $ and 
$\check{x}\left( T\right) =\check{s}$ in \ref{eq QExpCost}. We can write the
expected cost as 
\begin{eqnarray*}
\mathsf{J}_{r}(\varsigma ,\boldsymbol{\varkappa }_{r})
&=&\int_{r}^{T}\left\langle \varsigma \otimes \varrho _{r}|\gamma _{r,%
\boldsymbol{u}}^{t-r}(\check{c}(\boldsymbol{u}\left( t\right)
))\right\rangle \mathrm{d}t+\left\langle \varsigma \otimes \varrho
_{r}|\gamma _{r,\boldsymbol{u}}^{T-r}\left( \check{s}\right) \right\rangle \\
&=&\int_{r}^{T}\left\langle \varpi _{\lbrack r},\epsilon _{r}^{t}\left[
\gamma _{r,\boldsymbol{u}}^{t-r}\left( \check{c}(\boldsymbol{u}\left(
t\right) )\right) \right] \right\rangle \mathrm{d}t+\left\langle \varpi
_{\lbrack r}|\epsilon _{r}^{T}\left[ \gamma _{r,\boldsymbol{u}}^{T-r}\left( 
\check{s}\right) \right] \right\rangle \\
&=&\int_{r}^{T}\mathbb{E}_{\Omega }\left[ \left\langle \hat{\varsigma}_{r,%
\boldsymbol{u}}^{t-r}|\check{c}(\boldsymbol{u}\left( t\right) )\right\rangle %
\right] \mathrm{d}t+\mathbb{E}_{\Omega }\left[ \left\langle \hat{\varsigma}%
_{r,\boldsymbol{u}}^{T-r}|\check{s}\right\rangle \right] .
\end{eqnarray*}
\end{proof}

This allows us to treat the quantum control problem as a classical control
problem on the space of quantum states. We define the expected cost-to-go by
the classical expression (\ref{eq CosttogoEDB}) for a truncated admissible
strategy $\boldsymbol{\varkappa }_{r}\in \mathcal{K}_{r}$ when starting in
an arbitrary state $\varsigma $ at time $r$ where $\hat{\varsigma}_{r,%
\boldsymbol{\varkappa }}^{t-r}\left( \omega \right) $ is evaluated at $%
\omega \in \Omega _{t]}$ solution $\hat{\varsigma}\left( t\right) =\phi _{r,%
\boldsymbol{\varkappa }}^{t-r}\left( \varsigma \right) $ to the controlled
filtering equation for these initial conditions at $t=r$ and the initial
strategy $\boldsymbol{\varkappa }_{r}$ for $r<t\leq T$.

\begin{theorem}
Suppose that $\mathsf{S}(t,\varsigma )$ is a functional which is
continuously differentiable in $t$, has continuous Fr\`{e}chet derivatives
of all order with respect to $\varsigma $ and satisfies 
\begin{equation}
\inf_{\boldsymbol{u}\in \mathbb{U}(Y^{t})}\Big\{\langle \varsigma ,\check{c}(%
\boldsymbol{u})\rangle +\mathbb{E}_{\Omega }^{t}\big[\frac{\mathrm{d}}{%
\mathrm{d}t}\mathsf{S}(t,\hat{\varsigma})\big]\Big\}=0.  \label{eq VerThm}
\end{equation}%
for all $0<t<T$ and $\mathsf{S}(T,\varsigma )=\langle \varsigma ,\check{s}%
\rangle $ for all $\varsigma \in \mathcal{S}$. Suppose also that $%
\boldsymbol{\varkappa }^{o}$ is the strategy built from the control laws
attaining these minima within a convex space $\mathbb{U}$ of admissible
control values, then $\mathsf{S}(t,\varsigma )$ is the functional which
minimizes \eqref{eq CosttogoEDB} and $\boldsymbol{\varkappa }^{o}$ is the
optimal strategy for the control problem\footnote{%
Additional technical assumptions and mathematical rigour are required to
formalise the proof of this theorem when dealing with unbounded operators
which is beyond the scope of this paper. See recommended texts e.g.\ \cite%
{Ben92},\cite{KuV86} for a formal classical treatment.}.
\end{theorem}

\begin{proof}[Sketch proof]
Let $\{\boldsymbol{u}\left( t\right) \}$, $\{\hat{\varsigma}_{0}^{t}\}$ be
any control and state trajectories resulting from an admissible strategy $%
\boldsymbol{\varkappa }$ on the initial state $\varsigma $, then from 
\eqref{eq
VerThm}, we have the inequality 
\begin{equation*}
\langle \hat{\varsigma}_{0}^{t},\check{c}(\boldsymbol{u}\left( t\right)
)\rangle +\mathbb{E}_{\Omega }\big[\frac{\mathrm{d}}{\mathrm{d}t}\mathsf{S}%
(t,\hat{\varsigma}_{0}^{t})\big]\geq 0
\end{equation*}%
which we integrate over $[0,T]$ and take the expectation $\mathbb{E}_{\Omega
}$ to obtain due to convexity 
\begin{equation*}
\mathbb{E}_{\Omega }\big[\int_{0}^{T}\langle \hat{\varsigma}_{0}^{t},\check{c%
}(\boldsymbol{u}\left( t\right) )\rangle \mathrm{d}t+\mathsf{S}(T,\hat{%
\varsigma}_{0}^{T})-\mathsf{S}(0,\hat{\varsigma}_{0}^{0})\big]\geq 0
\end{equation*}%
Since we have $\hat{\varsigma}_{0}^{0}=\varsigma $ initially, we can
rearrange and use the terminal condition to obtain 
\begin{equation*}
\mathsf{S}(0,\varsigma )\leq \mathbb{E}_{\Omega }\big[\int_{0}^{T}\langle 
\hat{\varsigma}_{0}^{t},\check{c}(\boldsymbol{u}\left( t\right) )\rangle 
\mathrm{d}t+\langle \hat{\varsigma}_{0}^{T},\check{s}\rangle \big]=\mathsf{J}%
_{0}(\varsigma ,\boldsymbol{\varkappa }).
\end{equation*}%
with equality when $\boldsymbol{\varkappa }=\boldsymbol{\varkappa }^{o}$ and
so the lower bound $\mathsf{S}(0,\varsigma )$ is attained, proving
optimality of $\boldsymbol{\varkappa }^{o}$.
\end{proof}

A choice of controlled filtering equation is required to determine the
stochastic trajectories $\mathrm{d}\hat{\varsigma}_{0}^{t}$ along which to
differentiate candidate solutions. The next two sections are concerned with
the examples of feedback control with respect to QND measurements of the
diffusive process $\{\boldsymbol{Y}_{W}^{t}\}$ and the counting process $\{%
\boldsymbol{Y}_{N}^{t}\}$ respectively.

\subsection{Quantum state Bellman equations}

First let us introduce notations of differential calculus on the quantum
state space $\mathfrak{S}\subset \mathfrak{b}_{\star }$. Let $\mathsf{F}=%
\mathsf{F}\left[ \cdot \right] $ be a (nonlinear) functional $\varsigma
\mapsto \mathsf{F}\left[ \varsigma \right] $ on $\mathfrak{S}$, then we say
it admits a (Fr\`{e}chet) derivative if there exists an $\mathfrak{b}$%
-valued function $\nabla _{\varsigma }\mathsf{F}\left[ \cdot \right] $ on $%
\mathfrak{b}_{\star }$ such that%
\begin{equation}
\lim_{h\rightarrow 0}\frac{1}{h}\left\{ \mathsf{F}\left[ \cdot +h\tau \right]
-\mathsf{F}\left[ \cdot \right] \right\} =\left\langle \tau |\nabla
_{\varsigma }\mathsf{F}\left[ \cdot \right] \right\rangle =\left( \tau
,\nabla _{\varsigma }\mathsf{F}\left[ \cdot \right] \right)  \label{nabla_Q}
\end{equation}%
for each $\tau =\tau ^{\ast }\in \mathfrak{b}_{\star }$. In the same spirit,
a Hessian $\nabla _{\varsigma }^{\otimes 2}\equiv \nabla _{\varsigma
}\otimes \nabla _{\varsigma }$ can be defined as a mapping from the
functionals on to the $\mathfrak{b}_{sym}^{\otimes 2}$-valued functionals,
via 
\begin{gather}
\lim_{h,h^{\prime }\rightarrow 0}\frac{1}{hh^{\prime }}\left\{ \mathsf{F}%
\left[ \cdot +h\tau +h^{\prime }\tau ^{\prime }\right] -\mathsf{F}\left[
\cdot +h\tau \right] -\mathsf{F}\left[ \cdot +h^{\prime }\tau ^{\prime }%
\right] +\mathsf{F}\left[ \cdot \right] \right\}  \notag \\
=\left\langle \tau \otimes \tau ^{\prime }|\nabla _{\varsigma }\otimes
\nabla _{\varsigma }\mathsf{F}\left[ \cdot \right] \right\rangle .
\end{gather}%
and we say that the functional is twice continuously differentiable whenever 
$\nabla _{\varsigma }^{\otimes 2}\mathsf{F}\left[ \cdot \right] $ exists and
is continuous in the trace norm topology.

With the customary abuses of differential notation, we have for instance 
\begin{equation*}
\nabla _{\varsigma }f\left( \left\langle \varsigma |X\right\rangle \right)
=f^{\prime }\left( \left\langle \varsigma |X\right\rangle \right) X
\end{equation*}%
for any differentiable function $f$ of the scalar $x=\left\langle \varsigma
|X\right\rangle $.

\subsubsection{Diffusive Bellman equation}

We have the It\^{o} rule $\mathrm{d}Y_{i}^{t}\mathrm{d}Y_{k}^{t}=\delta
_{ik}I\mathrm{d}t$ with $i,k\in I_{e}=I_{W}$ for the increments of the
diffusive processes $Y_{i}^{t}=\upsilon _{t}\left( W_{i}^{t}\right) $ which
have the expectations $\mathbb{E}_{\Omega }^{t}[\mathrm{d}Y_{i}^{t}]=\langle
\varsigma |2\Re \left( L_{i}\right) \rangle \mathrm{d}t$, so using the
classical It\^{o} formula for the diffusive process $Y_{i}^{t}$ we can show 
\begin{equation}
\begin{split}
\mathbb{E}_{\Omega }^{t}[\frac{\mathrm{d}}{\mathrm{d}t}\mathsf{S}%
(t,\varsigma )]=& \frac{\partial }{\partial t}\mathsf{S}(t,\varsigma
)+\langle \varsigma |\check{\lambda}(\boldsymbol{u})[\nabla _{\varsigma }%
\mathsf{S}(t,\varsigma )]\rangle \\
& +\frac{1}{2}\sum_{i\in I_{W}}\langle \delta ^{i}(\varsigma )\otimes \delta
^{i}(\varsigma )|\nabla _{\varsigma }^{\otimes 2}\mathsf{S}(t,\varsigma
)\rangle
\end{split}
\label{eq Taylor1}
\end{equation}%
where $\nabla _{\varsigma }\mathsf{S}\in \mathfrak{b}$ denotes Fr\`{e}chet
derivative with respect to $\varsigma \in \mathfrak{b}_{\star }$, and $%
\nabla _{\varsigma }^{\otimes 2}\mathsf{S}\in \mathfrak{b}^{\bar{\otimes}2}:=%
\mathfrak{b}\bar{\otimes}\mathfrak{b}$ denotes Hessian applied to $\mathsf{S}
$. Observing$\ \frac{\partial }{\partial t}\mathsf{S}(t,\varsigma )$ is
independent of $\boldsymbol{u}$ leads to the following corollary.

\begin{corollary}
Suppose there exists a functional $\mathsf{S}(t,\varsigma )$ which is
continuously differentiable in $t$, has continuous first and second order Fr%
\`{e}chet derivatives with respect to $\varsigma $ and satisfies the
following Bellman equation 
\begin{eqnarray}
-\frac{\partial }{\partial t}\mathsf{S}(t,\varsigma ) &=&\inf_{\boldsymbol{u}%
\in \mathbb{U}}\Big\{\langle \varsigma |\check{c}(\boldsymbol{u})+\check{%
\lambda}(\boldsymbol{u})[\nabla _{\varsigma }\mathsf{S}(t,\varsigma
)]\rangle \Big\}  \label{eq Bellman1} \\
&&+\frac{1}{2}\sum_{i}\langle \delta ^{i}(\varsigma )\otimes \delta
^{i}(\varsigma )|\nabla _{\varsigma }^{\otimes 2}\mathsf{S}(t,\varsigma
)\rangle  \notag
\end{eqnarray}%
for all $t>0$, $\varsigma \in \mathcal{S}$ with the terminal condition $%
\mathsf{S}(T,\varsigma )=\langle \varsigma ,\check{s}\rangle $. Then the
strategy $\boldsymbol{\varkappa }^{o}\left( t,\boldsymbol{Y}_{W}^{t]}\right)
=\boldsymbol{u}^{o}\left( t,\hat{\varsigma}\right) $ built from the control
laws 
\begin{equation}
\boldsymbol{u}^{o}\left( t,\varsigma \right) =\arg \inf_{\boldsymbol{u}\in 
\mathbb{U}}\Big\{\langle \varsigma |\check{c}(\boldsymbol{u})+\check{\lambda}%
(\boldsymbol{u})[\nabla _{\varsigma }\mathsf{S}(t,\varsigma )]\rangle \Big\}
\label{eq ContLaw}
\end{equation}%
for $0\leq t<T$ is optimal for the feedback control problem based on
diffusive output measurements.
\end{corollary}

Note that last line in \ref{eq Bellman1} is precisely half of the Laplace
operator%
\begin{equation*}
\bigtriangleup \mathsf{S}\left( \varsigma \right) =\sum_{i\in
I_{W}}\left\langle \delta ^{i}\left( \varsigma \right) \otimes \delta
^{i}\left( \varsigma \right) |\nabla _{\varsigma }^{\otimes 2}\mathsf{S}%
(t,\varsigma )\right\rangle
\end{equation*}%
in the quantum state `coordinates' $\varsigma $ $\in \mathfrak{b}_{\star }$
as the sufficient coordinates from the preadjoint space of the algebra $%
\mathfrak{b}$.

\subsubsection{Counting Bellman equation}

We have the It\^{o} rules $\mathrm{d}Y_{i}^{t}\mathrm{d}Y_{k}^{t}=\delta
_{i}^{j}\delta _{k}^{j}\mathrm{d}Y_{j}^{t}$ for the increments of the
counting processes $Y_{j}^{t}=\upsilon \left( N_{j}^{t}\right) $, $j\in
I_{e}:I_{N}$ which have the expectations $\mathbb{E}_{\Omega }^{t}[\mathrm{d}%
Y_{i}^{t}]=\langle \varsigma |L_{i}\left( t\right) L_{i}^{\ast }\left(
t\right) \rangle \mathrm{d}t$, so using It\^{o} formula for the counting
processes $Y_{i}^{t}$ we we can show 
\begin{equation}
\begin{split}
\mathbb{E}_{\Omega }^{t}\left[ \frac{\mathrm{d}}{\mathrm{d}t}\mathsf{S}%
\left( t,\varsigma \right) \right] =& \frac{\partial }{\partial t}\mathsf{S}%
\left( t,\varsigma \right) +\left\langle \varsigma |\check{\lambda}\left( 
\boldsymbol{u}\right) \left[ \nabla _{\varsigma }\mathsf{S}\left(
t,\varsigma \right) \right] \right\rangle \\
& -\sum_{i\in I_{N}}\left\langle \varsigma |L_{i}L_{i}^{\ast }\right\rangle
\left\langle \alpha ^{i}\left( \varsigma \right) -\varsigma |\nabla
_{\varsigma }\mathsf{S}(t,\varsigma )\right\rangle \\
& +\sum_{i\in I_{N}}\left\langle \varsigma |L_{i}L_{i}^{\ast }\right\rangle
\left( \mathsf{S}\left( t,\alpha ^{i}\left( \varsigma \right) \right) -%
\mathsf{S}\left( t,\varsigma \right) \right)
\end{split}%
\end{equation}%
The last two lines can be written in the Feller form as $\frac{1}{2}\mu
_{i}\left( \varsigma \right) \bigtriangleup ^{i}\mathsf{S}\left( t,\varsigma
\right) $ as in the diffusive case in terms of doubled difference
combination $\mu _{i}\left( \varsigma \right) \bigtriangleup ^{i}\mathsf{S}%
\left( t,\varsigma \right) $ of the linear combination of differences 
\begin{equation*}
\mathsf{S}\left( t,\alpha ^{i}\left( \varsigma \right) \right) -\mathsf{S}%
\left( t,\varsigma \right) ,\;\;\mu _{i}\left( \varsigma \right)
=\left\langle \varsigma |L_{i}L_{i}^{\ast }\right\rangle
\end{equation*}%
and $\mu _{i}\left( \varsigma \right) \left\langle \delta ^{i}\left(
\varsigma \right) |\nabla _{\varsigma }\mathsf{S}\left( t,\varsigma \right)
\right\rangle $ in terms of $\delta ^{i}\left( \varsigma \right) =\alpha
^{i}\left( \varsigma \right) -\varsigma $. The formal Taylor expansion%
\begin{equation*}
\bigtriangleup ^{i}\mathsf{S}\left( t,\varsigma \right) =2\sum_{n=2}^{\infty
}\frac{1}{n!}\left\langle \delta ^{i}\left( \varsigma \right) ^{\otimes
n}|\nabla _{\varsigma }^{\otimes n}\mathsf{S}\left( t,\varsigma \right)
\right\rangle
\end{equation*}%
of each $\bigtriangleup ^{i}\mathsf{S}\left( t,\varsigma \right) =2\left( 
\mathsf{S}\left( \alpha \left( \varsigma \right) \right) -\mathsf{S}\left(
\varsigma \right) -\left\langle \delta ^{i}\left( \varsigma \right) |\nabla
_{\varsigma }\right\rangle \right) $ in terms of the higher order Fr\`{e}%
chet derivatives $\nabla _{\varsigma }^{\otimes n}$ starts from the Hessian $%
\nabla _{\varsigma }^{\otimes 2}=\nabla _{\varsigma }\otimes \nabla
_{\varsigma }$, determining the Laplace operators $\left\langle \delta
^{i}\left( \varsigma \right) ^{\otimes 2}|\nabla _{\varsigma }^{\otimes
2}\right\rangle $ for the diffusive approximation of this counting
measurement case.

Let us introduce the Pontryagin's `Hamiltonian' in the `coordinates' $\check{%
q}\left( \varsigma \right) =\varrho -\varsigma $ and `momenta' $\check{p}\in 
\mathfrak{b}$ as the Legendre-Fenchel transform%
\begin{equation*}
\mathsf{H}\left( \check{q},\check{p}\right) =\sup_{\boldsymbol{u}\in \mathbb{%
U}}\left\{ \left\langle \lambda \left( \boldsymbol{u}\right) \left[ \check{q}%
\right] \mid \check{p}\right\rangle -\mathsf{L}\left( \check{q},\boldsymbol{u%
}\right) \right\}
\end{equation*}%
of the `Lagrangian' $\mathsf{L}\left( \check{q},\boldsymbol{u}\right)
=\left\langle \varrho -\check{q}|\check{c}\left( \boldsymbol{u}\right)
\right\rangle $, where $\varrho =\check{q}-\varsigma $ is any stationary
element $\varrho =\varrho ^{\ast }$ in $\mathfrak{b}_{\star }$ such that $%
\lambda \left( \boldsymbol{u}\right) \left[ \varrho \right] =0$ for any $%
\boldsymbol{u}\in \mathbb{U}$ (e.g. $\varrho =0$). Then one can write
Bellman equation defining minimal expected cost-to-go (\ref{eq ActionS}) as
the action functional in the compact Jacobi-Feller form as follows, similar
as it was done for the diffusive case.

\begin{corollary}
Suppose there exists a functional $\mathsf{S}(t,\varsigma )$ which is
continuously differentiable in $t$, has continuous first order Fr\`{e}chet
derivatives with respect to $\varsigma $ and satisfies the following
Bellman-Jacobi-Feller equation 
\begin{eqnarray}
&&-\frac{\partial }{\partial t}\mathsf{S}(t,\varsigma )+\mathsf{H}\left( 
\check{q}\left( \varsigma \right) ,\nabla _{\varsigma }\mathsf{S}%
(t,\varsigma )\right)  \label{eq Bellman2} \\
&=&\frac{1}{2}\sum_{i\in I_{N}}\langle \varsigma \mid L_{i}L_{i}^{\ast
}\rangle \bigtriangleup ^{i}\mathsf{S}\left( t,\varsigma \right)
\end{eqnarray}%
for all $t>0$, $\varsigma \in \mathfrak{b}_{\star }$ with the terminal
condition $\mathsf{S}(T,\varsigma )=\langle \varsigma ,\check{s}\rangle $.
Then the strategy $\boldsymbol{\varkappa }^{o}\left( t,\boldsymbol{Y}%
_{N}^{t]}\right) =\boldsymbol{u}^{\mathrm{o}}\left( t,\hat{\varsigma}\right) 
$ built from the control laws 
\begin{equation}
\boldsymbol{u}^{\mathrm{o}}\left( t,\varsigma \right) =\arg \inf_{%
\boldsymbol{u}\in \mathbb{U}}\Big\{\langle \varsigma |\check{c}(\boldsymbol{u%
})+\check{\lambda}(\boldsymbol{u})[\nabla _{\varsigma }\mathsf{S}%
(t,\varsigma )]\rangle \Big\}
\end{equation}%
is optimal for the feedback control problem based on counting output
measurements.
\end{corollary}

Thus we have shown that without loss in optimality, one can reformulate the
unobservable quantum feedback control problem into a feedback problem based
on indirect QND measurements with feedback of the controlled conditional
density matrix. However, the corresponding Hamilton-Jacobi-Bellman equation
resulting from the minimization rarely has a regular solution $\mathsf{S}%
(t,\varsigma )$ from which to construct the optimal feedback laws. We now
study a specific quantum filtering and feedback case where such a control
solution can be explicitly found, which is familiar as the only such example
in the classical case.

\section{Application to a Linear Quantum Dynamical System}

Let $\check{x}_{\bullet }=\left( \check{x}_{1},\ldots ,\check{x}_{m}\right) $
be the row-vector of self-adjoint operators $\check{x}_{j}=\check{x}%
_{j}^{\ast }$, $j=1,...,m$ and $\mathbf{J}=(\mathrm{J}_{ik})$ be an
anti-symmetric real valued matrix defining the canonical commutation
relations (CCRs)%
\begin{equation}
\left[ \check{x}_{i},\check{x}_{k}\right] :=\check{x}_{i}\check{x}_{k}-%
\check{x}_{i}\check{x}_{i}=\mathrm{i}\hbar \mathrm{J}_{ik}\check{1},
\label{eq Xccr}
\end{equation}%
written in the matrix form as $\left[ \check{x}_{\bullet }^{\intercal },%
\check{x}_{\bullet }\right] =\mathrm{i}\hbar \mathbf{J}\check{1}$ where $%
\check{1}$ is the identity operator on $\mathfrak{h}$. Usually this is the
standard symplectic matrix $\mathbf{J}^{\intercal }=-\mathbf{J}=\mathbf{J}%
^{-1}$, but we may not assume that $\mathbf{J}$ is standard or
non-degenerate in order to include also the commuting random variables $%
\check{x}_{\bullet }=\left( \check{x}_{1},\ldots ,\check{x}_{m}\right) $ as
a special (classical) case. It is worth remarking at this point that the
noncommuting operators $\check{x}_{j}$ are secondary commuting (in the sense
of commutativity with all the commutants (\ref{eq Xccr})), and therefore
they must be unbounded in the Hilbert space $\mathfrak{h}$, affiliated to
the the generated algebra $\mathfrak{b}$.

We couple the open quantum system to $d_{e}=\left\vert I_{e}\right\vert $
estimation (side) channels with linear combinations $L_{i}=\sum_{j}\Lambda
_{ij}\check{x}_{j}$ indexed by a subset $I_{e}\subseteq \left\{ 1,\ldots
,d\right\} $, given by a complex-valued $d_{e}\times m$ matrix $\mathbf{%
\Lambda }_{e}=\left( \Lambda _{ij}\right) $ with $\Lambda _{ij}=0$ for $%
i\notin I_{e}$. In the $i$-th estimation channel, we perform a measurement
of the output $Y_{i}^{t}=\upsilon _{\infty }\left( 2\Re \left[
A_{i}^{-}\left( t\right) \right] \right) $ defined by $\upsilon _{t}|%
\mathcal{A}_{0}^{t}=\upsilon _{\infty }|\mathcal{A}_{0}^{t}$ where $2\Re %
\left[ A_{i}^{-}\right] =A_{i}^{+}+A_{-}^{i}$, $i\in I_{e}$ and the output
transformations is given by a quantum stochastic evolution $\upsilon _{t}$
on the algebra $\mathfrak{b}\otimes \mathcal{A}_{0}$ generated by the
canonical independent variables $\check{x}_{\bullet }$ and $A_{\circ }^{+}$.
The system is also coupled to $d_{f}=\left\vert I_{f}\right\vert $ feedback
(input) channels by the operators $K^{i}=L_{i}^{\ast }$ as linear
combinations of $\left( \check{x}_{j}\right) $, given by the row $%
\boldsymbol{L}_{f}=\check{x}_{\bullet }\mathbf{\Lambda }_{f}^{\intercal }$
of operators $L_{i}$, $i\in I_{f}$ for a complex-valued $d_{f}\times m$
matrix $\mathbf{\Lambda }_{f}=\left( \Lambda _{ij}\right) $ with $\Lambda
_{ij}=0$ for $i\notin I_{f}$, and we apply input controls with real-valued
components $u_{i}\left( t\right) \in \mathbb{R}$ in the row $\boldsymbol{u}%
=\left( u_{i}\right) _{i\in I_{f}}$ via the $i$-th feedback channel at time $%
t$. Both matrices $\mathbf{\Lambda }_{e}$, $\mathbf{\Lambda }_{f}$ may
depend on $t$, but they are always orthogonal such that $\mathbf{\Lambda }%
^{\dagger }\mathbf{\Lambda =\Lambda }_{e}^{\dagger }\mathbf{\Lambda }_{e}+%
\mathbf{\Lambda }_{f}^{\dagger }\mathbf{\Lambda }_{f}$, where $\mathbf{%
\Lambda }=\mathbf{\Lambda }_{e}+\mathbf{\Lambda }_{f}$ and $\mathbf{\Lambda }%
^{\dagger }$ is transposed to complex conjugated matrix $\mathbf{\Lambda }%
^{\ast }=\left( \overline{\Lambda _{ij}}\right) $.

Let the free dynamics of the quantum system be described by a quadratic
Hamiltonian $H_{0}=\frac{1}{2}\check{x}_{\bullet }\mathbf{M}^{-1}\check{x}%
_{\bullet }^{\intercal }$ for a symmetric real $m\times m$ matrix $\mathbf{M}%
^{-1}$. We now introduce a coherent control source \ separating the controls 
$\boldsymbol{u}\left( t\right) \mathbf{B}_{f}\mathrm{d}t$ in coherent
superposition with the noise $\hbar \Im \left( \mathrm{d}A_{\circ
}^{+}\right) \mathbf{B}_{f}$ in control channel, where $2\mathrm{i}\Im
\left( A_{i}^{+}\right) =A_{i}^{+}-A_{-}^{i}$, coming from the feedback
channel with $\mathbf{B}_{f}=2\func{Re}\mathbf{\Lambda }_{f}$, so that the
Hamiltonian in (\ref{eq Lindbcohcontr}) is modelled by 
\begin{equation}
H_{2\boldsymbol{u}}\left( t\right) =\frac{1}{2}\check{x}_{\bullet }\mathbf{M}%
^{-1}\check{x}_{\bullet }^{\intercal }+\boldsymbol{u}\left( t\right) \mathbf{%
B}_{f}\check{x}_{\bullet }^{\intercal }.  \label{eq lqH}
\end{equation}

Using CCR's (\ref{eq Xccr}) and assuming that the QS dynamics is purely
diffusive (no scattering, $S_{k}^{i}=\check{1}\delta _{k}^{i}$ in (\ref{eq
MultiHP})), we can easily evaluate Lindblad generator $\check{\lambda}_{%
\boldsymbol{u}}\left( \check{x}_{i}\right) $ for controlled CP hemigroup
dynamics $\check{\tau}_{\boldsymbol{u}}^{r}\left( t,\check{x}\right) =\check{%
\tau}_{t-r}^{r}\left( \boldsymbol{u},\check{x}\right) $ arising from time
dependent quadratic Hamiltonian (\ref{eq lqH}). Substituting $\check{x}_{j}$
into the decomposed generator (\ref{eq Lindbcohcontr}) with linear $L^{i}$
in $\check{x}_{j}$ and zero jump part, $\check{\rho}_{\nu }^{\mu }\left( 
\check{x}\right) =0$, we obtain $\check{\lambda}_{\boldsymbol{u}}\left( 
\check{x}_{i}\right) $ as linear transformation of $\check{x}_{\bullet
}=\left( \check{x}_{i}\right) $ written in vector form as 
\begin{equation*}
\check{\lambda}_{\boldsymbol{u}}\left( t,\check{x}_{\bullet }^{\intercal
}\right) =\mathbf{J}(\mathbf{M}^{-1}+\hbar \func{Im}\left( \mathbf{\Lambda }%
^{\dagger }\mathbf{\Lambda }\right) )\check{x}_{\bullet }^{\intercal }-%
\mathbf{C}_{f}\boldsymbol{u}^{\intercal }\left( t\right)
\end{equation*}%
where $\mathbf{C}_{f}^{\intercal }=\mathbf{B}_{f}\mathbf{J}$. From this, we
obtain the quantum Langevin equation for $X_{\bullet }\left( t\right)
=\upsilon _{t}\left( \check{x}_{\bullet }\otimes I\right) $ in the linear
form 
\begin{equation}
\mathrm{d}X_{\bullet }\left( t\right) +\left( X_{\bullet }\left( t\right) 
\mathbf{A}^{\intercal }+\boldsymbol{u}\left( t\right) \mathbf{C}%
_{f}^{\intercal }\right) \mathrm{d}t=\mathrm{d}{\normalsize U}_{\bullet }^{t}
\label{eq Langevin}
\end{equation}%
derived in \cite{Bel92a}, where $\mathbf{A}^{\intercal }:=(\hbar \func{Im}(%
\mathbf{\Lambda }^{\intercal }\mathbf{\Lambda }^{\ast })+\mathbf{M}^{-1})%
\mathbf{J}$ with quantum noise%
\begin{equation}
U_{\bullet }^{t}:=2\hbar \Im \left[ A_{\circ }^{+}\left( t\right) \mathbf{%
\Lambda }\right] \mathbf{J}=V_{\bullet }^{t}+W_{\bullet }^{t},
\label{eq qNoise}
\end{equation}%
for $\mathbf{\Lambda =\Lambda }_{e}+\mathbf{\Lambda }_{f}$. Here $V_{\bullet
}^{t}=V_{\circ }^{t}\mathbf{C}^{\intercal }$ is total Langevin force and $%
W_{\bullet }^{t}=W_{\circ }^{t}\mathbf{F}_{e}^{\intercal }$ is total Wiener
noise given by left action of matrices%
\begin{equation}
\mathbf{C}^{\intercal }=2\func{Re}\mathbf{\Lambda J},\;\;\mathbf{F}%
^{\intercal }=\hbar \func{Im}\mathbf{\Lambda J},  \label{eq C&F}
\end{equation}%
on rows $V_{\circ }^{t}=2\Im \left( A_{1}^{+},\ldots A_{d}^{+}\right) \left(
t\right) $, $W_{\circ }^{t}=2\Re \left( A_{1}^{+},\ldots A_{d}^{+}\right)
\left( t\right) $ of all quantum Langevin forces and all conjugate Wiener
noises respectively, coming both from the estimation and feedback channels.

\subsection{Quantum filtering of linear, Gaussian dynamics}

The linear output equation for the row $\boldsymbol{Y}_{e}^{t}=\left(
Y_{i}^{t}\right) _{i\in I_{e}}$ of observable processes in estimation
channel satisfies 
\begin{equation}
\mathrm{d}\boldsymbol{Y}_{e}^{t}=X_{\bullet }\left( t\right) \mathbf{B}%
_{e}^{\intercal }\mathrm{d}t+\mathrm{d}\boldsymbol{W}_{e}^{t}
\label{eq LinY}
\end{equation}%
where $\mathbf{B}_{e}=2\func{Re}\mathbf{\Lambda }_{e}$. The quantum
measurement noise is given by the row $\boldsymbol{W}_{e}=\left(
W_{i}\right) _{i\in I_{e}}$ having Gaussian independent increments on each
measurement channel with zero mean and standard variance given by $%
d_{e}\times d_{e}$ identity matrix $\mathbf{I}_{e}$. Considered alone, this
noise represents the standard $d_{e}$-dimensional classical Wiener process
which we measure in the field after interaction with the quantum object by
the output isomorphic transformation $\boldsymbol{Y}_{e}^{t}=2\Re \upsilon
_{\infty }\left[ \boldsymbol{A}_{e}^{+}\left( t\right) \right] $ for $%
\boldsymbol{A}_{e}^{+}=\left( A_{i}^{+}\right) _{i\in I_{d}}$. However, it
does not commute with the quantum Langevin force $V_{\bullet }^{t}$ as%
\begin{equation}
\left[ V_{\bullet }^{r\intercal },\boldsymbol{W}_{e}^{s}\right] =\left(
r\wedge s\right) \mathrm{i}\hbar \mathbf{C}_{e}I,  \label{eq VWccr}
\end{equation}%
due to the noncommutativity with $\boldsymbol{V}_{e}^{t}=\left(
V_{i}^{t}\right) _{i\in I_{e}}$ and commutativity with $\boldsymbol{V}%
_{f}^{t}=\hbar \Im \left[ \boldsymbol{A}_{f}^{+}\left( t\right) \right] $,
resulting form independence of $\boldsymbol{A}_{f}^{+}=\left(
A_{i}^{+}\right) _{i\in I_{f}}$. The fundamental CCR (\ref{eq VWccr}),
defined by nonzero (if $\mathbf{J}\neq 0)$ matrix $\mathbf{C}_{e}^{\intercal
}=\mathbf{B}_{e}\mathbf{J}$, was first derived in a complex form in \cite%
{Bel79},\cite{Bel80}, and in even more general infinite dimensional setting
in \cite{Bel92a}. It expresses the Heisenberg error-perturbation uncertainty
principle in a precise form%
\begin{equation}
\mathrm{d}V_{\bullet }^{\intercal }\mathrm{d}V_{\bullet }\geq \frac{\hbar
^{2}}{4}\mathbf{C}_{e}\mathbf{C}_{e}^{\intercal }\mathrm{d}t,\;\mathrm{d}%
\boldsymbol{W}_{e}\mathrm{d}\boldsymbol{W}_{e}^{\intercal }=\mathbf{I}_{e}%
\mathrm{d}t  \label{eq Huncertprinc}
\end{equation}%
derived by Belavkin in \cite{Bel80},\cite{Bel92a} as necessary and
sufficient condition for nondemolition causality of the observable past $%
\boldsymbol{Y}_{e}^{t]}=\left\{ \boldsymbol{Y}_{e}^{r}:r\leq t\right\} $ and
quantum future described by $\left\{ X_{\bullet }\left( s\right) :s\geq
t\right\} $.

Usually in classical filtering theory the process $U_{\bullet }^{t}$ and
measurement noises $W_{e}^{t}$ are considered to be independent, although in
the quantum setting the Heisenberg principle, implying the dependence of $%
U_{\bullet }^{t}=V_{\bullet }^{t}+W_{\bullet }^{t}$ and $\boldsymbol{W}%
_{e}^{t}$, may result in a nonzero covariance matrix $\mathbf{F}_{e}=\hbar 
\mathbf{J}\func{Im}\left( \mathbf{\Lambda }_{e}^{\dagger }\right) $
describing the real part of quantum It\^{o} table%
\begin{equation}
\mathrm{d}U_{\bullet }^{\intercal }\mathrm{d}\boldsymbol{W}_{e}=\left( 
\mathbf{F}_{e}+\frac{\mathrm{i}\hbar }{2}\mathbf{C}_{e}\right) \mathrm{d}t
\label{eq UWtable}
\end{equation}%
as the sum of $\mathrm{d}V_{\bullet }^{\intercal }\mathrm{d}\boldsymbol{W}%
_{e}=\frac{\mathrm{i}\hbar }{2}\mathbf{C}_{e}\mathrm{d}t$ and $\mathrm{d}%
W_{\bullet }^{\intercal }\mathrm{d}\boldsymbol{W}_{e}=\mathbf{F}_{e}\mathrm{d%
}t$. Note that although each component $U_{i}=U_{i}^{\ast }$ of the row $%
U_{\bullet }=\left( U_{1},\ldots ,U_{m}\right) $ for vector quantum noise (%
\ref{eq qNoise}) having the independent increments can also be realized as a
classical Wiener process, these components mutually do not commute, having
complex multiplication table $\mathrm{d}U_{\bullet }^{\intercal }\mathrm{d}%
U_{\bullet }=\hbar ^{2}\mathbf{\Lambda }^{\dagger }\mathbf{\Lambda }\mathrm{d%
}t$ with imaginary part defining the commutation relations%
\begin{equation*}
\left[ U_{\bullet }^{r}{}^{\intercal },U_{\bullet }^{s}\right] =\left(
r\wedge s\right) 2\mathrm{i}\hbar ^{2}\mathbf{J}^{\intercal }\func{Im}\left( 
\mathbf{\Lambda }^{\dagger }\mathbf{\Lambda }\right) \mathbf{J}.
\end{equation*}%
The symmetrized multiplication $\Re \left[ \mathrm{d}U_{\bullet }^{\intercal
}\mathrm{d}U_{\bullet }\right] =\hbar ^{2}\func{Re}\left[ \mathbf{\Lambda }%
^{\dagger }\mathbf{\Lambda }\right] \mathrm{d}t$ results in the symmetric
covariance%
\begin{equation*}
\func{Re}\left\langle V_{\bullet }^{r\intercal }V_{\bullet
}^{s}\right\rangle =\left( r\wedge s\right) \hbar ^{2}\mathbf{J}^{\intercal }%
\func{Re}\left( \mathbf{\Lambda }^{\dagger }\mathbf{\Lambda }\right) \mathbf{%
J}^{\intercal }
\end{equation*}%
defined by $\mathbf{\Lambda }^{\dagger }\mathbf{\Lambda =\Lambda }%
_{e}^{\dagger }\mathbf{\Lambda }_{e}\mathbf{+\Lambda }_{f}^{\dagger }\mathbf{%
\Lambda }_{f}$. It can be parametrized as $\mathbf{F}_{e}^{\intercal }%
\mathbf{F}_{e}+\mathbf{G}$ with positive matrix%
\begin{equation}
\mathbf{G}=\frac{\hbar ^{2}}{4}\mathbf{C}_{e}\mathbf{C}_{e}^{\intercal
}+\hbar ^{2}\mathbf{J}^{\intercal }\func{Re}\left( \mathbf{\Lambda }%
_{f}^{\dagger }\mathbf{\Lambda }_{f}\right) \mathbf{J}  \label{eq G}
\end{equation}%
implying the error-perturbation uncertainty relation 
\begin{equation*}
\Re \left[ \mathrm{d}U_{\bullet }^{\intercal }\mathrm{d}U_{\bullet }\right]
\geq \left( \mathbf{F}_{e}^{\intercal }\mathbf{F}_{e}+\left( \hbar /2\right)
^{2}\mathbf{C}_{e}\mathbf{C}_{e}^{\intercal }\right) \mathrm{d}t
\end{equation*}%
in terms of the total perturbative noise (\ref{eq qNoise}) in the Langevin
equation (\ref{eq Langevin}) with respect to the standard normalized error
noise in estimation channel (\ref{eq LinY}).

Let us denote the initial mean $\left\langle \check{x}_{\bullet
}\right\rangle $ of the phase space operator vector by the component wise
expectation $x_{\bullet }=\langle \varsigma ,\check{x}_{\bullet }\rangle $
and covariance matrix 
\begin{equation*}
\mathbf{\Sigma }:=\left( \func{Re}\langle \varsigma ,\check{x}_{i}\check{x}%
_{k}\rangle -x_{i}^{\intercal }x_{k}\right)
\end{equation*}%
where $\func{Re}\langle \varsigma ,\check{x}_{i}\check{x}_{k}\rangle =\frac{1%
}{2}\langle \varsigma ,\check{x}_{i}\check{x}_{k}+\check{x}_{k}\check{x}%
_{i}\rangle $, which is symmetric positive definite matrix $\mathbf{\Sigma }%
=(\Sigma _{ik})$ satisfying the Heisenberg uncertainty inequality 
\begin{equation}
\mathbf{\Sigma }\geq \pm \frac{\mathrm{i}\hbar }{2}\mathbf{J}
\label{eq HUP2}
\end{equation}

As it was shown by Belavkin in \cite{Bel88},\cite{Be_Sta89}, and even in
infinite dimensions in \cite{Bel92a}, the filtering equation (\ref{eq filt1}%
) preserves the Gaussian nature of the posterior state \cite{Bel92a}, so the
posterior mean $\hat{x}_{\bullet }^{t}=\langle \hat{\varsigma}_{0}^{t},%
\check{x}_{\bullet }\rangle $ and the matrix $\mathbf{\Sigma }\left(
t\right) $ of symmetric error covariances as real part of%
\begin{equation*}
\langle \hat{\varsigma}_{0}^{t},\left( \check{x}_{i}-\hat{x}_{i}^{t}\right)
\left( \check{x}_{k}-\hat{x}_{k}^{t}\right) \rangle =\langle \hat{\varsigma}%
_{0}^{t},\check{x}_{i}\check{x}_{k}\rangle -\hat{x}_{i}^{t}\hat{x}_{k}^{t}
\end{equation*}%
form a set of sufficient coordinates for the quantum LQG system which agree
with the initial mean and covariance for $\hat{\varsigma}\left( 0\right)
=\varsigma $. Applying (\ref{eq filt1}) to the first and second symmetrized
moments of $\check{x}_{\bullet }$ , i.e., rigorously speaking, to the
spectral projectors of these unbounded operators affiliated to $\mathfrak{b}=%
\mathcal{B}\left( \mathfrak{h}\right) $, provides posterior expectations of
these sufficient coordinates for diffusive non-demolition measurement of the
output operators $\boldsymbol{Y}_{e}^{t}$. These can be found as solutions
to Belavkin's Kalman filter equation \cite{Bel80},\cite{Bel92a} written in
vector form as 
\begin{equation}
\mathrm{d}\hat{x}_{\bullet }^{t}+(\hat{x}_{\bullet }^{t}\mathbf{A}%
^{\intercal }+\boldsymbol{u}\left( t\right) \mathbf{C}_{f}^{\intercal })%
\mathrm{d}t=\mathrm{d}\boldsymbol{\hat{W}}_{e}^{t}\mathbf{K}^{\mathbf{%
\intercal }}(t)  \label{eq qKalman}
\end{equation}%
with initial condition$\;\hat{x}_{\bullet }^{0}=x_{\bullet }$ for the
posterior mean, 
\begin{equation}
\mathbf{K}(t)=\mathbf{\Sigma }(t)\mathbf{B}_{e}^{\intercal }+\mathbf{F}%
_{e},\;\mathrm{d}\boldsymbol{\hat{W}}_{e}^{t}=\mathrm{d}\boldsymbol{Y}%
_{e}^{t}-\hat{x}_{\bullet }^{t}\mathbf{B}_{e}^{\intercal }\mathrm{d}t
\label{eq OptK}
\end{equation}%
and for the symmetric error covariance we have 
\begin{eqnarray}
\frac{\mathrm{d}}{\mathrm{d}t}\mathbf{\Sigma } &\mathbf{=}&\mathbf{G-\Sigma A%
}_{e}^{\mathbf{\intercal }}-\mathbf{A}_{e}\mathbf{\Sigma }-\mathbf{\Sigma B}%
_{e}^{\mathbf{\intercal }}\mathbf{B}_{e}\mathbf{\Sigma ,}
\label{eq FiltRiccati} \\
\mathbf{A}_{e} &=&\mathbf{A}+\mathbf{F}_{e}\mathbf{B}_{e},\mathbf{\;\;~\;%
\mathbf{\Sigma }(0)=\mathbf{\Sigma }.}
\end{eqnarray}

\subsection{Quantum LQG control and microduality}

We aim to control an \emph{output} quantum stochastic linear evolution $%
\boldsymbol{Z}\left( t\right) =\upsilon _{0}^{t}\left( \boldsymbol{\check{z}}%
\otimes I\right) $ of a $d_{f}$-dimensional linear combination $\boldsymbol{%
\check{z}}=\check{x}_{\bullet }\mathbf{E}_{f}^{\intercal }$, where $\mathbf{E%
}_{f}$ is a real,in general time dependent $d_{f}\times m$ matrix,
represented\ in the Heisenberg picture as a row $\left( Z_{i}\right) _{i\in
I_{f}}=X_{\bullet }\mathbf{E}_{f}^{\intercal }$ of $Z_{i}\left( t\right)
=\upsilon _{t}\left( \check{z}_{i}\right) $ by forcing $\boldsymbol{Z}\left(
t\right) $ to follow the classical input trajectory of the feedback
controlling force $\boldsymbol{u}\left( t\right) $ whilst constraining for
energy considerations a positive quadratic functional of phase space
operators $X_{\bullet }\left( t\right) =\upsilon _{t}\left( \check{x}%
_{\bullet }\otimes I\right) $. Thus, our control objectives and restraints
can be described by the general quadratic operator valued risk (\ref{eq
QExpCost}) in the canonical form 
\begin{equation}
\check{c}(\boldsymbol{u})=\left( \boldsymbol{u}-\boldsymbol{\check{z}}%
\right) \left( \boldsymbol{u}-\boldsymbol{\check{z}}\right) ^{\intercal }+%
\check{x}_{\bullet }\mathbf{H}\check{x}_{\bullet }^{\intercal }
\label{eq QuadrCost}
\end{equation}%
and $\check{s}=\check{x}_{\bullet }\mathbf{\Omega }\check{x}_{\bullet
}^{\intercal }$ for positive real symmetric $m\times m$ matrices $\mathbf{%
\Omega ,H}$.

Since $\hat{x}_{\bullet }$ and $\mathbf{\Sigma }$ are generators for the
full probability distribution given by the Gaussian posterior state $\hat{%
\varsigma}$, they form a set of sufficient coordinates, so we may consider
derivations of $\mathsf{S}(t,\varsigma )$ as partial derivative of $\mathsf{S%
}(t,x_{\bullet },\mathbf{\Sigma })$%
\begin{equation*}
\begin{split}
& \langle \delta \varsigma ,\nabla _{\varsigma }\mathsf{S}(t,\varsigma
)\rangle \\
& +\frac{1}{2}\langle \delta \varsigma \otimes \delta \varsigma ,\nabla
_{\varsigma }^{\otimes 2}\mathsf{S}(t,\varsigma )\rangle
\end{split}%
=%
\begin{split}
& \mathrm{d}x_{\bullet }\partial _{\bullet }^{\intercal }\mathsf{S}%
(t,x_{\bullet },\boldsymbol{\Sigma })+\big(\mathrm{d}\mathbf{\Sigma }%
,\partial ^{\bullet \bullet }\mathsf{S}(t,x_{\bullet },\boldsymbol{\Sigma })%
\big) \\
& +\frac{1}{2}\big(\mathrm{d}x_{\bullet }^{\intercal }\mathrm{d}x_{\bullet
},\partial _{\bullet }^{\intercal }\partial _{\bullet }\mathsf{S}%
(t,x_{\bullet },\boldsymbol{\Sigma })\big).
\end{split}%
\end{equation*}%
We use the notation $\big(\cdot ,\cdot \big)$ to denote the matrix trace
inner product $\big(\mathbf{D},\mathbf{F}\big)=\mathrm{Tr}[\mathbf{D}%
^{\intercal }\mathbf{F}]$ on the vector space of matrix configurations for
the multi-dimensional system. This gives the directional derivatives along $%
\mathrm{d}x_{j}$ and $\mathrm{d}\Sigma _{ik}$ as functionals of the column $%
\partial _{\bullet }^{\intercal }\mathsf{S}$ of partial derivatives $%
\partial ^{i}\mathsf{S}=\partial \mathsf{S}/\partial x_{i}$ and the matrices 
\begin{equation*}
\partial ^{\bullet \bullet }\mathsf{S}=\left( \frac{\partial }{\partial 
\boldsymbol{\Sigma }_{ik}}\mathsf{S}\right) ,\qquad \partial _{\bullet
}^{\intercal }\partial _{\bullet }\mathsf{S}=\left( \frac{\partial }{%
\partial x_{i}}\frac{\partial }{\partial x_{k}}\mathsf{S}\right)
\end{equation*}%
which are evaluated at $(\hat{x}_{\bullet },\boldsymbol{\Sigma }\left(
t\right) )$.

Inserting this parametrization into the Bellman equation (\ref{eq Bellman1})
and minimizing gives the optimal control strategy 
\begin{equation}
\boldsymbol{u}\left( t\right) =\frac{1}{2}\partial ^{\bullet }\mathsf{S}%
\left( t,\hat{x}_{\bullet }^{t},\mathbf{\Sigma }\right) \mathbf{C}+\hat{x}%
_{\bullet }^{t}\mathbf{E}_{f}^{\intercal }  \label{eq OptCont}
\end{equation}%
where $\mathsf{S}(t,x_{\bullet },\mathbf{\Sigma })$ at $x_{\bullet }=\hat{x}%
_{\bullet }^{t},\mathbf{\Sigma }=\mathbf{\Sigma }\left( t\right) $ now
satisfies the non-linear partial differential equation 
\begin{equation}
\begin{array}{rl}
-\frac{\partial }{\partial t}\mathsf{S}(t,x_{\bullet },\mathbf{\Sigma })= & 
\frac{1}{2}(x_{\bullet }\mathbf{A}^{\intercal }\partial _{\bullet
}^{\intercal }\mathsf{S}+\partial _{\bullet }\mathsf{S}\mathbf{A}x_{\bullet
}^{\intercal })+x_{\bullet }\mathbf{H}x_{\bullet }^{\intercal } \\ 
& +\left( \mathbf{G}-\mathbf{A\Sigma +\Sigma A}^{\intercal },\partial
^{\bullet \bullet }\mathsf{S}\right) +\left( \mathbf{\Sigma },\mathbf{H}%
\right) \\ 
& -(\frac{1}{2}\partial _{\bullet }\mathsf{S}\mathbf{C}+x_{\bullet }\mathbf{E%
}_{f}^{\intercal })(\frac{1}{2}\partial _{\bullet }\mathsf{S}\mathbf{C}%
+x_{\bullet }\mathbf{E}_{f}^{\intercal })^{\intercal } \\ 
& +\left( (\mathbf{\Sigma B}^{\intercal }+\mathbf{F}_{e}\mathbf{)}(\mathbf{%
\Sigma B}^{\intercal }+\mathbf{F}_{e}\mathbf{)^{\intercal }},\frac{1}{2}%
\partial _{\bullet }^{\intercal }\partial _{\bullet }\mathsf{S}-\partial
^{\bullet \bullet }\mathsf{S}\right)%
\end{array}
\label{eq HJBcont}
\end{equation}%
which is the Hamilton-Jacobi-Bellman equation for this example.

It is well known from classical control theory that LQG control has a
minimum cost-to-go which is quadratic in the state, so we try the candidate
solution 
\begin{equation*}
\mathsf{S}(t,x_{\bullet },\mathbf{\Sigma })=x_{\bullet }\mathbf{\Omega }%
(t)x_{\bullet }^{\intercal }+\big(\mathbf{\Omega }(t),\mathbf{\Sigma }\big)%
+\alpha (t)
\end{equation*}%
in the HJB equation (\ref{eq HJBcont}). This separates the HJB equation into
a set of coupled ordinary differential equations and gives the optimal
feedback control strategy 
\begin{equation}
\boldsymbol{u}\left( t\right) =\hat{x}_{\bullet }\left( t\right) \mathbf{L}%
^{\intercal }(t),\;\mathbf{L}^{\intercal }(t)=\mathbf{\Omega }(t)\mathbf{C}%
_{f}+\mathbf{E}_{f}^{\intercal }  \label{eq OptL}
\end{equation}%
which is linear in the solution to the filtering equation $\hat{x}_{\bullet
}^{t}$ at time $t$ where $\mathbf{\Omega }(t)$ satisfies the matrix Riccati
equation 
\begin{eqnarray}
-\frac{\mathrm{d}}{\mathrm{d}t}\mathbf{\Omega } &=&\mathbf{H-\Omega A}_{f}-%
\mathbf{A}_{f}^{\intercal }\mathbf{\Omega }-\mathbf{\Omega C}_{f}\mathbf{C}%
_{f}^{\mathbf{\intercal }}\mathbf{\Omega }  \label{eq ContRiccati} \\
\mathbf{A}_{f} &=&\mathbf{A+C}_{f}\mathbf{E}_{f},\;\;\mathbf{\Omega }(T)=%
\mathbf{\Omega }  \notag
\end{eqnarray}%
and $\alpha (t)$ satisfies 
\begin{eqnarray}
-\frac{\mathrm{d}}{\mathrm{d}t}\alpha (t) &=&\left( (\mathbf{\Omega }(t)%
\mathbf{C}_{f}+\mathbf{E}_{f}^{\intercal })(\mathbf{\Omega }(t)\mathbf{C}%
_{f}+\mathbf{E}_{f}^{\intercal })^{\intercal },\mathbf{\Sigma }(t)\right)
+\left( \mathbf{\Omega }(t),\mathbf{G}\right)  \label{eq alpha} \\
\alpha (T) &=&0.  \notag
\end{eqnarray}%
From this we obtain the total minimal cost for the control experiment 
\begin{eqnarray}
\mathsf{S}(0,x_{\bullet },\mathbf{\Sigma }) &=&x_{\bullet }\mathbf{\Omega }%
_{0}x_{\bullet }^{\intercal }+\text{$\mathrm{Tr}$}[\mathbf{\Omega }_{0}%
\mathbf{\Sigma }]+\int_{0}^{T}\text{$\mathrm{Tr}$}[\mathbf{\Omega }(t)%
\mathbf{G}]\mathrm{d}t  \label{eq LQGmincost} \\
&&+\int_{0}^{T}\text{$\mathrm{Tr}$}[(\mathbf{\Omega }(t)\mathbf{C}_{f}%
\mathbf{+E}_{f}^{\intercal })^{\intercal }\mathbf{\Sigma }(t)(\mathbf{\Omega 
}(t)\mathbf{C}_{f}\mathbf{+E}_{f}^{\intercal })]\mathrm{d}t  \notag
\end{eqnarray}%
where $\mathbf{\Omega }_{0}$ is the solution to (\ref{eq ContRiccati}) at
time $t=0$.

The equations (\ref{eq OptK})-(\ref{eq FiltRiccati}) and (\ref{eq OptL})-(%
\ref{eq ContRiccati}) demonstrate the intrinsic duality between optimal
quantum linear filtering and optimal classical linear control, which we call 
\emph{microduality}. To make this duality more transparent let us introduce
real matrices $\mathbf{E}$ and $\mathbf{F}$ defining the matrices $\mathbf{F}%
_{e}$ and $\mathbf{E}_{f}$ in (\ref{eq C&F}) and (\ref{eq QuadrCost}) by $%
\mathbf{EJ}=\mathbf{F}_{e}^{\intercal }$ and $\mathbf{E}_{f}\mathbf{J}=%
\mathbf{F}^{\intercal }$ in the similar to $\mathbf{BJ}=\mathbf{C}%
_{e}^{\intercal }$ and $\mathbf{B}_{f}\mathbf{J}=\mathbf{C}^{\intercal }$ in
terms of the estimation and feedback channel matrices $\mathbf{B}=\mathbf{B}%
_{e}$ and $\mathbf{C}=\mathbf{C}_{f}$. Then the microduality is summarized
in the table 
\begin{equation}
\begin{array}{c|c|c|c|c|c|c}
\text{Filt} & \mathbf{AJ} & \mathbf{BJ} & \mathbf{EJ} & \mathbf{K}\left(
T-t\right) & \mathbf{G}\left( T-t\right) & \mathbf{\Sigma }\left( T-t\right)
\\ \hline
\text{Con} & \mathbf{JA}^{\intercal } & \mathbf{C}^{\intercal } & \mathbf{F}%
^{\intercal } & \mathbf{JL}^{\mathbf{\intercal }}\left( t\right) & \mathbf{JH%
}\left( t\right) \mathbf{J^{\intercal }} & \mathbf{J\Omega }\left( t\right) 
\mathbf{J^{\intercal }}%
\end{array}
\label{tb Duality}
\end{equation}%
in which the duality notations are made in filtering-control alphabetical
order $\left( \mathbf{B,C}\right) $, $\left( \mathbf{E,F}\right) $, $\left( 
\mathbf{G,H}\right) $ and $\left( \mathbf{K,L}\right) $ and the matrices $%
\mathbf{A}$ $\mathbf{B,E}$ should be also taken at $t^{\intercal }=T-t$ if
they depend on $t$ for the duality with $\mathbf{A}^{\intercal },\mathbf{%
C^{\intercal },F^{\intercal }}$ evaluated at $t$. This duality allows us to
formulate and solve the dual classical control problem given the solution to
quantum filtering problem with dual parameters. The duality can be
understood when we examine the nature of each of the methods used. Both
methods involve the minimization of a quadratic function for linear,
Gaussian systems, (i.e.\ the least squares error for filtering and the
quadratic cost for control). The time reversal in the dual picture is
explained by the interchange of the input (feedback) and the output
(estimation) channels together with the linear canonical transformation
given by the symplectic matrix $\mathbf{J}$. Note that $\mathbf{E}=\hbar 
\func{Im}\mathbf{\Lambda }_{e}$, and $\mathbf{G}$ must be positive definite
satisfying the relation (\ref{eq G}) due to Heisenberg uncertainty relation
corresponding to the error-perturbation CCR's $\left[ \mathrm{d}\boldsymbol{W%
}_{e}^{\intercal },\mathrm{d}\boldsymbol{U}_{e}\right] =\mathrm{i}\hbar 
\mathbf{F}_{e}\mathbf{C}_{e}^{\intercal }\mathrm{d}t$ in the It\^{o}
multiplication table (\ref{eq UVtable}). In order to complete this
filtering-control microduality we may set $\mathbf{F}^{\intercal }=$ $\hbar 
\func{Im}\mathbf{\Lambda }_{f}\mathbf{J}$ to have the relation between $%
\mathbf{E}$ and $\mathbf{F}$ similar to the duality of $\mathbf{B}=2\func{Re}%
\mathbf{\Lambda }_{e}$ and $\mathbf{C}^{\intercal }=2\func{Re}\mathbf{%
\Lambda }_{f}^{\ast }\mathbf{J}$, and also assume that matrix $\mathbf{H}$
is also positive, satisfying%
\begin{equation}
\mathbf{H}=\left( \hbar /2\right) ^{2}\mathbf{B}_{f}^{\intercal }\mathbf{B}%
_{f}+\hbar ^{2}\func{Re}\left( \mathbf{\Lambda }_{e}^{\dagger }\mathbf{%
\Lambda }_{e}\right) ,  \label{eq H}
\end{equation}%
where $\mathbf{B}_{f}=2\func{Im}\mathbf{\Lambda }_{f}$. Note that although
the condition (\ref{eq H}) is not a requirement in this classical-quantum
setting, in which only positivity of the combination $\mathbf{H}+\mathbf{E}%
_{f}^{\intercal }\mathbf{E}_{f}$ suffices, this relation may be required for
the fully quantum\ setting when the both the input $\boldsymbol{u}$ and the
output $\boldsymbol{Y}_{e}$ are allowed to be noncommutative, which will be
studied and published elswhere.

\section{Discussion}

The Bellman equations for quantum systems having separate diffusive and
counting measurement schemes have been derived in continuous time under a
general setup. This presents original derivations of the results stated in 
\cite{Bel88} (a derivation of the diffusive case has also appeared in \cite%
{GBS05}) and allows us to reformulate the optimal control problem for a
fundamentally unobservable quantum system into a classical control problem
on the Banach space of observable filtered states.

The multi-dimensional quantum LQG problem which finishes the paper
demonstrates the first application of the general quantum Bellman equation
from which one can obtain the special cases of the Gaussian quantum
oscillator \cite{Bel99} and quantum free particle \cite{DoJ99,BeS92}. Note
that the fundamental difference between this example and the corresponding
well studied classical case is in the observability of the system. The
quantum noises introduced act only to account for the quantum backaction due
to the incompatibility of quantum events. No further restrictions on the
observability or additional classical noise are introduced. As such, the
corresponding classical problem (when $\hbar \rightarrow 0$) admits direct
observations of $\hat{x}_{\bullet }^{t}=\check{x}_{\bullet }^{t}$ for a
deterministic classical system and has an optimal direct feedback strategy $%
\boldsymbol{u}\left( t\right) =\hat{x}_{\bullet }^{t}\mathbf{L}\left(
t\right) $ and minimal cost $\mathsf{S}(0,x_{\bullet })=x_{\bullet }%
\boldsymbol{\Omega }_{0}x_{\bullet }^{\intercal }$. Also this example
clearly demonstrates the micro duality between quantum linear filtering and
classical feedback control as a \ more elaborated duality involving also the
linear symplectic transformation $\mathbf{J}$.

During the publication of this paper, there have appeared a number of recent
works on quantum filtering and feedback control by Bouten et al. \cite%
{BoH05a,BoH05b,BHJ06} to which we refer the interested reader for a more
detailed introduction to these subjects.

\section{Appendix}

\subsection{A. Some definitions and facts on W*-algebras}

\begin{enumerate}
\item A complex Banach algebra $\mathrm{A}$ with involution $a\mapsto
a^{\ast }$ such that $\left\Vert a^{\ast }a\right\Vert =\left\Vert
a\right\Vert ^{2}$ is called C*-algebra, and W*-algebra if it is dual to a
linear subspace $\mathrm{L}\subseteq \mathrm{A}^{\star }$ (called preadjoint
of $\mathrm{A}=\mathrm{L}^{\star }$ if it is closed, denoted as $\mathrm{L}=%
\mathrm{A}_{\star }$). They all can be realized as operator algebras on a
complex Hilbert space $\mathcal{H}$, and an operator W*-algebra is called
von Neumann algebra if its unit is the identity operator $I$ in $\mathcal{H}$%
. The simplest example of W*-algebra is the von Neumann algebra $\mathcal{B}%
\left( \mathcal{H}\right) $ of all bounded operators acting in a complex
Hilbert space $\mathcal{H}$. A von Neumann algebra $\mathcal{A}$ is called
semisimple if $\mathcal{H}$ has an orthogonal decomposition into invariant
subspaces $\mathcal{H}_{i}$ in which $\mathcal{A}$ is $\mathcal{B}\left( 
\mathcal{H}_{i}\right) $. Let $\left\{ Q_{i}\right\} $ (or $\left\{ \mathcal{%
A}_{i}\right\} $) be a family of self-adjoint operators (operator algebras $%
\mathcal{A}_{i}$) acting in $\mathcal{H}$, e.g. orthoprojectors $%
Q_{i}^{2}=Q_{i}=Q_{i}^{\ast }$.\ The W*-algebra generated by this family is
defined as the smallest weakly closed self-adjoint sub-algebra $\mathcal{A}%
\subseteq \mathcal{B}\left( \mathcal{H}\right) $ containing these operators,
or the spectral projectors of these operators if $Q_{i}$ are unbounded in $%
\mathcal{H}$ (or the algebras $\mathcal{A}_{i}$, $\mathcal{A}=\vee \mathcal{A%
}_{i}$). It is not necessarily semisimple but in the case $I\in \mathcal{A}$
it consists of all bounded operators that commute with the bounded commutant 
$\mathcal{B}=\left\{ B\in \mathcal{B}\left( \mathcal{H}\right) :BQ_{i}=Q_{i}B%
\text{ \ \ }\forall i\right\} \equiv \left\{ Q_{i}\right\} ^{\prime }$ (or
with $\mathcal{B}=\cap \mathcal{A}_{i}^{\prime }$), i.e., it is the second
commutant $\mathcal{A}=\left\{ Q_{i}\right\} ^{\prime \prime }$ of the
family $\left\{ Q_{i}\right\} $.\ The latter can be taken as the definition
of the von Neumann algebra generated by the family $\left\{ Q_{i}\right\} $.
Note that the commutant $\mathcal{B}$ is a von Neumann algebra such that $%
\mathcal{B}=\mathcal{A}^{\prime }$, called the commutant algebra of $%
\mathcal{A}$. \cite{Dix69}.

\item A (normal) state on a von Neumann algebra $\mathcal{A}$ is defined as
a linear ultraweakly continuous functional $\mathcal{A}\rightarrow \mathbb{C}
$, satisfying the positivity and normalization conditions 
\begin{equation}
\left( \varrho ,Q\right) \geq 0,\text{ \ \ \ \ }\forall Q\geqq 0\text{, \ \
\ \ }\left( \varrho ,I\right) =1  \label{A.1}
\end{equation}%
[$Q\geqq 0$ signifies the nonnegative definiteness $\left\langle \psi |Q\psi
\right\rangle \geqq 0$ $\forall \psi \in \mathcal{H}$ called Hermitian
positivity of $Q$].\ The linear span of all \emph{normal }states is
isometric with the preadjoint space $\mathcal{A}_{\star }$. The latter is
usually described as the space of density operators $\varsigma $ uniquely
defined as (generalized, or affiliated) elements of the algebra $\mathcal{A}$
with respect to a \emph{standard pairing} $\left( \varrho ^{\ast },Q\right)
:=\left\langle \varrho |Q\right\rangle \equiv \varrho ^{\star }\left(
Q\right) $ of $\mathcal{A}_{\star }$ and $\mathcal{A}$ given by the mass $%
\mu \left( \varrho \right) =\left\langle \varrho |I\right\rangle $ on the
positive $\varrho $ such that $\varrho ^{\ast }=\varrho \geq 0$ is state
density iff $\left\langle \varrho |I\right\rangle =1$.\ A state $\varrho $
is called vector state if $\left( \varrho ,Q\right) =\left\langle \psi
|Q\psi \right\rangle \equiv \left\langle \varrho _{\psi }|Q\right\rangle $
(denoted $\varrho =\varrho _{\psi }$) for some $\psi \in \mathcal{H}$, and
pure state if it is an extreme point of the convex set $\mathcal{S}\left( 
\mathcal{A}\right) $ of all normal states on $\mathcal{A}$.\ Every normal
state is in the closed convex hull of vector states $\varrho _{\psi }$ with $%
\left\Vert \psi \right\Vert =1$ but there might be no pure state in $%
\mathcal{S}\left( \mathcal{A}\right) $. If algebra $\mathcal{A}$ is
semifinite (there exists a faithful normal semi-finite trace $Q\mapsto 
\mathrm{tr}Q$, then the states on $\mathcal{A}$ can be described by unit
trace operators $\varrho \in \mathcal{A}$ (or $\varrho \vdash \mathcal{A}$
if the are only affiliated to $\mathcal{A}$), by means of the tracial
pairing $\left\langle \varrho |Q\right\rangle =\mathrm{tr}\left[ \varrho
^{\ast }Q\right] $.\ In the simple case $\mathcal{A}=\mathcal{B}\left( 
\mathcal{H}\right) $ the density operator $\varrho $ is any nuclear positive
operator normalized with respect to the usual trace \cite{Dix69}.

\item Let $\mathcal{A}$, $\mathcal{B}$ be von Neumann algebras in respective
Hilbert spaces $\mathcal{H}_{0}$ and $\mathcal{H}_{1}$, and let $\Phi :%
\mathcal{B}\rightarrow \mathcal{A}$ be a linear map that transforms the
operators $B\in \mathcal{B}$ into operators $A\in \mathcal{A}$ (called
sometimes superoperator). The map $\Phi $ is called a \emph{transfer map} if
it is ultraweakly continuous, completely positive (CP) in the sense 
\begin{equation}
\sum\limits_{i,k=1}^{\infty }\left\langle \psi _{i}|\Phi \left( B_{i}^{\ast
}B_{k}\right) \psi _{k}\right\rangle \geqq 0\text{, \ \ \ \ }\forall
B_{j},\psi _{j}  \label{A.2}
\end{equation}%
($i=1,\ldots ,d_{e}<\infty $), and unity-preserving: $\Phi \left(
I_{1}\right) =I_{0}$ (or $\Phi \left( I_{1}\right) \leq I_{0}$). The CP
condition is obviously satisfied if $\Phi $ is normal homomorphism (or
W*-representation) $\pi :\mathcal{B}\rightarrow \mathcal{A}$, which is
defined by the additional multiplicativity property $\pi \left( B^{\ast
}B\right) =\pi \left( B\right) ^{\ast }\pi \left( B\right) $.\ A composition 
$\left( \varrho _{0},\Phi \left( B\right) \right) =\left\langle \varrho
_{0}|\Phi \left( B\right) \right\rangle $ with any state $\varrho
_{0}=\varrho _{0}^{\ast }$ is a state $\varrho _{1}$ on $\mathcal{B}$
described by the adjoint action of the superoperator $\Phi $ on $\varrho
_{0} $: 
\begin{equation*}
\left\langle \varrho _{0}|\Phi \left( B\right) \right\rangle =\left\langle
\Phi ^{\star }\left( \varrho _{0}\right) |B\right\rangle ,\forall B\in 
\mathcal{B},\varrho _{0}\in \mathcal{A}_{\star }.
\end{equation*}%
\ A transfer map $\Phi $ is called spatial if 
\begin{equation}
\Phi \left( B\right) =FBF^{\ast }\text{ \ \ \ or \ \ \ }\Phi ^{\star }\left(
\varrho _{0}\right) =F_{\star }\varrho _{0}F_{\star }^{\ast },  \label{A.3}
\end{equation}%
where $F$ is a linear coisometric operator $\mathcal{H}_{1}\rightarrow 
\mathcal{H}_{0}$, $FF^{\ast }=I_{0}$ (or $FI_{1}F^{\ast }\leq I_{0}$) called
the \emph{propagator} and $F_{\star }=F^{\sharp }$ is defined as left
adjoint $\left\langle F^{\sharp }\varrho |Q\right\rangle =\left\langle
\varrho |FQ\right\rangle $ with respect to the standard pairings (which is
usual adjoint, $F^{\sharp }=F^{\ast }$, in the case of tracial pairing $%
\left\langle \varrho |Q\right\rangle =\mathrm{tr}\left[ \varrho ^{\ast }Q%
\right] $).\ Every transfer map is in the closed convex hull of spatial
transfer maps, but there might be no extreme point in this hull.

\item Let $\mathbb{V}$ be a measurable space, and $\mathfrak{B}$ its Borel $%
\sigma $-algebra.\ A mapping $\Pi :\mathrm{d}v\in \mathfrak{B}\mapsto \Pi
\left( \mathrm{d}v\right) $ with values $\Pi \left( \mathrm{d}v\right) $ in
ultraweakly continuous, completely positive superoperators $\mathcal{B}%
\rightarrow \mathcal{A}$ is called a \emph{transfer measure} if for any $%
\varphi \in \mathcal{A}_{\star }$, $B\in \mathcal{B}$ the $\mathbb{C}$%
-valued function 
\begin{equation*}
\left\langle \Pi \left( \mathrm{d}v\right) ^{\star }\varrho |B\right\rangle
=\left\langle \varrho |\Pi \left( \mathrm{d}v\right) B\right\rangle
\end{equation*}%
of the set $\mathrm{d}v\subseteq \mathbb{V}$ is a countably additive measure
normalized to unity for $B=I$. \ In other words, $\Pi \left( \mathrm{d}%
v\right) $ is a CP map valued measure that is $\sigma $-additive in the weak
(strong) operator sense and for $\mathrm{d}v=\mathbb{V}$ is equal to some
transfer-map $\Phi $. In particular, $\Pi \left( \mathrm{d}v,B\right)
=M\left( \mathrm{d}v\right) \Phi \left( B\right) $ with $M\left( \mathrm{d}%
v\right) =\Pi \left( \mathrm{d}v,I\right) $ is transfer map iff $\left[
M\left( \mathrm{d}v\right) ,\Phi \left( B\right) \right] =0$ for all $%
\mathrm{d}v\in \mathfrak{B}$ and $B\in \mathcal{B}$ as it is the case of the 
\emph{nondemolition measurements} given by representations $M=E$ of $%
\mathfrak{B}$ in $\mathcal{A}$ and $\Phi =\pi $ in $E\left( \mathfrak{B}%
\right) ^{\prime }\cap \mathcal{A}$.\ The quantum state transformations $%
\varrho \mapsto \Pi ^{\star }\left( \Delta ,\varrho \right) $ corresponding
to the results $v\in \Delta $ of an \emph{ideal measurement} are described
by transfer-operator measures of the form 
\begin{equation}
\Pi \left( \Delta ,B\right) =\int_{\Delta }F\left( v\right) BF\left(
v\right) ^{\ast }\mathrm{d}v,  \label{A.4}
\end{equation}%
where $F\left( v\right) $ denote linear operators $\mathcal{H}%
_{1}\rightarrow \mathcal{H}_{0}$, the integral with respect to a positive
Borel measure $\lambda $ on $\mathbb{V}$ is interpreted in strong operator
topology, and $\int F\left( v\right) F\left( v\right) ^{\ast }\mathrm{d}%
v=I_{0}$. Every transfer-operator $\Phi :\mathcal{B}\rightarrow \mathcal{A}$
can be represented by the integral (\ref{A.4}) on $\Delta =\mathbb{V}$ of an
ideal measurement $\Pi \left( \mathrm{d}v\right) $ as the compression%
\begin{equation}
\Pi \left( \Delta ,B\right) =FE\left( \Delta \right) \pi \left( B\right)
F^{\ast }  \label{A.5}
\end{equation}%
of the nondemolition measurement $E\left( \Delta \right) \pi \left( B\right)
=\hat{1}_{\Delta }\otimes B$ on the extended Hilbert space $\mathcal{H}%
=\int_{\mathbb{V}}^{\oplus }F\left( v\right) ^{\ast }\mathcal{H}_{0}\mathrm{d%
}v$ with the isometric embedding $F^{\ast }\psi _{0}=\int_{\mathbb{V}%
}^{\oplus }F\left( v\right) ^{\ast }\psi _{0}\lambda \left( \mathrm{d}%
v\right) $ of $\psi _{0}\in \mathcal{H}_{0}$ into $\mathcal{H}$ adjoint to
the coisometry $F\psi =\int_{\mathbb{V}}F\left( \nu \right) \psi \left( \nu
\right) \lambda \left( \mathrm{d}v\right) $ of $\psi \in \mathcal{H}$ into $%
\mathcal{H}_{0}$.
\end{enumerate}

\subsection{B. Notations of quantum stochastic calculus}

\begin{enumerate}
\item Guichardet Fock space $\mathcal{F}_{t}^{s}=\Gamma \left( \mathcal{E}%
_{t}^{s}\right) $ over the space $\mathcal{E}_{t}^{s}=L^{2}($\textrm{I}$%
_{t}^{s})$ of square integrable complex functions $\xi \left( r\right) $ on
the interval $\mathrm{I}_{t}^{s}=[t,t+s)$ is built as the Hilbert sum $%
\oplus _{n=0}^{\infty }\mathcal{E}_{n}$ of the spaces $\mathcal{E}_{n}=%
\mathrm{L}^{2}(\Gamma _{n})$ of square integrable functions $\chi _{n}:\tau
\in \Gamma _{n}\mapsto \mathbb{C}$ of \emph{finite chains} $\tau =\left\{
t_{1}<\ldots <t_{n}\right\} \subset \mathrm{I}_{t}^{s}$ for all $%
n=0,1,\ldots $ where $\chi _{0}\left( \emptyset \right) =c$ is a constant
corresponding to only one -- empty chain $\tau _{0}=\emptyset $ of $\Gamma
_{0}$, with $c=1$ for the vacuum vector state $\chi _{n}=\delta _{0}^{n}$.
It is $\mathrm{L}^{2}\left( \Gamma \left( \mathrm{I}_{t}^{s}\right) \right) $%
, where $\Gamma \left( \mathrm{I}_{t}^{s}\right) $ is disjoint union $%
\sum_{n=0}^{\infty }\Gamma _{n}\left( \mathrm{I}_{t}^{s}\right) $ of the $n$%
-simplices $\Gamma _{n}\left( \mathrm{I}_{t}^{s}\right) $. The integration
on $\Gamma $ is assumed over the Lebesgues sum $\mathrm{d}\tau =\sum \mathrm{%
d}\tau _{n}$ of measures $\mathrm{d}\tau _{n}=\mathrm{d}t_{1}\ldots \mathrm{d%
}t_{n}$ on the simplices $\Gamma _{n}$ of chains $\tau _{n}:\left\vert \tau
_{n}\right\vert =n$ with the only atom $\mathrm{d}\tau _{0}=1$ at $\Gamma
_{0}=\left\{ \emptyset \right\} $ such that%
\begin{equation*}
\sum_{n=0}^{\infty }\int_{\Gamma _{n}}\left\Vert \xi ^{\otimes n}\left( \tau
_{n}\right) \right\Vert ^{2}\mathrm{d}\tau _{n}=\exp \left[ \int_{0}^{\infty
}\left\Vert \xi \left( t\right) \right\Vert ^{2}\mathrm{d}t\right]
\end{equation*}%
for the exponential vector-functions $\xi ^{\otimes n}\left( \tau
_{n}\right) =\xi \left( t_{1}\right) \ldots \xi \left( t_{n}\right) $ given
by a single-point function $\xi \in \mathcal{E}$. It is isomorphic to both
usual Fock spaces $\oplus _{n=0}^{\infty }\mathcal{E}_{n}^{\pm }$ of
symmetric and antisymmetric functions $\chi _{n}\left( r_{1},\ldots
,r_{n}\right) $ extending $\chi \left( \tau _{n}\right) $ on $\left( \mathrm{%
I}_{t}^{s}\right) ^{n}$ with respect to the measure $\left( n!\right) ^{-1}%
\mathrm{d}r_{1}\cdots \mathrm{d}r_{n}$. Fock space is infinitely divisible
in the multiplicative sense $\mathcal{F}_{t-r}^{r+s}\sim \mathcal{F}%
_{t-r}^{r}\otimes \mathcal{F}_{t}^{s}$ for any $r,s>0$, which is a
reflection of the additive divisibility $\mathcal{E}_{t-r}^{r+s}\sim 
\mathcal{E}_{t-r}^{r}\oplus \mathcal{E}_{t}^{s}$ of $L^{2}\left( \mathrm{I}%
_{t-r}^{r+s}\right) $. The generalization to the multiple Fock-Guichardet
case over the space $\mathcal{E}_{t}^{s}=\mathrm{L}^{2}\left( \mathrm{I}%
_{t}^{s}\rightarrow \mathfrak{k}\right) $ of vector-valued functions $\xi
\left( t\right) $ in a Hilbert space $\mathfrak{k}$ ($=\mathbb{C}^{d}$, say)
is straight forward and can be found in \cite{Be88},\cite{Bel92b}. All
properties remain the same, and the only difference is that $\mathcal{F}%
_{0}^{t}$ is not $\mathrm{L}^{2}$-space of scalar-valued functions $\chi $
on $\Gamma $ but is Hilbert integral $\int_{\tau \in \Gamma \left( \mathrm{I}%
_{t}^{s}\right) }^{\oplus }\mathcal{K}\left( \tau \right) \mathrm{d}\tau $
of $\mathcal{K}\left( \tau \right) \sim \mathfrak{k}^{\otimes \left\vert
\tau \right\vert }\otimes \mathrm{L}^{2}\left( \mathrm{I}_{t}^{s}\right) $,
the spaces of square integrable tensor-valued functions $\chi :\Gamma
_{n}\rightarrow \mathfrak{k}^{\otimes n}$.

\item Four basic \emph{integrators} $A_{-}^{+}$,\ $A_{-}^{\circ }$, $%
A_{\circ }^{+}$ and\ $A_{\circ }^{\circ }$ of the universal quantum
stochastic (QS) calculus \cite{Be88},\cite{Bel92b} are operator-valued
measures $A_{\mu }^{\nu }\left( \mathrm{I}\right) $ of preservation,
annihilation, creation and exchange respectively, defining the basic QS
integrals of the total QS integral as sum-integral 
\begin{equation}
\mathfrak{i}_{0}^{t}\left( \boldsymbol{K}\right) =\sum_{\mu =-,\circ ;\nu
=\circ ,+}\int_{0}^{t}K_{\nu }^{\mu }\left( r\right) A_{\mu }^{\nu }\left( 
\mathrm{d}r\right)
\end{equation}%
of four basic integrants $K_{+}^{-}$, $K_{\circ }^{-}$, $K_{+}^{\circ }$ and 
$K_{\circ }^{\circ }$ as measurable operator-valued functions $K_{\nu }^{\mu
}\left( r\right) $ in $\mathcal{F}$ by the following explicit formulas:%
\begin{eqnarray}
\left[ \mathfrak{i}_{0}^{t}\left( K\right) _{-}^{+}\right] \chi \left( \tau
\right) &=&\int_{0}^{t}\left[ K\left( r\right) \chi \right] \left( \tau
\right) \mathrm{d}r,\; \\
\left[ \mathfrak{i}_{0}^{t}\left( K\right) _{-}^{\circ }\right] \chi \left(
\tau \right) &=&\int_{0}^{t}\left[ K\left( r\right) \mathring{\chi}\left(
r\right) \right] \left( \tau \right) \mathrm{d}r, \\
\left[ \mathfrak{i}_{0}^{t}\left( K\right) _{\circ }^{+}\chi \right] \left(
\tau \right) &=&\sum_{r\in \tau _{0}^{t}}\left[ K\left( r\right) \chi \right]
\left( \tau \backslash r\right) ,\; \\
\left[ \mathfrak{i}_{0}^{t}\left( K\right) _{\circ }^{\circ }\right] \chi
\left( \tau \right) &=&\sum_{r\in \tau _{0}^{t}}\left[ K\left( r\right) 
\mathring{\chi}\left( r\right) \right] \left( \tau \backslash r\right) .
\end{eqnarray}%
Here $\left[ \mathring{\chi}\left( r\right) \right] \left( \tau \right)
=\chi \left( r\sqcup \tau \right) $, where $r\sqcup \tau $ is union of
disjoint $\tau \in \Gamma $ and $r\notin \tau $, $\tau \backslash r$ is
difference of $\tau $ and a singleton $r\equiv \left\{ r\right\} \subseteq
\tau $ and $\tau _{0}^{t}=\tau \cap \mathrm{I}_{0}^{t}$. The functions $%
K_{\nu }^{\mu }\left( r\right) $ should be $\mathrm{L}^{p}$-integrable in a
uniform operator topology \cite{Be88},\cite{Bel92b}, with $p=2/\left( \nu
-\mu \right) $ where $-=-1$, $\circ =0$, $+=1$. Note that these definitions
do not assume adaptedness of integrants as they generalize Hitsuda-Skorochod
extended stochastic integral. The multiple version of this explicit
QS-integration is straight forward and can be found also in \cite{Be88},\cite%
{Bel92b}, and the adapted version based on coherent vectors is in \cite%
{Par92}.

\item It\^{o} rule (\ref{eq ItoRule}) for QS integrals $M\left( t\right)
=M\left( 0\right) +$ $\mathfrak{i}_{0}^{t}\left( \boldsymbol{K}\right) $
with adapted four-integrant $\boldsymbol{K}\left( t\right) =\left( K_{\nu
}^{\mu }\right) \left( t\right) $ is based on the noncommutative It\^{o}
table (\ref{eq QSItotable}) which uses $\star $-matrix algebra of the
canonical triangular representation%
\begin{equation}
\boldsymbol{K}=\left[ 
\begin{array}{ccc}
0 & K_{\circ }^{-} & K_{+}^{-} \\ 
0 & K_{\circ }^{\circ } & K_{+}^{\circ } \\ 
0 & 0 & 0%
\end{array}%
\right] ,\;\boldsymbol{K}^{\star }=\left[ 
\begin{array}{ccc}
0 & R_{\circ }^{-} & R_{+}^{-} \\ 
0 & R_{\circ }^{\circ } & R_{+}^{\circ } \\ 
0 & 0 & 0%
\end{array}%
\right] ,
\end{equation}%
where $R_{-\nu }^{\mu }=K_{-\mu }^{\nu \ast }$, for any \emph{noncommutative
It\^{o} algebra} \cite{Be88},\cite{Bel92b}. It was derived in \cite{HuP84}
for simple bounded integrants, and therefore can not be rigorously applied
for multiple integration of quantum stochastic equations except the special
unitary case. In the general form presented here QS It\^{o} formula was
proved for unbounded integrants in \cite{Be91} where it was also extended to
nonadapted integrants, and the \emph{functional noncommutative It\^{o}
formula} was also obtained in the pseudo-Poisson form as 
\begin{equation}
\mathrm{d}f\left( M\left( t\right) \right) =\left( f\left( \boldsymbol{M}%
\left( t+\right) \right) -f\left( M\left( t\right) \right) \otimes 
\boldsymbol{I}\right) _{\nu }^{\mu }A_{\mu }^{\nu }\left( \mathrm{d}t\right)
,
\end{equation}%
where $\boldsymbol{M}\left( t+\right) =M\left( t\right) \otimes \boldsymbol{I%
}+\boldsymbol{K}\left( t\right) $ is QS \emph{germ} \cite{Be96} of the QS
integral $M\left( t\right) $ which is defined by its four QS-derivatives $%
K_{\nu }^{\mu }\left( t\right) $ and unite matrix $\boldsymbol{I}=\left(
\delta _{\nu }^{\mu }\right) $, and the summation convention over $\mu ,\nu
=-,\circ ,+$ is applied. Using this formula the HP differential conditions (%
\ref{eq HPunicond}) of QS unitarity of QS interaction evolution $U_{t}$ were
obtained as pseudo-unitarity condition in terms of germ $\boldsymbol{U}%
_{t+}=U_{t}\otimes \boldsymbol{S}$, and also QS differential conditions of
complete positivity, contractivity and projectivity were found in \cite{Be96}%
,\cite{Be97} respectively as its pseudo complete positivity, pseudo
contractivity and pseudoprojectivity of the corresponding QS germs $%
\boldsymbol{M}\left( t+\right) $.

\item Using quantum It\^{o} formula the general QS \emph{evolution equation}
for a quantum stochastic density operator $\hat{\varrho}\left( t\right) $
was obtained in the form of quantum stochastic Master equation \cite{Be95a},%
\cite{Be96a} 
\begin{equation}
\mathrm{d}\hat{\varrho}\left( t\right) =\left( G_{\nu }^{\iota }\hat{\varrho}%
\left( t\right) G_{\kappa }^{\star \nu }-\hat{\varrho}\left( t\right) \delta
_{\kappa }^{\iota }\right) \mathrm{d}A_{\iota }^{\kappa },~\hat{\varrho}%
\left( 0\right) =\varrho  \label{eq QSmaster}
\end{equation}%
Here $\boldsymbol{G}=\boldsymbol{I}+\boldsymbol{L}$ is the germ of QS
evolution equation \textrm{d}$V_{t}=V_{t}L_{\nu }^{\mu }\mathrm{d}A_{\mu
}^{\nu }$ which is not assumed to be pseudo-unitary. In the case of
normalization condition $S_{\iota }^{-}{\normalsize S}_{+}^{\star \iota }=O$
in terms of\ left adjoint operators $S_{-\nu }^{\mu }=G_{-\mu }^{\nu \sharp
} $ with respect to a standard pairing $\left\langle \mathfrak{b}_{\star }|%
\mathfrak{b}\right\rangle $, this equation is called QS \emph{decoherence,
or entangling equation} for quantum states satisfying normalization $%
\left\langle \left( \hat{\varrho}\left( t\right) ,I\right) \right\rangle
_{\emptyset }=\left( \varrho ,\check{1}\right) $ with respect to the pairing 
$\left( \hat{\varrho}^{\ast },A\right) _{\emptyset }=\left\langle \hat{%
\varrho}\delta _{\emptyset }|A\delta _{\emptyset }\right\rangle \equiv
\left\langle \hat{\varrho}|A\right\rangle _{\emptyset }$ on the noise
algebra $\mathcal{A}$ given by the vacuum vector $\delta _{\emptyset
}=0^{\otimes }$. This is the most general QS equation preserving complete
positivity and normalization in this mean form. Denoting $K_{\iota
}^{-}=-G_{\iota }^{-}\equiv K_{\iota }$ such that $G_{+}^{\star \iota
}=-K_{\iota }^{\ast }$, this can be written \cite{Be95a},\cite{Be97} as 
\begin{equation*}
\mathrm{d}\hat{\varrho}\left( t\right) +2\Re \left[ K_{\iota }\hat{\varrho}%
\left( t\right) \mathrm{d}A_{-}^{\iota }\right] =\left( \sum_{j}G_{\kappa
}^{j}\hat{\varrho}\left( t\right) G_{\iota }^{j\ast }-\hat{\varrho}\left(
t\right) \delta _{\kappa }^{\iota }\right) \mathrm{d}A_{\iota }^{\kappa }.
\end{equation*}%
More explicitly this Belavkin equation can be written in terms of $K=K_{+},$ 
$L_{+}^{i}=G_{+}^{i}\equiv L^{i}$ such that $G_{+}^{j\ast }=L^{j\ast }$ as 
\begin{eqnarray*}
&&\mathrm{d}\hat{\varrho}\left( t\right) +\left( 2\Re \left[ K\hat{\varrho}%
\left( t\right) \right] -\sum_{j}L^{j}\hat{\varrho}\left( t\right) L^{j\ast
}\right) \mathrm{d}t \\
&=&\sum_{k}2\Re \left[ \left( \sum_{j}G_{k}^{j}\hat{\varrho}\left( t\right)
L^{j\ast }-K_{k}\hat{\varrho}\left( t\right) \right) \mathrm{d}A_{-}^{k}%
\right] \\
&&+\sum_{ik}\left( \sum_{j}G_{k}^{j}\hat{\varrho}\left( t\right)
G_{i}^{j\ast }-\hat{\varrho}\left( t\right) \delta _{k}^{i}\right) \mathrm{d}%
A_{i}^{k}.
\end{eqnarray*}%
The weak normalization condition can be written as $L+L^{\ast
}+\sum_{i}L_{i}L_{i}^{\ast }=0$ in terms of left adjoint $L,L_{i}$ to $%
-K,L^{i}$ such that $K=-L^{\sharp },L^{i}=L_{i}^{\sharp }$ ($K=-L^{\ast
},L^{i}=L_{i}^{\ast }$ in the case of trace pairing) for any number of $i$%
's, and arbitrary $K_{i}$, $G_{k}^{i}$, $i,k=1,\ldots ,d$. This is QS
generalization of Lindblad equation \cite{Lin76} given by the generator (\ref%
{eq Lindblad}) corresponding to the case $d=0$.
\end{enumerate}


\begin{thebibliography}{99}
\bibitem{AASDM02} M.~A. Armen, J.~K. Au, J.~K. Stockton, A.~C. Doherty, and
H.~Mabuchi, ``Adaptive homodyne measurement of optical phase,'' \emph{%
Phys.~Rev.~Lett.}, vol.~89, p. 133602, 2002.

\bibitem{GSDM03} J.~M. Geremia, J.~K. Stockton, A.~C. Doherty, and
H.~Mabuchi, ``Quantum {Kalman} filtering and the {Heisenberg} limit in
atomic magnetometry,'' \emph{Phys. Rev. Lett.}, vol.~91, p. 250801, 2003.

\bibitem{Bel83} V.~P. Belavkin, ``On the theory of controlling observable
quantum systems,'' \emph{Automatica and Remote Control}, vol.~44, no.~2, pp.
178--188, 1983.

\bibitem{Be85} ------, ``Reconstruction theorem for quantum stochastic
processes,'' \emph{Theoret Math Phys}, vol.~3, pp. 409--431, 1985.

\bibitem{Bel88} ------, ``Nondemolition stochastic calculus in {Fock} space
and nonlinear filtering and control in quantum systems,'' in \emph{%
Proceedings {XXIV} Karpacz winter school}, ser. Stochastic methods in
mathematics and physics, R.~Guelerak and W.~Karwowski, Eds.\hskip 1em plus
0.5em minus 0.4em\relax
World Scientific, Singapore, 1988, pp. 310--324.

\bibitem{Bel99} ------, ``Measurement, filtering and control in quantum open
dynamical systems,'' \emph{Rep on Math Phys}, vol.~43, no.~3, pp. 405--425,
1999.

\bibitem{DJJ01} A.~C. Doherty, K.~Jacobs, and G.~Jungman, ``Information,
disturbance and {Hamiltonian} quantum feedback control,'' \emph{Phys. Rev. A}%
, vol.~63, p. 062306, 2001.

\bibitem{BEB05} L.~M. Bouten, S.~C. Edwards, and V.~P. Belavkin, ``Bellman
equations for optimal feedback control of qubit states,'' \emph{J. Phys. B},
vol.~38, pp. 151--160, 2005.

\bibitem{Jac03} K.~Jacobs, ``How to project qubits faster using quantum
feedback,'' \emph{Phys. Rev. A.}, vol.~67, p. 030301(R), 2003.

\bibitem{Jac04} ------, ``Optimal feedback control for the rapid preparation
of a single qubit,'' in \emph{Proceedings of the SPIE}, vol. 5468, 2004, pp.
355--364.

\bibitem{Jam04} M.~R. James, ``Risk sensitive optimal control of quantum
systems,'' \emph{Phys. Rev. A.}, vol.~69, p. 032108, 2004.

\bibitem{Jam05} ------, ``A quantum {Langevin} formulation of risk sensitive
optimal contol,'' \emph{J. Opt. B: Quantum Semiclass. Opt.}, vol.~7, pp.
S198--S207, 2005, special issue on quantum control.

\bibitem{ADL02} C.~Ahn, A.~C. Doherty, and A.~J. Landahl, ``Continuous
quantum error correction via quantum feedback control,'' \emph{Phys.~Rev.~A}%
, vol.~65, p. 042301, 2002.

\bibitem{GrW04} M.~Gregoratti and R.~F. Werner, ``On quantum
error-correction by classical feedback in discrete time,'' \emph{J. Math.
Phys.}, vol.~45, pp. 2600--2612, 2004.

\bibitem{HSM05} R.~van Handel, J.~K. Stockton, and H.~Mabuchi, ``Feedback
control of quantum state reduction,'' \emph{IEEE Trans. Automat. Control},
vol.~50, no.~6, pp. 768--780, 2005.

\bibitem{DHJMT00} A.~C. Doherty, S.~Habib, K.~Jacobs, H.~Mabuchi, and S.~M.
Tan, ``Quantum feedback control and classical control theory,'' \emph{Phys.
Rev. A}, vol.~62, no. 012105, 2000.

\bibitem{Be78} V.~P. Belavkin, ``Optimal quantum filtration of markovian
signals,'' \emph{Problems of Control and Information Theory}, vol.~7, no.~5,
pp. 345--360, 1978.

\bibitem{Bel79} ------, ``Optimal measurement and control in quantum
dynamical systems,'' Copernicus University, Torun,'' Preprint Instytut
Fizyki, 1979.

\bibitem{Be80} ------, ``Optimal filtering of markov signals with quantum
white noise,'' \emph{Radio Eng Electron Physics}, vol.~25, pp. 1445--1453,
1980.

\bibitem{Be87a} ------, ``Non-demolition measurement and control in quantum
dynamical systems,'' in \emph{Proc of {C I S M} Seminar on Information
Complexity and Control in Quantum Physics, Udine 1985.}, A.~Blaquiere, Ed.%
\hskip 1em plus 0.5em minus 0.4em\relax Wien--New York: Springer--Verlag,
1987, pp. 311--329.

\bibitem{Be90a} ------, ``A stochastic posterior {S}chr\"odinger equation
for counting non-demolition measurement,'' \emph{Letters in Math Phys},
vol.~20, pp. 85--89, 1990.

\bibitem{Be90b} ------, ``A posterior stochastic equations for quantum
brownian motion,'' in \emph{Proc of the 1989 Conference on Stochastic
Methods in Experimental Sciences}.\hskip 1em plus 0.5em minus 0.4em\relax %
Singapore: World Scientific, 1990, pp. 26--42.

\bibitem{Be91d} ------, ``Nondemolition observation and a new stochastic
equation for quantum photon emission,'' in \emph{Proc of the Conf on
Foundations and Phylosophical Aspects of Physics}, Moscow, 1991.

\bibitem{Bel92b} ------, ``Quantum stochastic calculus and quantum nonlinear
filtering,'' \emph{Journal of Multivariate Analysis}, vol.~42, pp. 171--201,
1992.

\bibitem{BeS92} V.~P. Belavkin and P.~Staszewski, ``Nondemolition
observation of a free quantum particle,'' \emph{Phys. Rev. A}, vol.~45,
no.~3, pp. 1347--1356, 1992.

\bibitem{Be93d} V.~P. Belavkin, ``Quantum diffustion, measurement and
filtering.'' \emph{Probability Theory and its Application}, vol.~38, no.~4,
pp. 742--757, 1993.

\bibitem{Bel94} ------, ``Nondemolition principle of quantum measurement
theory,'' \emph{Foundation of Physics}, vol.~24, no.~5, pp. 685--714, 1994.

\bibitem{Bel92a} ------, ``Quantum continual measurements and a posteriori
collapse on {CCR},'' \emph{Commun. Math. Phys.}, vol. 146, pp. 611--635,
1992.

\bibitem{Be95b} ------, ``The interplay of classical and quantum
stochastics: Diffusion, measurement and filtering,'' in \emph{Chaos -- The
Interplay Between Stochastic and Deterministic Behaviour}, ser. Lecture
Notes in Physics.\hskip
1em plus 0.5em minus 0.4em\relax Springer, 1995, pp. 21--41.

\bibitem{Be95} ------, ``A dynamical theory of quantum continuous
measurement and spontaneous localization,'' \emph{Russian Journal of
Mathematical Physics}, vol.~3, no.~1, pp. 3--24, 1995.

\bibitem{Be_Mel96} V.~P. Belavkin and O.~Melsheimer, ``A stochastic
hamiltonian approach for quantum jumps, spontaneous localizations, and
continuous trajectories,'' \emph{Quantum Semiclass. Opt.}, vol.~8, pp.
167--187, 1996.

\bibitem{Bel89} V.~P. Belavkin, ``A new wave equation for a continuous
nondemolition measurement,'' \emph{Physics Letters A}, vol. 140, pp.
355--358, 1989.

\bibitem{Be90} ------, ``A posterior {S}chr\"odinger equation for continuous
non-demolition measurement,'' \emph{J of Math Phys}, vol.~31, no.~12, pp.
2930--2934, 1990.

\bibitem{Be89a} ------, ``A continuous counting observation and posterior
quantum dynamics,'' \emph{J Phys A Math Gen}, vol.~22, pp. L 1109--L 1114,
1989.

\bibitem{Be_Sta90} V.~P. Belavkin and P.~Staszewski, ``A continuous
observation of photon emission,'' \emph{Reports in Mathematical Physics},
vol.~29, pp. 213--225, 1990.

\bibitem{Be90e} V.~P. Belavkin, ``Stochastic equations of quantum
filtering,'' in \emph{Proc of 5th International Conference on Probability
Theory and Mathematical Statistics}, Vilnius, 1990.

\bibitem{BaB91} A.~Barchielli and V.~P. Belavkin, ``Measurements continuous
in time and a posteriori states in quantum mechanics,'' \emph{J. Phys. A},
vol.~24, pp. 1495--1514, 1991.

\bibitem{Be94e} V.~P. Belavkin, ``A stochastic model of quantum
observation,'' in \emph{Proc of Symposium on the Foundations of Modern
Physics}, P.~B. et~al, Ed.\hskip 1em plus 0.5em minus 0.4em\relax World
Scientific, 1994, pp. 38--54.

\bibitem{Be87} ------, ``Ordered *- semirings and generating functionals of
quantum statistics,'' \emph{Soviet Math. Dokl.}, vol.~35, no.~2, pp.
246--249, 1987.

\bibitem{Str62} R.~.L.Stratonovich, ``On the theory of optimal control.
sufficient coordinates.'' \emph{Automation and Remote Control.}, vol.~23,
no.~7, pp. 910--917, 1962.

\bibitem{Str68} R.~L. Stratonovich, \emph{Conditional {Markov} processes and
their application to the theory of optimal control}.\hskip 1em plus 0.5em
minus 0.4em\relax
American Elsevier Publishing Company, Inc, New-York, 1968.

\bibitem{Mor66} R.~E. Mortensen, ``Stochastic optimal control with noisy
observations,'' \emph{Int. J. Control}, vol.~4, pp. 455--464, 1966.

\bibitem{GBS05} J.~Gough, V.~P. Belavkin, and O.~G. Somolyanov, ``{%
Hamilton-Jacobi-Bellman} equations for quantum filtering and control,'' 
\emph{J. Opt. B: Quantum Semiclass. Opt.}, vol.~7, pp. S237--S244, 2005,
special issue on quantum control.

\bibitem{Bel80} V.~P. Belavkin, ``Quantum filtering of {Markov} signals with
white quantum noise,'' \emph{Radiotechnika i Electronika}, vol.~25, pp.
1445--1453, 1980, english translation in: Quantum communications and
measurement, Belavkin, V.P., Hiroto, O., Hudson, R.L., eds, Plenum Press,
1994, 381-392.

\bibitem{Acc_Fri_Lew82} {Accardi L.}, F.~A., and L.~J.T., ``Quantum
stochastic processes,'' \emph{Publ. of RIMS Kyoto Univ.}, vol.~18, no.~1,
pp. 97--133, 1982.

\bibitem{GaZ00} C.~Gardiner and P.~Zoller, \emph{Quantum Noise}.\hskip 1em
plus 0.5em minus 0.4em\relax Springer, Berlin, 2000.

\bibitem{Hol82} A.~S. Holevo, \emph{Probabilistic and Statistical Aspects of
Quantum Theory}.\hskip 1em plus 0.5em minus 0.4em\relax North Holland,
Amsterdam, 1982.

\bibitem{HuP84} R.~L. Hudson and K.~R. Parthasarathy, ``Quantum {It\^o's}
formula and stochastic evolutions,'' \emph{Commun. Math. Phys.}, vol.~93,
pp. 301--323, 1984.

\bibitem{von32} J.~von Neumann, \emph{Mathematische Grundlagen der
Quantenmechanik}.\hskip 1em plus 0.5em minus 0.4em\relax Springer, Berlin,
1932.

\bibitem{Par92} K.~R. Parthasarathy, \emph{An Introduction to Quantum
Stochastic Calculus}.\hskip 1em plus 0.5em minus 0.4em\relax Birkh\"auser,
Boston, 1992.

\bibitem{Be92a} V.~P. Belavkin, ``Kernel representations of *-semigroups
associated with infinitely divisible states,'' in \emph{Quantum Probability
and Related Topics}.\hskip 1em plus 0.5em minus 0.4em\relax World
Scientific, 1992, vol.~7, pp. 31--50.

\bibitem{Be92c} ------, ``Chaotic states and stochastic integrations in
quantum systems,'' \emph{Usp. Math Nauk (Russian Math Surveys)}, vol.~47,
pp. 47--106, 1992.

\bibitem{Be91} ------, ``A quantum nonadapted {I}to formula and stochastic
analysis in {F}ock scale,'' \emph{J of Funct Analysis}, vol. 102, no.~2, pp.
414--447, 1991.

\bibitem{Be88} ------, ``A new form and $\ast$-algebraic structure of
quantum stochastic integrals in fock space,'' in \emph{Rendiconti del
Seminario Matematico e Fisico di Milano LVIII}, 1988, pp. 177--193.

\bibitem{Be92} ------, ``Quantum stochastic calculus and quantum nonlinear
filtering,'' \emph{Journal of Multivariate Analysis}, vol.~42, no.~2, pp.
171--201, 1992.

\bibitem{GKS76} V.~Gorini, A.~Kossakowski, and E.~Sudarshan, ``Completely
positive dynamical semigroups of {N}-level systems,'' \emph{J. Math Phys.},
vol.~17, no.~5, pp. 821--825, 1976.

\bibitem{Lin76} G.~Lindblad, ``On the generators of quantum dynamical
semigroups,'' \emph{Commun. Math. Phys.}, vol.~48, pp. 119--130, 1976.

\bibitem{Be97} V.~P. Belavkin, ``Quantum stochastic positive evolutions:
Characterization, construction, dilation,'' \emph{Commun. Math. Phys.}, vol.
184, pp. 533--566, 1997.

\bibitem{Be95a} ------, ``On stochastic generators of completely positive
cocycles,'' \emph{Russ Journ of Math Phys}, vol.~3, no.~4, pp. 523--528,
1995.

\bibitem{Be96a} ------, ``On the general form of quantum stochastic
evolution equation,'' in \emph{Stochastic Analysis and Applications},
I.~M.~D. at~al, Ed.\hskip 1em plus 0.5em minus 0.4em\relax World Scientific,
1996, pp. 91--106.

\bibitem{Kal60} R.~E. Kalman, ``A new approach to linear filtering and
prediction problems,'' \emph{J. Basic Eng}, vol.~82, pp. 34--45, 1960.

\bibitem{KaB61} R.~E. Kalman and R.~S. Bucy, ``New results in linear
filtering and prediction theory,'' \emph{J. Basic Eng.}, pp. 95--108, 1961.

\bibitem{Be88a} V.~P. Belavkin, ``Non-demolition measurements, nonlinear
filtering and dynamic programming of quantum stochastic processes,'' in 
\emph{Proc of Bellmann Continuum Workshop `Modelling and Control of
Systems', Sophia--Antipolis 1988}, ser. Lecture notes in Control and Inform
Sciences, A.Blaquiere, Ed., vol. 121.\hskip 1em plus 0.5em minus 0.4em\relax %
Berlin--Heidelberg--New York--London--Paris--Tokyo: Springer--Verlag, 1988,
pp. 245--265.

\bibitem{Ben92} A.~Bensoussan, \emph{Stochastic Control of Partially
Observable Systems}.\hskip
1em plus 0.5em minus 0.4em\relax Cambridge University Press, 1992.

\bibitem{KuV86} P.~R. Kumar and P.~Varaiya, \emph{Stochastic Systems:
Estimation, Identification and Adaptive Control}.\hskip 1em plus 0.5em minus
0.4em\relax
Prentice-Hall, NJ, 1986.

\bibitem{Be_Sta89} V.~P. Belavkin and P.~Staszewski, ``A quantum particle
undergoing continuous observation,'' \emph{Phys Letters A}, vol. 140, pp.
359--362, 1989.

\bibitem{DoJ99} A.~C. Doherty and K.~Jacobs, ``Feedback-control of quantum
systems using continuous state-estimation,'' \emph{Phys. Rev. A}, vol. 129,
p. 419, 1999.

\bibitem{BoH05a} L.~M. Bouten and R.~van Handel, ``Quantum filtering: a
reference probability approach, {arXiv}:math-ph/0508006,'' 2005.

\bibitem{BoH05b} ------, ``On the separation principle of quantum control, {%
arXiv}:math-ph/0511021,'' 2005.

\bibitem{BHJ06} L.~M. Bouten, R.~van Handel, and M.~James, ``An introduction
to quantum filtering, {arXiv}:math.OC/0601741,'' 2006.

\bibitem{Dix69} J.~Dixmer, \emph{Les Algebres {D'Operateurs} Dans {L'Espace}
Hilbertien}.\hskip
1em plus 0.5em minus 0.4em\relax Gauthier-Villars, Paris, 1969.

\bibitem{Be96} V.~P. Belavkin, ``Positive definite germs of quantum
stochastic processes,'' \emph{C. R. Acad. Sci. Paris}, vol. 322, no.~1, pp.
385--390, 1996.
\end{thebibliography}
\end{document}